\documentclass[fleqn,usenatbib]{mnras}

\usepackage[T1]{fontenc}

\DeclareRobustCommand{\VAN}[3]{#2}
\let\VANthebibliography\thebibliography
\def\thebibliography{\DeclareRobustCommand{\VAN}[3]{##3}\VANthebibliography}

\usepackage{graphicx}	
\usepackage{amsmath}	
\usepackage{amssymb}	
\usepackage{multirow}

\usepackage{ulem}
\usepackage{xcolor}
\usepackage{soul}

\usepackage{newtxtext,newtxmath}

\title[CPD observations with JWST \& ELT]{Observability of Forming Planets and their Circumplanetary Disks IV. -- with JWST \& ELT}

\author[Chen \& Szul\'agyi]{
Xueqing Chen,$^{1}$\thanks{E-mail: xuechen@ethz.ch}
Judit Szul\'agyi,$^{1}$
\\
$^{1}$Institute for Particle Physics and Astrophysics, ETH Z\"urich, Wolfgang-Pauli-Strasse 27, CH-8093, Z\"urich, Switzerland}

\date{Accepted XXX. Received YYY; in original form ZZZ}

\pubyear{2021}

\begin{document}
\providecolor{added}{rgb}{0,0.15,0}
\providecolor{deleted}{rgb}{0.2,0,0}
\providecolor{addednew}{rgb}{0,0.55,0}
\providecolor{deletednew}{rgb}{0.85,0,0}
\newcommand{\add}[1]{{\color{black}{#1}}}
\newcommand{\del}[1]{{}}
\newcommand{\addnew}[1]{{\color{black}{#1}}}
\newcommand{\delnew}[1]{{}}

\label{firstpage}
\pagerange{\pageref{firstpage}--\pageref{lastpage}}
\maketitle

\begin{abstract}
To understand the potential for observing forming planets and their circumplanetary disks (CPDs) with \textit{JWST} and ELT, we created mock observations from 3D radiative hydrodynamic simulations and radiative transfer post-processing for planets with 10, 5, 1 Jupiter and 1 Saturn masses with orbital separation of 50 and 30 AU in 0$^{\circ}$, 30$^{\circ}$ and 60$^{\circ}$ inclinations. Instrumental effects were then simulated with \texttt{Mirage} for \textit{JWST}/NIRCam and NIRISS, \texttt{MIRISim} for \textit{JWST}/MIRI and \texttt{SimCADO} \& \texttt{SimMETIS} for ELT/MICADO and METIS. 
We found that the longer wavelengths (mid-IR and beyond) are the best to detect CPDs\add{, since they allow CPD of planet with smaller mass to} be detected. MIRI on \textit{JWST} and METIS on ELT offers the best possibility on these telescopes. Specifically, below 3 $\mu$m, only 10 $M_{\mathrm{Jup}}$ planets with their CPDs are detectable with NIRCam and MICADO. 5 $M_{\mathrm{Jup}}$ planets are only detectable if at 30 AU (i.e. closer) orbital separation. Planets above 5 $M_{\mathrm{Jup}}$ with their CPDs are detectable between 3-5 $\mu$m with NIRCam and METIS L/M band, or above 10 $\mu$m with MIRI and METIS N band. For $\leq$ 1 $M_{\mathrm{Jup}}$ planets > 15 $\mu$m are needed, where MIRI uniquely offers imaging capability.
We present magnitudes and spectral energy distributions for separate components of the planet+CPD+CSD system, to differentiate the extinction rates of CPDs and CSDs and to provide predictions for observational proposals. Because the CPD turns out to be the main absorber of the planet's emission, especially <10 $\mu$m, this makes the detection of forming planets quite challenging. 
\end{abstract}

\begin{keywords}
planets and satellites : detection -- hydrodynamics -- radiative transfer -- infrared: planetary systems -- instrumentation: high angular resolution
\end{keywords}



\section{Introduction}
Young, forming giant planets embedded in circumstellar disks (CSDs) form their own sub-disks around themselves, the so-called circumplanetary disks (CPDs). It is shown from 3D hydrodynamic simulations that the accretion from CSD to the CPD \del{are}\add{is} dominated by a vertical inflow \citep{Tanigawa2012,Szulagyi2014}. This high-velocity vertical inflow of gas creates a shock front on the CPD surface which can alter observations of forming planets and their CPDs \citep{SzulagyiMordasini2017, Szulagyi2020}. Because we cannot see the forming planet directly due to the surrounding circumplanetary material, detecting a forming planet is actually detecting its CPD or circumplanetary envelope \citep{Szulagyi2016}.

There are only a few detections of forming planet candidates and their circumplanetary material to date. In the LkCa 15 system a planet candidate was detected in near-IR by \citet{Kraus2012} and \del{confirmed} \add{in} H-${\alpha}$ emission \add{by} \citet{Sallum2015}. The  H-${\alpha}$ detection is debated, due to non-detections in the follow-up observations \citep{Mendigutia2018,Currie2019}. However, it is possible that due to line variability the H-${\alpha}$ signal disappears and might appear again. There were multiple planet candidates suggested around HD 169142 \citep{Biller2014, Reggiani2014,Osorio2014} from near-infrared imaging. In the PDS 70 system, the planet candidate, PDS 70 b, was discovered in multiple near IR bands \citep{Keppler2018,Muller2018}, followed by confirmation in H-$\alpha$ \citep{Wagner2018, Haffert2019}. A second planet candidate, PDS 70 c, was later discovered in H-$\alpha$ \citep{Haffert2019}. Around this planet, \citet{Isella2019} reported sub-mm continuum emission with the Atacama Large Millimeter/submillimeter Array (ALMA), making this the first likely detection of a dusty CPD around a forming planet. Evidence of a CPD around PDS 70 b was also observed in K band spectrum \citep{Christiaens2019}, via near-IR excess. In a follow-up observation by \citet{Benisty2021}, a higher resolution image of the CPD around PDS 70 c was reported. 

\add{The} observability of forming planets \del{have}\add{has} been studied with synthetic observations combining hydrodynamic simulations with post-processing Monte Carlo radiative transfer calculations. This method has been used to interpret existing observations and to make predictions for future observations.
Disk substructures, such as gaps, cavities, and spirals have been studied through mock observations in sub-mm \& radio with ALMA \citep{Gonzalez2012, Dong2015a, Dong2015b, Dipierro2015, Mayer2016, Zhang2018, Szulagyi2018}. Predictions were also made in near-IR scattered light for various current instruments such as Subaru/HiCIAO \citep{Ovelar2013}, VLT/SPHERE and GPI \citep{Szulagyi2021}, as well as  morphology studies, including the influence of inclination on the disk features \citep{Dong2016, Szulagyi2021}. Several previous works have focused on the potential CPD detections themselves, showing that CPDs can be detected in ALMA bands \citep{Perez2015, Zhu2016, Szulagyi2018} or near-IR in polarized scattered light \citep{Stolker2016,Szulagyi2021}. \citet{Sanchis2020} predicted the magnitudes of protoplanets with and without extinction by the circumplanetary material. The detectability of hydrogen recombination lines (such as H-$\alpha$) from accreting protoplanets \del{were}\add{was} studied by \citet{Aoyama2018, Szulagyi2020, Aoyama2021, Marleau2021}.

The spectral energy distribution (SED) of CPDs and accreting protoplanets have also been studied with analytical models and via hydrodynamical simulations in the past. \citet{Zhu2015} calculated the SEDs of viscous heating-dominated CPDs using atmospheric radiative transfer models and showed that moderately accreting CPDs can be much brighter than the planet itself. \citet{Eisner2015} reported a SED study of planets at various evolutionary stages. \citet{Szulagyi2019} derived SEDs of CPDs from radiative hydrodynamic simulations and identified that the best contrast between the CPD and CSDs for a priori unknown planetary masses can be found in the sub-mm and radio wavelengths. 

With the new generation telescopes, such as \textit{James Webb Space Telescope (JWST)} and the Extremely Large Telescope (ELT), the observations of forming planets and their CPDs will further improve. In the near-IR and mid-IR wavelength-regime, NIRCam \citep{Rieke2005nrc}, NIRISS \citep{Doyon2012nrs} and MIRI \citep{Rieke2015miri} will provide high-contrast imaging capabilities on \textit{JWST}, and  MICADO \citep{Davies2010micado} \& METIS \citep{Brandl2008metis} on the ELT.
During the design and testing phase of these instruments, studies have been made to predict their performance with various science goals, such as imaging circumstellar disks. \citet{Ertel2012} made predictions for imaging planet-disk interactions in debris disks with \textit{JWST}/MIRI using simulated MIRI point-spread function (PSF). In a study of the MIRI coronagraphs, \citet{Boccaletti2015} simulated an observation for the HD 181327 debris disk, showing that MIRI will more successfully reduce the stellar PSF residuals compared to HST in near-IR. \citet{Lebreton2016} modeled the $\eta$ Corvi debris disk and created synthetic images of the outer dust belt with the \textit{JWST}/MIRI coronagraph, and potential belt-shaping planets observations with the \textit{JWST}/NIRCam coronagraph. The above works focused on somewhat different disk observations than our science goal, which is the detectability of forming planets and their CPDs embedded in gaseous CSDs \add{(i.e. before the class III phase)}. With the start of operation of \textit{JWST}, it is very timely to study this problem with mock observations of various \textit{JWST} \& ELT instruments. 

In the previous papers in the series, the observability of forming planets and their CPDs were studied in sub-mm and radio wavelengths with ALMA \citep{Szulagyi2018}, with near-IR imaging and SEDs \citep{Szulagyi2019} for NaCo \& ERIS on VLT, in polarized scattered light for SPHERE and GPI \citep{Szulagyi2021}, and with hydrogen recombination lines \citep{Szulagyi2020}. In this paper, we study the observability of forming planets and their CPDs with NIRCam, NIRISS and MIRI on \textit{JWST} and MICADO+METIS on the ELT. 

\section{Methods}

The three-step method we used to create synthetic observations for \textit{JWST} and the ELT is described in this section. First, 3D radiative hydrodynamic simulations of circumstellar disks with embedded giant planets were run. Then, the simulation results were post-processed by the 3D radiative transfer code \textsc{radmc3d} to create wavelength-dependent intensity images. \del{Finally, i}Instrumental effects (such as PSF convolution\del{ and addition of noise}) were simulated by running the intensity images through specific instrument simulators for \textit{JWST} and the ELT. \add{Noise from the stellar photon was included by including the star in the radiative transfer calculation and telescope simulations, and the sky background noise and instrumental noise were added by the instrument simulators. For the analysis of the simulated images, noise subtraction and aperture photometry were performed to obtain predicted magnitudes for forming planet detection.} SEDs are also created to identify the wavelengths where \add{the} best contrast between CSD and CPD is obtained.

\subsection{Hydrodynamic Simulations}
\label{sec:hydro}
\add{We performed} radiative, 3D hydrodynamic simulations \del{in this paper are same as in the \citet{Szulagyi2019} and \citet{Szulagyi2021} of the paper series, but here we use additional ones too, to enlarge the parameter space. We }us\add{ing} the 3D grid-based code \textsc{jupiter}\del{ with nested meshes to resolve the planet vicinity}, which was developed by F. Masset and J. Szul\'agyi \citep{Szulagyi2016}. The code solves the hydrodynamic equations for mass, momentum and total energy conservation, as well as radiative transfer with the flux-limited diffusion approximation using the two-temperature approach \citep{Kley1989, Commerson2011}. The gas heating mechanisms include adiabatic compression, viscous heating, shock heating and stellar irradiation, while the cooling mechanisms are adiabatic expansion and radiative diffusion. The circumplanetary disk is mainly heated up by gas accretion onto the planet which leads to adiabatic compression. The accretion shock front on the surface of CPD can also be heated to high temperature when the planet is massive enough. Viscous heating in the CPD also plays a role, but with \add{the} low viscosity value chosen, it is one of the least important heating mechanisms. 

The simulations were performed in 7 separate runs where the forming planet was set to the mass of 10, 5, 1 $M_{\mathrm{Jup}}$ and 1 $M_{\mathrm{Sat}}$ at 50 AU from the star\add{ (same as in \citealt{Szulagyi2019} and \citealt{Szulagyi2021} of the paper series)}, and 10, 5, 1 $M_{\mathrm{Jup}}$ at 30 AU from the star. \add{The parameters of the simulations are summarized in Table \ref{tab:simus}.} The base grid extended radially between 0.4-2.4 times of the planet’s semi-major axis (30 or 50 AU), which was sampled in 215 cells\addnew{ in equidistant spacing}. Azimuthally the circumstellar disk had 680 cell resolution over 2$\pi$. Vertically, the CSD opening angle was 7.4 degrees from the midplane, sampled in 20 cells. However, this opening angle \del{were}\add{was} further extended in the radiative transfer post-processing step, see Sect. \ref{sec:rad}. Four levels of mesh refinements were used around the planet, each doubling the resolution such that the circumplanetary region was well resolved. \add{We stopped the simulations, when a steady state has been reached.}

\begin{table}
\caption{\add{Parameters of the 7 simulations used in this work.}}
\label{tab:simus}
\setlength{\tabcolsep}{0pt} 
\begin{tabular}{cccccc}
\hline
Simulation &
  \begin{tabular}[c]{@{}c@{}}$r_{p}$\\ \relax[AU]\end{tabular} &
  \begin{tabular}[c]{@{}c@{}}$M_{p}$\\ \relax[$M_{\mathrm{Jup}}$]\end{tabular} &
  \begin{tabular}[c]{@{}c@{}}Inner and outer\\ grid edge [AU]\end{tabular} &
  \begin{tabular}[c]{@{}c@{}}CSD mass\\ \relax[$M_\odot$]\end{tabular} &
  \begin{tabular}[c]{@{}c@{}}Viscosity\\ \relax[$a_{P}^{2}\Omega_{P}$]\end{tabular} \\ \hline
\texttt{10jup50au} & 50 & 10  & 20-120 & 10$^{-2}$ & 10$^{-5}$ \\
\texttt{5jup50au}  & 50 & 5   & 20-120 & 10$^{-2}$ & 10$^{-5}$ \\
\texttt{1jup50au}  & 50 & 1   & 20-120 & 10$^{-2}$ & 10$^{-5}$ \\
\texttt{1sat50au}  & 50 & 0.3 & 20-120 & 10$^{-2}$ & 10$^{-5}$ \\
\texttt{10jup30au} & 30 & 10  & 12-72  & 10$^{-2}$ & 10$^{-5}$ \\
\texttt{5jup30au}  & 30 & 5   & 12-72  & 10$^{-2}$ & 10$^{-5}$ \\
\texttt{1jup30au}  & 30 & 1   & 12-72  & 10$^{-2}$ & 10$^{-5}$ \\ \hline
\end{tabular}
\end{table}

The dust-to-gas ratio was set to a constant \add{of} 0.01. Although not explicitly treated in these hydrodynamic simulations, the effect of dust \addnew{on the temperature of the disk} was taken into account through the dust opacity table. \delnew{, which was taken from \citet{Szulagyi2017}.} \addnew{The Rossland and Planck mean opacities in the code were constructed from dust opacities computed by the Mie code of \citet{Bohren1984}, where the dust consists of 40\% silicate, 40\% water 20\% carbonaceous material. To account for dust evaporation, the opacity for cells above the carbon evaporation temperature (2000 K) are given by the gas opacities, which were taken from \citet{BellLin1994}.} The stellar properties were chosen to be Solar values. The circumstellar disk had a mass of 10$^{-2}$ $M_\odot$ and an initial surface density slope of -0.5. The mean molecular weight was chosen to be 2.3 according to solar abundances. A constant kinematic viscosity of 10$^{-5}$ $a_{p}^{2}\Omega_{p}$ was used, where $a_{p}$ was the planet semi-major axis and $\Omega_{p}$ its orbital frequency.

\subsection{Radiative transfer post-processing}
\label{sec:rad}
To produce intensity images from the hydrodynamic simulations, the \textsc{radmc-3d}\footnote{\url{http://www.ita.uni-heidelberg.de/~dullemond/software/radmc-3d/}} radiative transfer was used, developed by Cornelis Dullemond \citep{Dullemond2012}. Images at different wavelengths and resolution\add{s} were created to match the filters and pixel scale of each instrument\addnew{, considering the radiation from both thermal re-emission and scattering}. \delnew{Flux conservation was taken care of regardless of the image resolution with the command \texttt{fluxconv}.} We studied 0$^\circ$, 30$^\circ$ and 60$^\circ$ CSD inclinations. To reduce the photon noise from the CPD's shock front \citep{SzulagyiMordasini2017, Szulagyi2020} in the 10- and 5- Jupiter mass cases, every result image was averaged from 5 repeated runs with varying randomized seed number\add{s}. 

Since we are dealing with micron-sized, well-coupled dust in these short wavelengths, the dust density was created from gas density\del{with} assuming a dust-to-gas ratio of 0.01, consistent with the hydrodynamic simulations dust-to-gas value. The dust temperature was also taken from the hydrodynamic simulations using the gas temperature and assuming thermal equilibrium between dust and gas, i.e. $T_{dust} = T_{gas}$. This is a realistic assumption since it accounts for the shock heating and accretion heating in the CPD region, which cannot be captured by ray tracing temperature calculation methods such as the thermal Monte-Carlo provided by \textsc{radmc-3d} \citep{Szulagyi2021}. Dust evaporation was included by setting dust density to 0 for regions with temperatures above 1500 K (the silicate sublimation temperature).

In the near IR, scattering of photons off dust grains in the disk atmosphere is an important process affecting the disk brightness. Because the hydrodynamic simulations could only simulate a modest disk opening angle of 7.4$^\circ$,  we extended the disk vertically. For this step, the density fields were fitted with a Gaussian function to each vertical cell column and extrapolated from the original 40 cells to 100 cells with this fitted Gaussian function. For the temperature field vertical extension, we took the temperature of the last cell (vertically) and kept it constant in the extended CSD atmosphere. This upper, optically thin layer is heated by\del{ the} stellar irradiation, which was included in the hydrodynamic simulations.

We used $10^{7}$ photons for the scattering Monte-Carlo runs with \textsc{radmc-3d}. \delnew{A scattering mode = 3 was used, which corresponds to }\addnew{We used }anisotropic scattering with a tabulated phase function. In the opacity table, the wavelength-dependent absorption and scattering opacities $\kappa_{abs}$ and $\kappa_{scat}$, and scattering matrix elements $Z_{ij}$ were calculated by the Mie code from \citet{Bohren1984}. The dust species were assumed to be 70\% silicates and 30\% carbon, with grain sizes ranging from 0.1 to 1000 microns with a -3.5 power-law index. 

All fluxes in the images were scaled to 100 pc distance. Spectral Energy Distributions (SEDs) were also created for each simulations\delnew{ using the \texttt{sed} command in \textsc{radmc-3d}}. The initial parameters for the SED runs were kept the same as those for intensity images, except that the scattering photon number was set to 10$^{4}$ due to the high computational cost for the SED runs. The photon noise was removed by replacing a data point $n$ with the average of data points $n-1$ and $n+1$ if data point $n$ is larger than 1.5 times of point $n-1$ or $n+1$. 

In addition to the nominal case where the synthetic images and SEDs were calculated on the entire system (i.e. planet + CPD + CSD), we tested several cases where we cropped out different components of the system (CSD, or CPD, or planet itself). These runs were done to calculate the brightness contribution of each component and to be able to compare them in order to calculate the extinction by the CSD and CPD separately. In summary, these cases were investigated:
\begin{enumerate}
  \item \textit{CSD only} -  density and temperature inside the CPD region were set to 0, i.e. the CPD region \del{were}\add{was} cut out from the CSD
  \item  \textit{CPD + planet only} - density and temperature outside the CPD region were set to 0, i.e. the CSD was cropped off from the CPD region
  \item  \textit{planet only} - density and temperature outside the planet surface were set to 0, i.e. only the planet present, without any disks around it to absorb the planet's emission (note: the planet emission was coming purely from the radiative hydrodynamic simulations, not by planet interior modeling)
  \item  \textit{CPD + planet + perfect absorber CSD} - only temperature outside the CPD region was set to 0
  \item  \textit{planet + perfect absorber CPD} - \add{the} temperature outside planet surface was set to 0, as well as the density outside CPD was set to 0 (CSD cropped off)
\end{enumerate}
In the above cuts of the temperature/density fields, the CPD was defined as a cylinder with both radius and height equal to 0.5 $r_{Hill}$. The choice of this value is consistent with previous hydrodynamic simulations, which showed that CPDs have a radius of 0.3-0.5 $r_{Hill}$ \citep{Tanigawa2012, Szulagyi2014}. The planet radius was defined at 0.002 $r_{Hill}$, again informed by the hydrodynamic simulations. For the planet's emission, we did not use planet interior modeling, the temperature at the planet location was calculated by the radiative module of the hydrodynamic code. However, it is important to report the magnitude values we got from this hydrodynamic simulation generated planet, in order to calculate the extinction rate of the CPD self-consistently (mag$_{\mathrm{cpd+planet}}$-mag$_{\mathrm{planet}}$). Coincidentally perhaps, but the effective temperatures of our planets are not far off from planet interior models calculated effective temperatures of young planets (see Sect. \ref{sec:sed}).  

\subsection{Telescope/instrument simulators}

\subsubsection{\textit{JWST} instruments}
The 6.5m \textit{James Webb Space Telescope (JWST)} \citep{Gardner2006} is going to operate in the near-mid infrared with unprecedented sensitivity, offering great potential for imaging circumstellar disk substructures, circumplanetary disks with forming planets with its three cameras. The Near-InfraRed Camera (NIRCam; \citealt{Rieke2005nrc}), the workhorse imager in near-IR, will cover 0.6–2.3 $\mu$m with \add{a} pixel size of 0.031" in the short-wavelength mode and 2.4–5.0 $\mu$m with pixel size 0.063" in the long-wavelength mode. The Mid-InfraRed Imager (MIRI; \citealt{Rieke2015miri}) will work in 5-28 $\mu$m with pixel size 0.11". The Near-InfraRed Imager and Slitless
Spectrograph (NIRISS; \citealt{Doyon2012nrs}) will have a direct imaging mode, but only designated to be used in parallel operation. The NIRISS aperture masking interferometry (AMI) mode \citep{Sivaramakrishnan2010nrsami} will offer high spatial resolution imaging at  2.8, 3.8, 4.3, and 4.8 $\mu$m with \add{a} zoomed field of view at separations between $\sim $70–400 mas. \textit{JWST} offers three options for high-contrast imaging: NIRCam \& MIRI coronagraphs, NIRISS AMI, or non-coronagraphic PSF subtraction with any instrument. 

In this paper, we simulated the performance of NIRCam, MIRI and NIRISS AMI. Since coronagraphy simulation for user input images is not supported in the current public \textit{JWST} simulators, we only used the non-coronagraphic imaging modes of NIRCam and MIRI. The effect of applying a coronagraph was mimicked in our mock images by masking the brightness of the star to 1\% of \add{its} original value before inputting the intensity image into the telescope simulators. 

\textbf{\textit{JWST/NIRCam.}} The \texttt{Mirage}\footnote{\url{https://www.stsci.edu/jwst/science-planning/proposal-planning-toolbox/mirage}} package \citep{Hilbert2019} was used for simulating NIRCam images. The \textit{JWST} Astronomer’s Proposal Tool (APT)\footnote{\url{https://www.stsci.edu/scientific-community/software/astronomers-proposal-tool-apt}} was used to generate the input files for \texttt{Mirage}. The \textsc{radmc-3d} intensity images were inputted using the extended source catalog. \delnew{For NIRCam, we used the B2 module for the short-wavelength channel and \add{the} B5 module for the long-wavelength channel.} Five filters spanning NIRCam's operation range were chosen: F115W, F150W, F210M, F360M and F480M. \del{The SUB160 subarray was chosen to reduce saturation. }\add{We used the \texttt{SHALLOW4} readout pattern with 10 groups per integration and a total of 10 integrations, which leads to an integration time of 139 seconds. The background was set to medium and the cosmic ray level is set to \texttt{SUNMIN}. During testing it is found that for very massive planets, the source can already saturate the FULL array at the end of the first group in each integration, so the SUB160 subarray was adopted to reduce saturation, as recommended by the \textit{JWST} user guide. The combination of detector subarray, readout pattern and integration time was chosen to balance between the saturation limit and signal-to-noise level.} 

\textbf{\textit{JWST/NIRISS.}} NIRISS imaging is not recommended as a prime observing mode and is designed to be used in parallel when \del{other}\add{another} \textit{JWST} instrument is in operation. It does not come with a coronagraph or subarray, which makes direct imaging with NIRISS impossible for our science case. On the other hand, the aperture masking interferometry (AMI) mode is suited for this task. We again used \texttt{Mirage} for NIRISS AMI mode simulation. All four available filters are simulated: F277W, F380M, F430M and F480M. \add{In APT we set detector subarray to SUB80 and readout pattern to NISRAPID. We simulated for 10 integrations, each with 50 groups, leading to a total exposure time of 37.7 seconds.} The PSF reference observation step was omitted since this is highly target-dependent. It was also found during testing that the \textit{JWST} reduction pipeline currently does not handle correctly some planet-star mass combinations during the AMI processing step.

\textbf{\textit{JWST/MIRI.}} The \texttt{MIRISim}\footnote{\url{https://wiki.miricle.org/Public/MIRISimPublicRelease2dot4}} package \citep{Klaassen2021} was used for simulating MIRI images. To meet the flux conservation requirements of \texttt{MIRISim}, cube images were created with \textsc{radmc-3d} with an extra spectral dimension. Three filters were simulated: F1000W, F1500W and F2100W. \add{We used a detector subarray SUB256 and readout pattern \texttt{FAST} (During testing it was found that a 10 $M_{\mathrm{Jup}}$ planet can already saturate the FULL array at the end of first group). We set the background level to \texttt{low} with gradient = 5 and pa = 45. We used 50 groups per integration with 10 integrations in total, which resulted in an integration time of 150 seconds.}

The output images of \texttt{Mirage} and \texttt{MIRISim} resembled the uncalibrated raw data from \textit{JWST} detectors. They were run through the first two stages of the \textit{JWST} calibration pipeline (\texttt{calwebb\_detector1} and \texttt{calwebb\_image2}) for basic data reduction including ramp fitting, photometric calibration and background subtraction. 

\subsubsection{ELT instruments}
The Extremely Large Telescope (ELT), currently under construction in Cerro Armazones, Chile, will become the world’s largest optical/infrared telescope with its approx. 39 m diameter. Its unprecedented spatial and spectral resolution with the aid of AO systems makes it the perfect platform for ground-based imaging of embedded forming planets. The Multi-adaptive optics Imaging CamerA for Deep Observations (MICADO; \citealt{Davies2010micado}) will work in the near-IR, covering 0.8 - 2.45 $\mu$m with a zoomed field-of-view of 18"$\times$18" and 1.5 mas pixel scale. The Mid-infrared ELT Imager and Spectrograph (METIS; \citealt{Brandl2008metis}) will image in two channels (L/M band at 3-5 $\mu$m and N band at 7.5–13.5 $\mu$m) with $\sim$10" field-of-view and 5-7 mas pixel scale.

\textbf{\textit{ELT/MICADO and METIS.}} We used \texttt{SimCADO}\footnote{\url{https://simcado.readthedocs.io/en/latest/}} \citep{Leschinski2019} and \texttt{SimMetis}\footnote{\url{https://metis.strw.leidenuniv.nl/simmetis/}} for simulating MICADO and METIS images. Coronagraphy is not supported by the two simulators, so to mimic that, the stellar flux was reduced to 1\% on the intensity images before inputting into the simulator. We simulated the filters J, H and Ks for MICADO. For \texttt{SimCADO} to run, a spectrum of the source covering the full bandwidth of the simulated filter is created from \textsc{radmc-3d} as input along with the intensity image. \delnew{We set the imager mode to \texttt{zoom} and detector layout to \texttt{centre}.} The number of integration was set to 1 and the integration time was set to 3600 seconds. For METIS, we simulated images for the filters L, M$_{p}$ and N$_{2}$. \delnew{We set \texttt{small\_FOV=True} for the detector.} The number of integration was set to 1 and the integration time was set to 7200 seconds. The other simulation parameters were adopted from the default configuration.

Since there is no existing calibration pipeline for ELT instruments, the flux of each image was converted from ADU (Analogue-to-Digital Unit) to Jansky with\footnote{\url{https://pixinsight.com/doc/tools/FluxCalibration/FluxCalibration.html}}

\begin{equation}\label{eq:fluxconv}
    \qquad F_{\lambda, \mathrm{ADU}}=F{_{\lambda, \mathrm{Jy}} \cdot \frac{1}{E_{ph}} \cdot \pi  (\frac{D}{2})^2 \cdot \Delta \lambda \cdot QE \cdot \frac{1}{G}}
\end{equation}

where $D$ is the diameter of telescope primary mirror, $E_{ph}$ is photon energy, $\Delta \lambda$ is \add{the} filter bandwidth, $QE$ is the quantum efficiency of the instrument and $G$ is sensor gain. \add{The values for $\Delta\lambda$, $QE$ and $G$ were obtained from the filter transmission data files and detector data files in the \texttt{SimCADO} and \texttt{SimMETIS} packages for the corresponding filters. The values used for flux conversion for the two ELT instruments are summarized in Table \ref{tab:phot_specs}. In the final step, the background rate was evaluated over a patch on the corner for each image and} subtracted from the image. 

\begin{table}
\centering
\caption{\add{ELT Instrument parameters for flux conversion.}}
\label{tab:phot_specs}
\begin{tabular}{cccccc}
\hline
\textbf{}           instrument    & filter & $\Delta \lambda$ {[}$\mu$m{]} & QE   & Gain                 & $D$ {[}m{]}           \\ \hline
\multirow{3}{*}{MICADO} & J      & 0.195                         & 0.88 & \multirow{3}{*}{1.0} & \multirow{3}{*}{39} \\
                        & H      & 0.29                          & 0.90 &                      &                     \\
                        & Ks     & 0.35                          & 0.85 &                      &                     \\ \hline
\multirow{3}{*}{METIS}  & L      & 0.6                           & 0.86 & \multirow{3}{*}{1.0} & \multirow{3}{*}{39} \\
                        & Mp     & 0.5                           & 0.71 &                      &                     \\
                        & N2     & 2.95                          & 0.80 &                      &                     \\ \hline
\end{tabular}
\end{table}

\subsubsection{Flux determination}

The brightness at the planet location in the simulated images \del{were}\add{was} obtained with aperture photometry. 
An aperture centered at the planet location was defined with \add{a} diameter equal to 2-3 times the FWHM of the instrument PSF, this way encircling 50\% - 70\% of the PSF energy but not touching the PSF core of the central star. 
The CPD brightness was determined by integrating the flux over this aperture. Since the CSD background significantly contributes to the measured CPD flux, we also determined the CSD flux at 180$^{\circ}$ azimuthally from the planet, i.e. the anti-planet location. Using the same aperture, we integrated the flux of the CSD at this anti-planet location, to be subtracted from the reported CPD brightness. This step also removed any stellar PSF extended to the location of CPD. \add{By such a method, we assume that the flux at the planet and anti-planet location are equal. This assumption valid since we found during testing that the flux difference between the two locations in the CSD is usually within 4\%, which is much lower than the uncertainty level caused by other steps in our method.}

We calculated Signal-to-noise (SNR) values as the ratio between CPD brightness (without subtracting anti-planet contribution) and the anti-planet brightness of the CSD. (Notice the difference \add{in} how SNR is calculated in our method from the usual definition of SNR, which is the signal with respect to the general background noise. The reason for the difference is \del{because}\add{that} CPD is part of the CSD, and CPD detection actually means detecting its signal on top of the CSD background. This makes CPD detection more difficult than the usual object vs background noise detection since we are dealing with a CPD vs. CSD contrast problem.)
Depending on our SNR value, we marked each simulated image as detection or non-detection. For cases where detection is not clear, but only disk features caused by \add{the} planet can be seen, we marked it as an "asymmetry". In those cases advanced PSF subtraction techniques would be needed to determine if there is detection at the CPD location. Strictly speaking, the asymmetry cases are also non-detections, since a disk feature would not necessarily \add{be} caused by planets.

The fluxes were converted to apparent magnitudes at 100 pc. The zero magnitude fluxes used for conversion are: F$_{0}$ = 1587 Jy for J band/F115W, 1074 Jy for H/F150W, 653 Jy for Ks/F210M, 253 Jy for L/F360M, 150 Jy for Mp/F480M, 34.9 Jy for N/F1000W, 18 Jy for F1500W and 8 Jy for F2100W\footnote{\url{https://www.gemini.edu/observing/resources/magnitudes-and-fluxes/conversions-between-magnitudes-and-flux}}.


\section{Results}

\subsection{Simulated images}

\begin{figure*}
  \includegraphics[width=\textwidth]{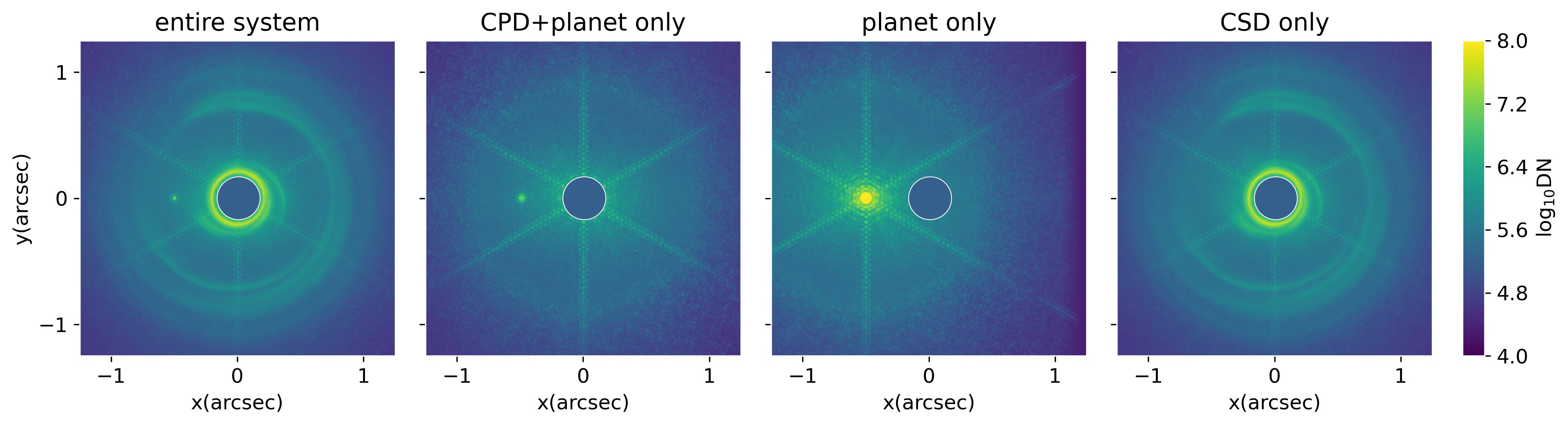}
  \caption{ELT/MICADO Ks band synthetic images of the four extinction cases projected to 100 pc. The planet is 5 M$_{\mathrm{Jup}}$ and located at 30 AU from the star at 9 o’clock direction on the images. Although the central star is included in the \textsc{radmc-3d} and telescope simulations, it is masked out in the final figure with a circle to allow focus on the CPD region. The 4 panels in this figure correspond to cases: entire system (planet + CPD + CSD); CPD + planet only; planet only case; CSD only (CPD region cropped out from the CSD). Although the images \del{is}\add{are} shown for one example simulation only, the behavior is similar for the other simulations and instruments.}
  \label{fig:cutouts}
\end{figure*}

Fig. \ref{fig:cutouts} illustrates an example of the synthetic images separately for the entire system, for the CPD+planet case, the planet-only case and the CSD only case described in section \ref{sec:rad}. The images were done in \add{the} J band of ELT/MICADO. In this case, the planet  was 5 $M_{\mathrm{Jup}}$ and located at 30 AU. It can be seen that the CSD absorbs the CPD flux, and slightly reduce the planet brightness when compared to the CPD+planet only case (second image from left). Comparing the first and third panels in Fig \ref{fig:cutouts}, it can be seen that the CPD strongly absorbs the planet photons and the planet would be much brighter if not suffering extinction by the CPD. While this is just one example image, this behavior is generic for all cases.

\subsubsection{\textit{JWST}/NIRCam}\label{sec:nrc}

\begin{figure*}
  \includegraphics[width=\textwidth]{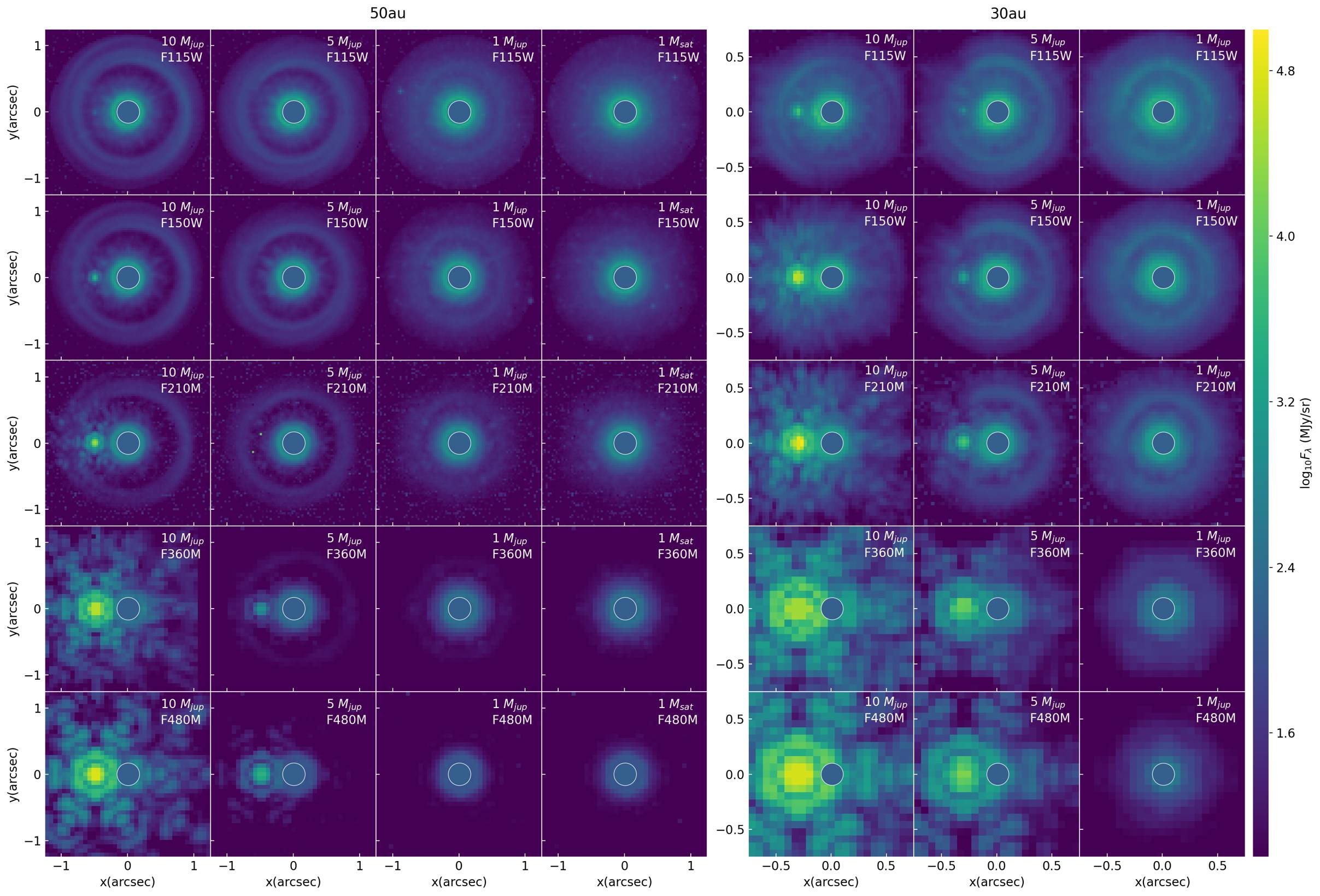}
  \caption{\textit{JWST}/NIRCam synthetic images of face-on systems projected to 100 pc. The columns represent hydrodynamic simulations with different planetary masses (10, 5, 1 $M_{\mathrm{Jup}}$ and 1 $M_{\mathrm{Sat}}$), and the rows represent the 5 filters simulated. The planet is located at 50 AU (\textit{left}) or 30 AU (\textit{right}) from the star at 9 o’clock direction on the images. Although the central star is included in the \textsc{radmc-3d} and telescope simulations, it is masked out in the final figure with a circle to allow focus on the CPD region. 10 $M_{\mathrm{Jup}}$ planets with their CPDs are detected in all 5 filters. 5 $M_{\mathrm{Jup}}$ planets with their CPDs are detected in all 5 filters if at 30 AU separation, while at 50 AU they are only detected in filters above 3 $\mu$m. }
  \label{fig:nrc0}
\end{figure*}

\begin{table*}
\caption{\textit{JWST}/NIRCam predicted magnitudes at 100 pc in 0$^{\circ}$, 30$^{\circ}$, 60$^{\circ}$ inclinations. The 4 column groups correspond to cases: \textit{CPD in the entire system} - planet + CPD + CSD; \textit{CPD+planet only brightness}; \textit{planet only brightness}; \textit{CSD only brightness} - CPD region cropped out from the CSD. The magnitude values for the first 3 groups (CPD in the entire system, CPD+planet only, planet only) are given by flux within an aperture at planet location minus the flux within a same-sized aperture 180$^{\circ}$ from the planet (i.e. anti-planet location). The magnitude values for the last group (CSD only) are given by the flux within a same-sized aperture at the anti-planet location. Non-detections are marked with a '/'. For saturated cases, only upper limits of magnitude are reported.}
\label{tab:nrc_full}
\begin{tabular}{ccccccccccc@{\quad \vline \quad}ccc}
\hline
\textbf{model} &
  \textbf{filter} &
  \multicolumn{3}{c}{\textbf{\begin{tabular}[c]{@{}c@{}}CPD brightness in the \\ entire system [mag]\end{tabular}}} &
  \multicolumn{3}{c}{\textbf{\begin{tabular}[c]{@{}c@{}}CPD+planet only \\ brightness [mag]\end{tabular}}} &
  \multicolumn{3}{c|}{\textbf{\begin{tabular}[c]{@{}c@{}}planet only brightness\\\relax [mag]\end{tabular}}}\quad &
  \multicolumn{3}{c}{\textbf{\begin{tabular}[c]{@{}c@{}}CSD only brightness\\\relax [mag]\end{tabular}}} \\ \hline
\multicolumn{2}{c}{\textit{inclination}} &
  0$^{\circ}$ &
  30$^{\circ}$ &
  60$^{\circ}$ &
  0$^{\circ}$ &
  30$^{\circ}$ &
  60$^{\circ}$ &
  0$^{\circ}$ &
  30$^{\circ}$ &
  60$^{\circ}$ &
  0$^{\circ}$ &
  30$^{\circ}$ &
  60$^{\circ}$ \\
\multirow{5}{*}{10jup50au} & F115W & 20.58  & 19.56  & /     & 18.99  & 18.16  & 18.99 & <12.77 & <12.75 & <13.23 & 21.18 & 21.05 & 20.25 \\
                           & F150W & 16.11  & 16.70  & /     & 16.14  & 15.63  & 18.58 & <11.74 & <11.71 & <12.29 & 20.44 & 20.27 & 19.50 \\
                           & F210M & 13.00  & 13.42  & 17.42 & 13.51  & 13.98  & 17.28 & <9.69  & <9.72  & <9.92  & 19.72 & 19.59 & 18.80 \\
                           & F360M & <9.96  & 10.55  & 13.27 & <10.01 & <10.76 & 13.88 & <9.66  & <9.62  & <9.49  & 17.60 & 17.46 & 16.93 \\
                           & F480M & <8.60  & 9.44   & 12.45 & <8.88  & 9.02   & 12.40 & <7.64  & <7.66  & <7.82  & 16.94 & 16.79 & 16.14 \\[1ex]
\multirow{5}{*}{5jup50au}  & F115W & /      & /      & /     & 19.67  & 19.52  & 19.49 & 17.86  & 17.40  & 17.38  & 21.11 & 20.93 & 20.12 \\
                           & F150W & /      & /      & /     & 19.19  & 19.06  & 18.97 & 14.07  & 13.87  & 13.87  & 20.36 & 20.19 & 19.34 \\
                           & F210M & /      & /      & /     & 19.04  & 18.90  & 18.95 & 11.91  & 12.13  & 12.03  & 19.64 & 19.47 & 18.61 \\
                           & F360M & 14.29  & 15.65  & /     & 13.65  & 14.60  & 18.69 & <9.78  & <9.92  & <9.91  & 17.43 & 17.28 & 16.67 \\
                           & F480M & 11.89  & 12.93  & /     & 11.99  & 12.62  & 17.19 & <8.05  & <8.10  & <8.10  & 16.79 & 16.63 & 15.87 \\[1ex]
\multirow{5}{*}{1jup50au}  & F115W & /      & /      & /     & 20.65  & 20.87  & 20.38 & 21.23  & 21.28  & 22.05  & 20.64 & 20.47 & 19.69 \\
                           & F150W & /      & /      & /     & 20.03  & 20.35  & 20.07 & 16.90  & 16.91  & 17.53  & 19.87 & 19.62 & 18.86 \\
                           & F210M & /      & /      & /     & 20.35  & 19.58  & 20.09 & 13.84  & 13.83  & 14.35  & 19.08 & 18.91 & 18.09 \\
                           & F360M & /      & /      & /     & /      & /      & /     & <10.71 & <10.72 & 10.66  & 16.97 & 16.79 & 16.13 \\
                           & F480M & /      & /      & /     & /      & /      & /     & 9.18   & 9.20   & 9.44   & 16.37 & 16.22 & 15.46 \\[1ex]
\multirow{5}{*}{1sat50au}  & F115W & /      & /      & /     & 21.75  & /      & 22.29 & /      & /      & /      & 20.34 & 20.12 & 19.28 \\
                           & F150W & /      & /      & /     & /      & 21.46  & /     & /      & /      & /      & 19.54 & 19.29 & 18.15 \\
                           & F210M & /      & /      & /     & /      & /      & 19.85 & /      & /      & /      & 18.83 & 18.63 & 17.78 \\
                           & F360M & /      & /      & /     & /      & /      & /     & /      & /      & /      & 16.83 & 16.65 & 15.93 \\
                           & F480M & /      & /      & /     & /      & /      & /     & /      & /      & /      & 16.28 & 16.08 & 15.31 \\[1ex]
\multirow{5}{*}{10jup30au} & F115W & 16.48  & 16.55  & 19.23 & 15.97  & 16.38  & 18.80 & <12.48 & <12.47 & <12.95 & 19.99 & 19.87 & 19.31 \\
                           & F150W & 13.06  & 14.18  & 17.55 & 13.62  & 13.47  & 16.93 & <11.70 & <11.65 & <11.66 & 19.21 & 19.07 & 18.60 \\
                           & F210M & 11.56  & 11.91  & 14.74 & 11.77  & 12.06  & 15.33 & <9.79  & <9.79  & <9.86  & 18.25 & 18.12 & 17.67 \\
                           & F360M & <10.06 & <10.44 & 11.88 & <9.89  & <10.34 & 12.79 & <9.70  & <9.66  & <9.74  & 15.83 & 15.69 & 15.43 \\
                           & F480M & <8.31  & <8.65  & 11.15 & <8.39  & <8.52  & 10.90 & <7.90  & <7.87  & <7.94  & 15.46 & 15.34 & 14.97 \\[1ex]
\multirow{5}{*}{5jup30au}  & F115W & 19.37  & 19.75  & /     & 18.84  & 18.79  & 19.37 & <13.23 & <13.27 & <13.71 & 19.84 & 19.68 & 18.96 \\
                           & F150W & 16.89  & 17.53  & /     & 16.93  & 17.47  & 19.06 & <11.73 & <11.62 & <11.67 & 19.06 & 18.90 & 18.15 \\
                           & F210M & 14.54  & 15.73  & /     & 14.77  & 15.47  & 18.11 & <10.11 & <10.14 & <10.39 & 18.03 & 17.88 & 17.18 \\
                           & F360M & 11.40  & 11.78  & /     & 11.32  & 11.46  & 14.94 & <9.65  & <9.70  & <9.69  & 15.70 & 15.53 & 15.03 \\
                           & F480M & 10.24  & 10.09  & 12.88 & 10.02  & 10.16  & 14.08 & <8.69  & <8.55  & <8.80  & 15.32 & 15.15 & 14.49 \\[1ex]
\multirow{5}{*}{1jup30au}  & F115W & /      & /      & /     & /      & 20.74  & 19.72 & 14.68  & 14.61  & 16.03  & 19.46 & 19.23 & 18.40 \\
                           & F150W & /      & /      & /     & /      & /      & /     & <12.50 & <12.50 & 13.41  & 18.60 & 18.43 & 17.53 \\
                           & F210M & /      & /      & /     & /      & /      & /     & <10.68 & <10.71 & 12.15  & 17.67 & 17.46 & 16.60 \\
                           & F360M & /      & /      & /     & /      & /      & /     & <9.88  & <9.85  & <9.85  & 15.55 & 15.34 & 14.66 \\
                           & F480M & /      & /      & /     & /      & /      & /     & <8.59  & <8.57  & <8.57  & 15.13 & 14.91 & 14.11 \\ \hline

\end{tabular}
\end{table*}

\begin{figure}
  \includegraphics[width=\columnwidth]{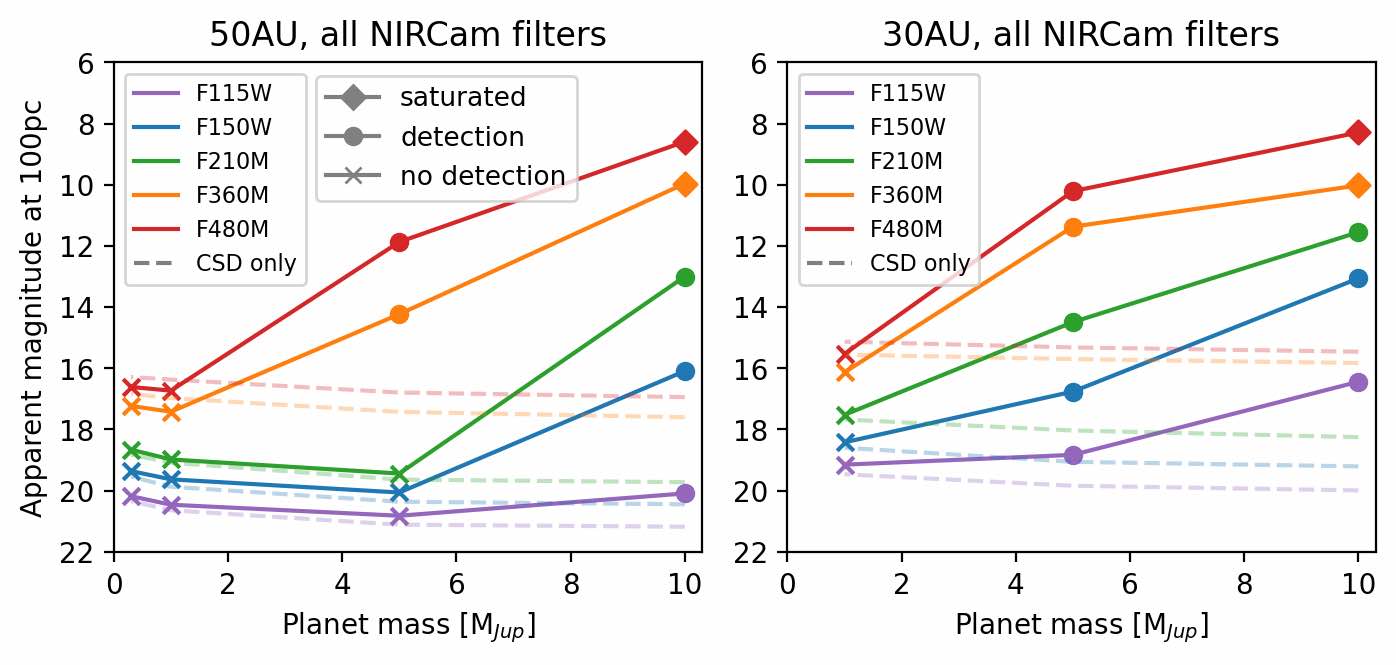}
  \caption{Apparent magnitudes at 100 pc for the different planetary masses and\del{ for} the 5 filters in NIRCam. \textit{Left}: 50 AU orbital separation, \textit{right}: 30 AU separation. The dashed lines represent the magnitude of the circumstellar disk background in the corresponding filter. A circle symbol represents detection, and a cross symbol represents non-detection. A diamond symbol indicates detection but the CPD flux has saturated the detector. The CPD brightness scales with higher planetary mass and longer wavelengths, and saturation is seen for 10 $M_{\mathrm{Jup}}$ planets at the two longest wavelengths.}
  \label{fig:nrc_mass_mag}
\end{figure}

\begin{figure}
  \includegraphics[width=\columnwidth]{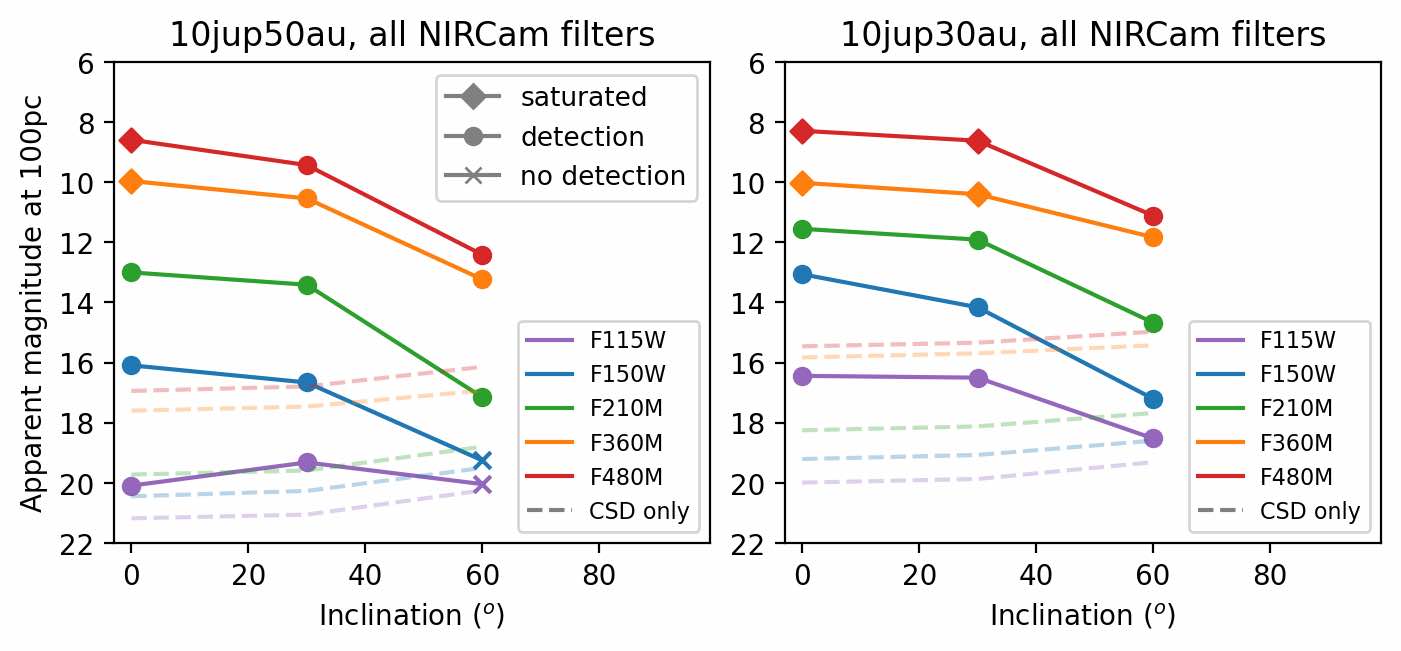}
  \caption{Apparent magnitude at 100 pc for the 10 $M_{\mathrm{Jup}}$ system with different inclinations for the 5 filters in NIRCam. \textit{Left}: 50 AU separation, \textit{right}: 30 AU separation. The dashed lines represent the magnitude of the circumstellar disk background in the corresponding filter. A circle symbol represents detection, and a cross symbol represents non-detection. A diamond symbol indicates detection but the CPD flux has saturated the detector. The CPD region brightness decreases with higher inclinations, so systems with < 30$^{\circ}$ are easiest to detect.}
  \label{fig:nrc_inc_mag}
\end{figure}

\begin{figure}
\centering
  \includegraphics[width=0.8\columnwidth]{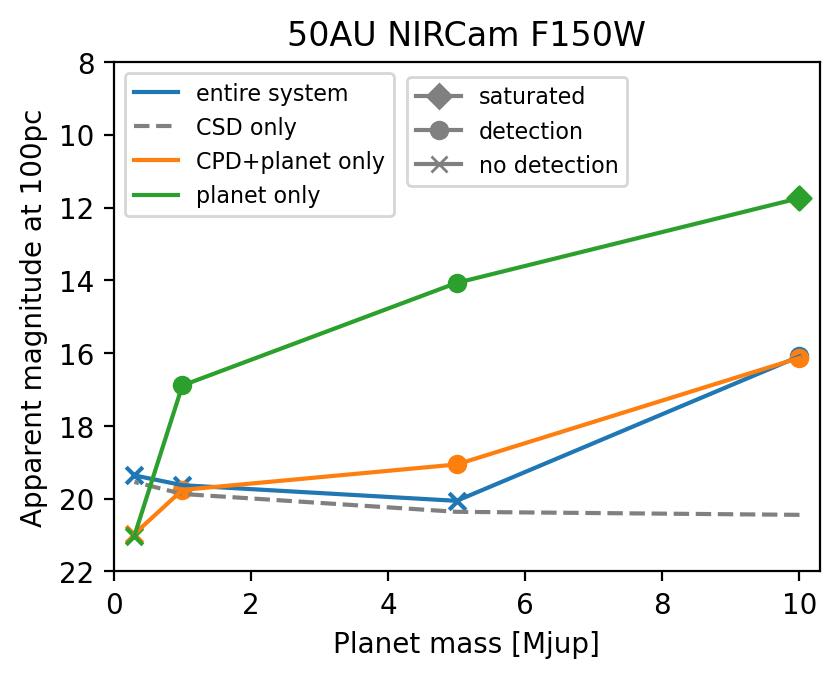}
  \caption{Apparent magnitude at 100 pc in NIRCam F150W filter with different planetary masses for 4 extinction cases: entire system; CPD+planet only; planet only; CSD only. A circle symbol represents detection, and a cross symbol represents non-detection. A diamond symbol indicates detection but the CPD flux has saturated the detector. This plot indicates that the main absorber of planet's emission is the CPD.}
  \label{fig:nrc_mass_mag_cases}
\end{figure}

The simulated images of NIRCam filters F115W, F150W, F210M, F360M and F480M in 0$^{\circ}$ inclination at 100 pc are presented in Fig. \ref{fig:nrc0}, while the 30$^{\circ}$ and 60$^{\circ}$ inclination images are in Appendix \ref{sec:app} (Figs. \ref{fig:nrc30} \& \ref{fig:nrc60}). The magnitude of the CPD in the entire system, CPD+planet only, planet only and CSD only for all three inclinations are presented in Table \ref{tab:nrc_full}. It can be seen from Figs. \ref{fig:nrc0} and Table \ref{tab:nrc_full} that a 10 $M_{\mathrm{Jup}}$ planet with its CPD can be detected in all 5 NIRCam filters. A 5 $M_{\mathrm{Jup}}$ planet and its CPD can also be detected in all 5 filters if the planet is located at 30 AU from the star, but they can only be detected in filters above 3 $\mu$m if the orbital separation is 50 AU. 

There are two reasons contributing to the detectability of the 5 and 10 $M_{\mathrm{Jup}}$ planets and their CPDs. First, only massive planets can open a deep gap in the CSD, which makes the planetary gap region optically thin outside the Hill-sphere. Smaller mass planets that do not open gas gaps in CSDs are subject to higher extinction. This can be seen in Table \ref{tab:nrc_full}, comparing the "CPD brightness in the entire system" and "CPD+planet only" columns. The second reason why higher mass planets can be more easily detected with their CPDs is \del{because}\add{that} the accretion shock front formed on the CPD surface is hotter and more luminous for the 5 and 10 $M_{\mathrm{Jup}}$ cases than for the less massive ones \citep{SzulagyiMordasini2017, Szulagyi2020}.

We compared the CPD brightness with increasing planetary mass for all NIRCam filters in Fig. \ref{fig:nrc_mass_mag}. The brightness of the CSD backgrounds \del{were}\add{was} plotted in dashed lines as a reference, since our science case is to detect the CPD on top of the background CSD emission (i.e. the contrast between the two disks). Although we reported the CSD-subtracted CPD brightness in Table \ref{tab:nrc_full}, in Fig. \ref{fig:nrc_mass_mag} we plotted the CPD brightness before CSD subtraction (which are slightly different from Table \ref{tab:nrc_full}) in order to compare the contrast between the two disks. In Fig. \ref{fig:nrc_mass_mag}, a data point marked with a circle symbol represents detection, and a cross symbol represents non-detection. A diamond symbol indicates detection but the CPD flux has saturated the detector with the used integration times. As seen in Fig. \ref{fig:nrc_mass_mag}, all detection points are well above the CSD background. Fig. \ref{fig:nrc_mass_mag} shows that the observed CPD flux increases with planetary mass, but not in a linear relation. This is due to the presence of the CPD (+ CSD) around the forming planet; their absorption and emission depend on the different disk characteristics and the disk features created by the different planetary masses (e.g. how deep gas gap the planet is opening in the CSD). This adds to the complexity of detecting forming planets and their CPDs. The CSD background (dashed lines in Fig. \ref{fig:nrc_mass_mag}) slightly decreases with planetary mass, which indicates the deeper planetary gap opening with increasing planet mass. Since the CSD background was defined in an aperture at the anti-planet location, this region is located in the planetary gap. From Fig. \ref{fig:nrc0} it can be seen that for 5 and 10 $M_{\mathrm{Jup}}$ planets, a deep gas gap is carved by the massive planets. In the case of 1 $M_{\mathrm{Jup}}$ and 1 $M_{\mathrm{Sat}}$ planets, the gap is shallower, almost not visible on the images.

We also studied the detectability of CPD with various inclinations of the planetary system, in order to determine how the inclination of the CSD affects the detectability of CPDs. Fig. \ref{fig:nrc_inc_mag} shows the change of CPD brightness with increasing inclination for a 10 $M_{\mathrm{Jup}}$ planet at 50 AU and 30 AU. In most filters, the trend shows an expected decrease of CPD brightness due to extinction by the circumstellar disk. Naturally, to detect the CPD the face-on systems are the best, however, the <30 degrees inclination systems are all well detectable. The higher inclinations have a more significant extinction from the CSD material, therefore reducing their observability. 

The orbital separation of the planet also affects its observability. It can be seen in Figs. \ref{fig:nrc0}, \ref{fig:nrc_mass_mag} and \ref{fig:nrc_inc_mag} that planet brightness goes up with smaller orbital separation. Comparing 50 AU observations to 30 AU, the latter results in a higher planet brightness in all cases, due to somewhat higher CPD temperature. 

Moreover, we compared the brightness of CPD in the entire system, the CPD+planet only brightness, the planet only case brightness. Fig. \ref{fig:nrc_mass_mag_cases} shows the variation of brightness with increasing planetary mass for these three cases, while the CSD only brightness is shown as a dashed line. The difference between the blue and orange curves in Fig. \ref{fig:nrc_mass_mag_cases} shows that the CSD absorbs the planet photons, especially at 5 $M_{\mathrm{Jup}}$. At 10 $M_{\mathrm{Jup}}$ the extinction is not that obvious, which indicates that gap opening is more efficient. For 1 $M_{\mathrm{Sat}}$ and 1 $M_{\mathrm{Jup}}$ planets, the CPD alone is even fainter than the CSD, so it is not detectable in the CSD. The difference between \add{the} green and \add{the} orange curves shows that the CPD strongly absorbs the planet emission for planets above 1 $M_{\mathrm{Jup}}$. 

The planet itself is very hot during its forming phase, therefore it is very bright. From Table \ref{tab:nrc_full} it can be seen that from our hydrodynamic simulations a 1 $M_{\mathrm{Sat}}$ planet (or below) is so faint in near-IR wavelengths that it is not detectable even if there is no absorption by surrounding material. While the brightness of the planet would be different with using a planet interior and evolution model, hence also the planetary mass which would be detectable or not, we simply wanted to show the amount of absorption by the CPD, hence we report the planet only magnitudes as well. Proper interior modeling is needed to estimate the forming planet's brightness in the first $\sim$ 3 Myrs.

The CPD brightness and its contrast against the background CSD both increase with longer NIRCam filters, as can be seen in Figs. \ref{fig:nrc0}, \ref{fig:nrc_mass_mag} and \ref{fig:nrc_inc_mag}. Therefore, the best wavelengths to detect CPDs with NIRCam are the longer wavelength channels. However, there is always \del{the}\add{a} balance between the CPD signal and resolution, since the long-wavelength channels of NIRCam have reduced pixel scale compared to the short-wavelength channels, and \del{that }the PSF FWHM also increases with longer wavelengths. For massive planets at close separation (e.g. 10 $M_{\mathrm{Jup}}$ at 30 AU), short-wavelength channels could work better. In addition, it should be noted that a 10 $M_{\mathrm{Jup}}$ planet with its CPD can already saturate the NIRCam SUB160 subarray in wavelengths above 3 $\mu$m (for these cases, only lower limits of the magnitude values are reported in Table \ref{tab:nrc_full} with a '<' symbol in front). Thus, our recommendation is to use detector subarrays to avoid saturation when the forming planet is expected to have a mass above $\sim$ 10 $M_{\mathrm{Jup}}$. Due to the limitation of \texttt{Mirage} simulator, our results \del{does}\add{do} not realistically include the effect of a coronagraph. In future observation planning, \add{a} real coronagraphy simulator should be used.


\subsubsection{\textit{JWST}/NIRISS Aperture Masking Interferometry}\label{sec:nrsami}

\begin{figure*}
  \includegraphics[width=\textwidth]{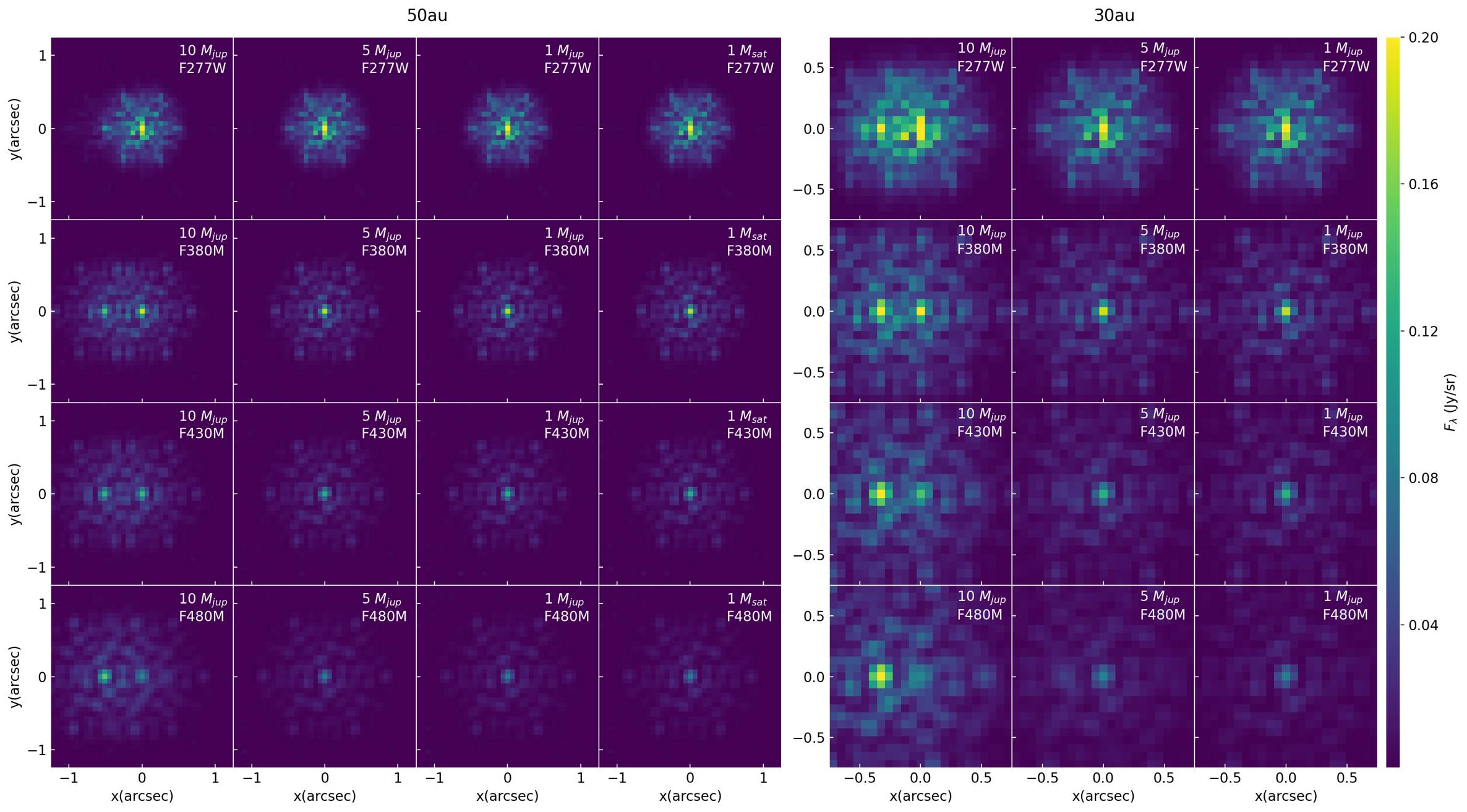}
  \caption{\textit{JWST}/NIRISS AMI synthetic images of face-on systems projected to 100 pc. The columns represent hydrodynamic simulations with different planetary masses (10, 5, 1 $M_{\mathrm{Jup}}$ and 1 $M_{\mathrm{Sat}}$), and the rows represent the 5 filters simulated. The planet is located at 50 AU (\textit{left}) or 30 AU (\textit{right}) from the star at 9 o’clock direction on the images. 10 $M_{\mathrm{Jup}}$ planets with their CPDs are detected in all filters except the case at 50 AU separation in F277W. 5 $M_{\mathrm{Jup}}$ planets with their CPDs are only detected at 30 AU in F480M.}
  \label{fig:nrsami0}
\end{figure*}

\begin{table*}
\caption{\textit{JWST}/NIRISS AMI predicted magnitudes at 100 pc for CPD in the entire system, CPD+planet only, planet only and CSD only in 0$^{\circ}$, 30$^{\circ}$, 60$^{\circ}$ inclinations. The magnitude values for the first 3 groups (CPD in the entire system, CPD+planet only, planet only) are given by flux within an aperture at planet location minus the flux within a same-sized aperture 180$^{\circ}$ from the planet (i.e. anti-planet location). The magnitude values for the last group (CSD only) are given by the flux within a same-sized aperture at the anti-planet location. Non-detections are marked with a '/'. For saturated cases, only upper limits of magnitude are reported.}
\label{tab:nrsami_full}
\begin{tabular}{ccccccccccc@{\quad \vline \quad}ccc}
\hline
\textbf{model} &
  \textbf{filter} &
  \multicolumn{3}{c}{\textbf{\begin{tabular}[c]{@{}c@{}}CPD brightness in the\\ entire system [mag]\end{tabular}}} &
  \multicolumn{3}{c}{\textbf{\begin{tabular}[c]{@{}c@{}}CPD+planet only\\ brightness [mag]\end{tabular}}} &
  \multicolumn{3}{c}{\textbf{\begin{tabular}[c]{@{}c@{}}planet only brightness\\\relax [mag]\end{tabular}}}\quad  &
  \multicolumn{3}{c}{\textbf{\begin{tabular}[c]{@{}c@{}}CSD only brightness\\\relax [mag]\end{tabular}}} \\ \hline
\multicolumn{2}{c}{\textit{inclination}} &
  0$^{\circ}$ &
  30$^{\circ}$ &
  60$^{\circ}$ &
  0$^{\circ}$ &
  30$^{\circ}$ &
  60$^{\circ}$ &
  0$^{\circ}$ &
  30$^{\circ}$ &
  60$^{\circ}$ &
  0$^{\circ}$ &
  30$^{\circ}$ &
  60$^{\circ}$ \\
\multirow{4}{*}{10jup50au} & F277W & /    & /     & / & /    & /    & / & <7.70 & <7.63 & <8.01 & 11.07 & 11.07 & 11.07 \\
                           & F380M & 9.09 & 9.95  & / & 9.22 & 9.81 & / & 5.47  & 5.44  & 5.92  & 10.29 & 10.29 & 10.27 \\
                           & F430M & 8.96 & 10.08 & / & 9.26 & 9.33 & / & 5.24  & 5.21  & 5.68  & 10.14 & 10.14 & 10.12 \\
                           & F480M & 8.28 & 9.29  & / & 8.58 & 8.87 & / & 5.11  & 5.05  & 5.54  & 10.24 & 10.23 & 10.22 \\[1ex]
\multirow{4}{*}{5jup50au}  & F277W & /    & /     & / & /    & /    & / & 10.16 & 10.19 & 10.23 & 11.07 & 11.07 & 11.06 \\
                           & F380M & /    & /     & / & /    & /    & / & 7.83  & 7.94  & 7.93  & 10.29 & 10.29 & 10.26 \\
                           & F430M & /    & /     & / & /    & /    & / & 7.44  & 7.54  & 7.53  & 10.14 & 10.14 & 10.12 \\
                           & F480M & /    & /     & / & /    & /    & / & 7.17  & 7.25  & 7.26  & 10.23 & 10.23 & 10.22 \\[1ex]
\multirow{4}{*}{1jup50au}  & F277W & /    & /     & / & /    & /    & / & /     & /     & /     & 11.07 & 11.07 & 11.06 \\
                           & F380M & /    & /     & / & /    & /    & / & 9.77  & 9.77  & 10.04 & 10.29 & 10.28 & 10.26 \\
                           & F430M & /    & /     & / & /    & /    & / & 9.49  & 9.50  & 9.82  & 10.14 & 10.14 & 10.12 \\
                           & F480M & /    & /     & / & /    & /    & / & 9.04  & 9.05  & 9.29  & 10.23 & 10.23 & 10.21 \\[1ex]
\multirow{4}{*}{1sat50au}  & F277W & /    & /     & / & /    & /    & / & /     & /     & /     & 11.07 & 11.07 & 11.06 \\
                           & F380M & /    & /     & / & /    & /    & / & /     & /     & /     & 10.29 & 10.28 & 10.26 \\
                           & F430M & /    & /     & / & /    & /    & / & /     & /     & /     & 10.14 & 10.13 & 10.11 \\
                           & F480M & /    & /     & / & /    & /    & / & /     & /     & /     & 10.23 & 10.22 & 10.21 \\[1ex]
\multirow{4}{*}{10jup30au} &
  F277W &
  9.83 &
  10.61 &
  / &
  10.39 &
  10.50 &
  / &
  <7.66 &
  <7.61 &
  <7.70 &
  10.37 &
  10.37 &
  10.36 \\
                           & F380M & 8.87 & 9.59  & / & 9.04 & 9.44 & / & 6.29  & 6.27  & 6.33  & 9.67  & 9.66  & 9.66  \\
                           & F430M & 8.44 & 8.95  & / & 8.55 & 8.57 & / & 6.05  & 6.04  & 6.08  & 9.70  & 9.70  & 9.69  \\
                           & F480M & 7.91 & 8.45  & / & 7.96 & 8.16 & / & 5.88  & 5.85  & 5.90  & 9.69  & 9.69  & 9.68  \\[1ex]
\multirow{4}{*}{5jup30au}  & F277W & /    & /     & / & /    & /    & / & <8.46 & <8.46 & <8.57 & 10.37 & 10.37 & 10.35 \\
                           & F380M & /    & /     & / & /    & /    & / & 7.10  & 6.94  & 7.18  & 9.66  & 9.66  & 9.65  \\
                           & F430M & /    & /     & / & /    & /    & / & 6.88  & 6.76  & 6.96  & 9.70  & 9.69  & 9.69  \\
                           & F480M & 9.99 & /     & / & 9.94 & /    & / & 6.67  & 6.55  & 6.76  & 9.69  & 9.69  & 9.68  \\[1ex]
\multirow{4}{*}{1jup30au}  & F277W & /    & /     & / & /    & /    & / & 9.93  & 9.94  & 9.90  & 10.37 & 10.37 & 10.35 \\
                           & F380M & /    & /     & / & /    & /    & / & 9.04  & 9.01  & 9.01  & 9.66  & 9.66  & 9.64  \\
                           & F430M & /    & /     & / & /    & /    & / & 8.73  & 8.74  & 8.72  & 9.70  & 9.70  & 9.68  \\
                           & F480M & /    & /     & / & /    & /    & / & 8.38  & 8.39  & 8.37  & 9.69  & 9.68  & 9.66   \\ \hline
\end{tabular}
\end{table*}

\begin{figure}
  \includegraphics[width=\columnwidth]{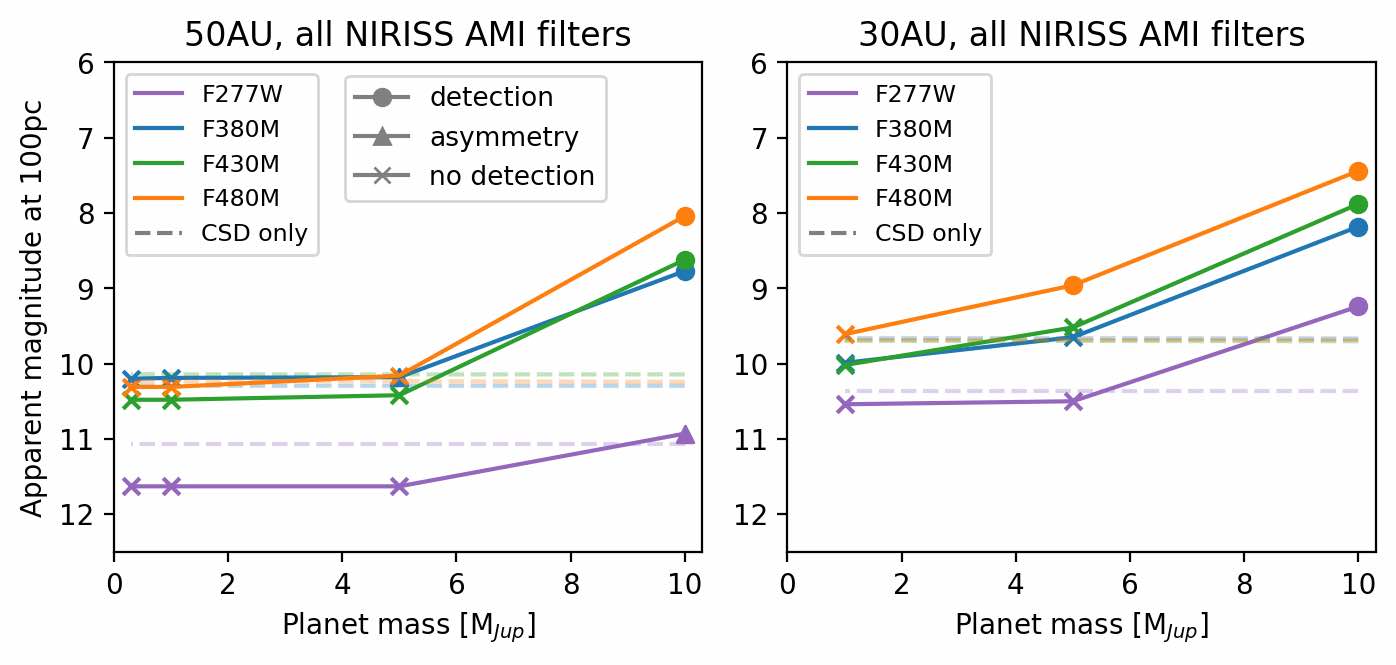}
  \caption{Apparent magnitudes at 100 pc for the different planetary masses and for the 4 filters in NIRISS AMI mode. \textit{Left}: 50 AU orbital separation, \textit{right}: 30 AU separation. The dashed lines represent the magnitude of the circumstellar disk background in the corresponding filter. A circle symbol represents detection, and a cross symbol represents non-detection. A triangle symbol indicates that an asymmetry of flux is observed, but the SNR is not high enough to count as a detection. PSF subtraction is not included in the images. Only 10 $M_{\mathrm{Jup}}$ planets and the 5 $M_{\mathrm{Jup}}$ planet at 30 AU in F480M are detected.}
  \label{fig:nrsami_mass_mag}
\end{figure}

The simulated images of NIRISS AMI filters F277W, F380M, F430M and F480M in 0$^{\circ}$ inclination at 100 pc are presented in Fig. \ref{fig:nrsami0}, while the 30$^{\circ}$ and 60$^{\circ}$ inclination images are in Appendix \ref{sec:app} (Figs. \ref{fig:nrsami30} \& \ref{fig:nrsami60}). The trend for CPD brightness with increasing planetary mass for all NIRISS AMI filters is compared in Fig. \ref{fig:nrsami_mass_mag}. It can be seen from Figs. \ref{fig:nrsami0} and \ref{fig:nrsami_mass_mag} that at orbital separation of 50 AU, only 10 $M_{\mathrm{Jup}}$ planets with their CPDs can be detected in the 3 long wavelength filters of NIRISS AMI. At 30 AU separation, 10 $M_{\mathrm{Jup}}$ planets with theirs CPDs can be detected in all 4 filters, and 5 $M_{\mathrm{Jup}}$ planets with theirs CPDs can only be detected in F480M. 

We also studied the brightness of the CPD in the entire system, the CPD+planet only brightness without the CSD, planet only brightness and CSD only brightness for varying inclinations in Table \ref{tab:nrsami_full}. The change of CPD brightness with increasing inclinations follows the same trend as the NIRCam observations, as can be seen from Table \ref{tab:nrsami_full}: The CPD region brightness decreases with higher inclinations, due to higher extinction by the CSD. At 60$^{\circ}$, even 10 $M_{\mathrm{Jup}}$ planets are not detectable. The comparison between different extinction cases in Table \ref{tab:nrsami_full} also shows a similar result as NIRCam observations: The CSD does not absorb significantly for massive planets. The CPD acts as the main absorber for the planet's photons. 

It should be noted that the stellar PSF significantly contributes to the circumstellar brightness of the simulated AMI images, making their CSD-only magnitude values a few orders smaller in Table \ref{tab:nrsami_full} compared to other direct imaging instruments. This is because to mimic the effect of coronagraphy for other instruments, we reduced the stellar brightness to 1\% before inputting the intensity image to the instrument simulators. However, this step was not included in \add{the }NIRISS AMI simulation, since it will not include a coronagraph. Moreover, because \del{that }we calculated the CSD-only flux at the anti-planet location, the CSD-only brightness might exceed the CPD brightness in the entire system in some cases (e.g. the F277W curves in Fig. \ref{fig:nrsami_mass_mag}), but one should be reminded that this is only a result from the asymmetric stellar PSF. We stress again that in real observations, the science target observation should be accompanied by a reference star PSF observation for proper PSF subtraction during AMI data reduction, which is essential to access the full potential of NIRISS AMI imaging. Despite this, one can already see by comparing Fig. \ref{fig:nrc0} and Fig. \ref{fig:nrsami0} that NIRISS AMI imaging has \add{a} significant advantage over regular direct imaging for resolving closely separated planets with its sharper PSF.


\subsubsection{\textit{JWST}/MIRI}\label{sec:miri}

\begin{figure*}
  \includegraphics[width=\textwidth]{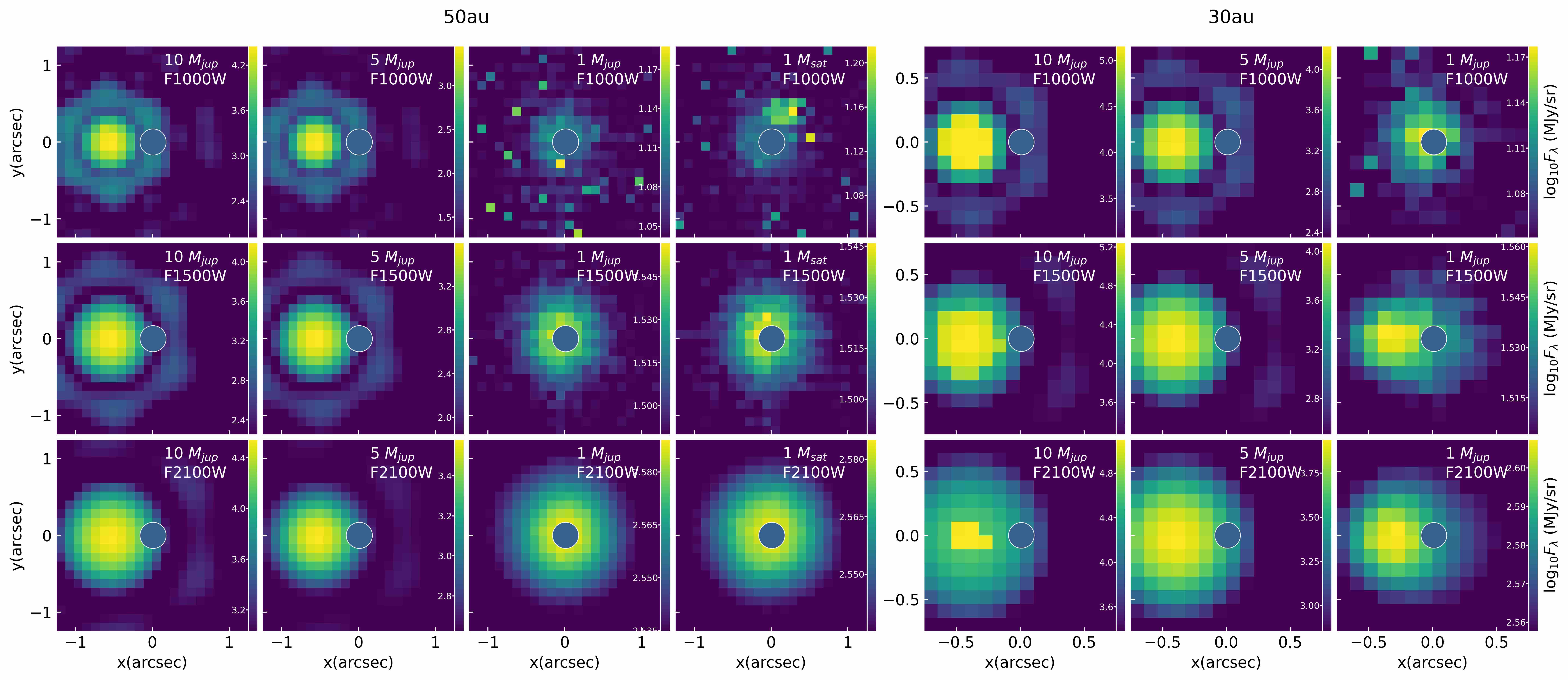}
  \caption{\textit{JWST}/MIRI synthetic images of face-on systems projected to 100 pc. The columns represent hydrodynamic simulations with different planetary masses (10, 5, 1 $M_{\mathrm{Jup}}$ and 1 $M_{\mathrm{Sat}}$), and the rows represent the 3 filters simulated. The planet is located at 50 AU (\textit{left}) or 30 AU (\textit{right}) from the star at 9 o’clock direction on the images. Although the central star is included in the \textsc{radmc-3d} and telescope simulations, it is masked out in the final figure with a circle to allow focus on the CPD region. 10 and 5 $M_{\mathrm{Jup}}$ planets with their CPDs are detected in all filters.}
  \label{fig:miri0}
\end{figure*}

\begin{table*}
\caption{\textit{JWST}/MIRI predicted magnitudes at 100 pc for CPD in the entire system, CPD+planet only, planet only and CSD only in 0$^{\circ}$, 30$^{\circ}$, 60$^{\circ}$ inclinations. The magnitude values for the first 3 groups (CPD in the entire system, CPD+planet only, planet only) are given by flux within an aperture at planet location minus the flux within a same-sized aperture 180$^{\circ}$ from the planet (i.e. anti-planet location). The magnitude values for the last group (CSD only) are given by the flux within a same-sized aperture at the anti-planet location. Non-detections are marked with a '/'. For saturated cases, only upper limits of magnitude are reported.}
\label{tab:miri_full}
\begin{tabular}{ccccccccccc@{\quad \vline \quad}ccc}
\hline
\textbf{model} &
  \textbf{filter} &
  \multicolumn{3}{c}{\textbf{\begin{tabular}[c]{@{}c@{}}CPD brightness in the\\ entire system [mag]\end{tabular}}} &
  \multicolumn{3}{c}{\textbf{\begin{tabular}[c]{@{}c@{}}CPD+planet only\\ brightness [mag]\end{tabular}}} &
  \multicolumn{3}{c}{\textbf{\begin{tabular}[c]{@{}c@{}}planet only brightness\\\relax [mag]\end{tabular}}}\quad  &
  \multicolumn{3}{c}{\textbf{\begin{tabular}[c]{@{}c@{}}CSD only brightness\\\relax [mag]\end{tabular}}} \\ \hline
\multicolumn{2}{c}{\textit{inclination}} &
  0$^{\circ}$ &
  30$^{\circ}$ &
  60$^{\circ}$ &
  0$^{\circ}$ &
  30$^{\circ}$ &
  60$^{\circ}$ &
  0$^{\circ}$ &
  30$^{\circ}$ &
  60$^{\circ}$ &
  0$^{\circ}$ &
  30$^{\circ}$ &
  60$^{\circ}$ \\
\multirow{3}{*}{10jup50au} & F1000W & 6.54 & 6.23 & 10.23 & 6.83  & 6.30  & 10.35 & <3.81 & <3.80 & <3.88 & 14.87 & 15.15 & 14.71 \\
                           & F1500W & 5.67 & 4.65 & 6.82  & 4.06  & 4.76  & 6.60  & <3.12 & <3.08 & <3.15 & 12.87 & 12.66 & 12.57 \\
                           & F2100W & 3.44 & 3.83 & 5.16  & 3.60  & 3.74  & 5.02  & 3.79  & 3.80  & 3.83  & 9.43  & 9.38  & 9.23  \\[1ex]
\multirow{3}{*}{5jup50au}  & F1000W & 9.04 & 9.55 & 12.45 & 9.25  & 9.19  & 12.39 & <4.65 & <4.59 & <4.59 & 14.85 & 15.03 & 14.77 \\
                           & F1500W & 7.16 & 7.44 & 9.81  & 7.26  & 7.21  & 9.56  & 5.27  & 5.16  & 5.36  & 13.04 & 12.91 & 12.77 \\
                           & F2100W & 5.93 & 6.23 & 8.16  & 6.00  & 6.26  & 7.99  & 5.16  & 5.03  & 5.19  & 9.40  & 9.38  & 9.25  \\[1ex]
\multirow{3}{*}{1jup50au}  & F1000W & /    & /    & /     & /     & /     & /     & 7.00  & 7.00  & 7.00  & 15.03 & 14.75 & 14.73 \\
                           & F1500W & /    & /    & /     & /     & /     & /     & 6.91  & 6.92  & 6.90  & 13.39 & 13.06 & 12.98 \\
                           & F2100W & /    & /    & /     & /     & /     & /     & 6.77  & 6.77  & 6.78  & 9.50  & 9.50  & 9.38  \\[1ex]
\multirow{3}{*}{1sat50au}  & F1000W & /    & /    & /     & /     & /     & /     & /     & /     & /     & 14.79 & 14.95 & 14.72 \\
                           & F1500W & /    & /    & /     & /     & /     & /     & /     & /     & /     & 13.23 & 12.98 & 12.73 \\
                           & F2100W & /    & /    & /     & /     & /     & /     & /     & /     & /     & 9.51  & 9.52  & 9.37  \\[1ex]
\multirow{3}{*}{10jup30au} & F1000W & 4.95 & 5.10 & 8.95  & <4.58 & <4.69 & 8.27  & <3.99 & <4.09 & <4.16 & 14.79 & 15.38 & 14.99 \\
                           & F1500W & 3.22 & 3.88 & 6.13  & 3.22  & 3.62  & 5.55  & 3.81  & 4.21  & 4.28  & 12.17 & 12.21 & 12.17 \\
                           & F2100W & 2.97 & 3.41 & 4.75  & 2.51  & 3.43  & 4.63  & 4.46  & 4.59  & 4.67  & 8.19  & 8.20  & 8.25  \\[1ex]
\multirow{3}{*}{5jup30au}  & F1000W & 7.13 & 7.82 & 12.24 & 7.72  & 7.81  & 12.37 & <4.30 & <4.30 & <4.38 & 15.01 & 15.19 & 14.83 \\
                           & F1500W & 5.99 & 6.23 & 8.87  & 5.81  & 6.26  & 8.30  & 5.12  & 5.12  & 5.16  & 12.98 & 12.50 & 12.80 \\
                           & F2100W & 5.13 & 5.33 & 7.04  & 5.21  & 5.36  & 6.83  & 5.25  & 5.24  & 5.28  & 9.30  & 9.25  & 9.19  \\[1ex]
\multirow{3}{*}{1jup30au}  & F1000W & /    & /    & /     & 13.84 & 13.47 & 14.52 & 7.44  & 7.46  & 7.45  & 14.46 & 15.00 & 14.80 \\
                           & F1500W & /    & /    & /     & 10.05 & 9.38  & 11.46 & 7.58  & 7.58  & 7.56  & 13.23 & 12.98 & 12.76 \\
                           & F2100W & /    & /    & /     & 7.54  & 7.52  & 9.17  & 7.67  & 7.69  & 7.68  & 9.72  & 9.59  & 9.42  \\ \hline
\end{tabular}
\end{table*}

\begin{figure}
  \includegraphics[width=\columnwidth]{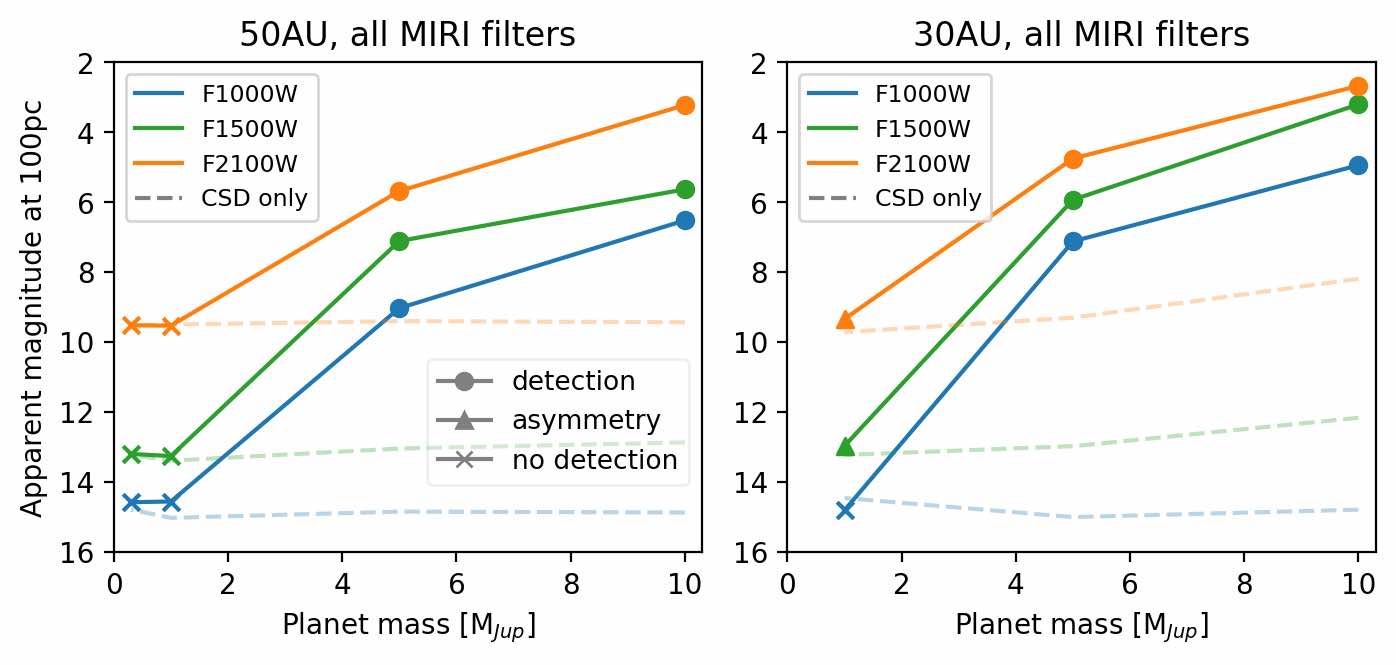}
  \caption{Apparent magnitudes at 100 pc for the different planetary masses and for the 5 filters in \textit{JWST}/MIRI. \textit{Left}: 50 AU orbital separation, \textit{right}: 30 AU separation. The dashed lines represent the magnitude of the circumstellar disk background in the corresponding filter. A circle symbol represents detection, and a cross symbol represents non-detection. A diamond symbol indicates detection but the planet flux has saturated the instrument detector. A triangle symbol indicates that an asymmetry of flux is observed, but the CPD is not resolved from the background CSD. All planets above 5 $M_{\mathrm{Jup}}$ with their CPDs are detected, and the CPD brightness scales with planetary mass and wavelength. This would not be the case if the CPDs are located in CSDs with different properties (e.g. a different CSD mass would lead to a different CPD mass and optical thickness, therefore affecting the observed brightness through extinction \citealt{Szulagyi2017}, \citealt{Szulagyi2019}).}
  \label{fig:miri_mass_mag}
\end{figure}

The simulated images of MIRI filters F1000W, F1500W, F2100W in 0$^{\circ}$ inclination at 100 pc are presented in Fig. \ref{fig:miri0}, while the 30$^{\circ}$ and 60$^{\circ}$ inclination images are in Appendix \ref{sec:app} (Figs. \ref{fig:miri30} and \ref{fig:miri60}). The change of CPD brightness with increasing planetary mass for the three MIRI filters is shown in Fig. \ref{fig:miri_mass_mag}. Fig. \ref{fig:miri0} and Fig. \ref{fig:miri_mass_mag} together shows that 10 and 5 $M_{\mathrm{Jup}}$ planets with their CPDs are detectable in all three MIRI filters. While the CPD brightness is higher in longer filters, the contrast between CPD and CSD goes down. In these simulations, a bright PSF core (the CPD) with an offset to 9 o'clock direction from the center dominates the image (e.g. the first 2 columns in both the left and right panel of Fig. \ref{fig:miri0}). The circumstellar disk is faint compared to the CPD and does not appear in the images. For 1 $M_{\mathrm{Jup}}$ and 1 $M_{\mathrm{Sat}}$ planets at 50 AU orbital separation (3rd and 4th columns in the left panel of Fig. \ref{fig:miri0}), we see only a continuum of flux concentrated in a circular shape at the image center. Although the circumstellar disk structure is unresolved, the fact that the height-to-width ratio of this circular-shaped object decreases with increasing inclination (see Figs. \ref{fig:miri30} and \ref{fig:miri60} in Appendix \ref{sec:app}) confirms that it is indeed the circumstellar disk. At 30 AU separation, a 1 $M_{\mathrm{Jup}}$ planet with its CPD can lead to a flux asymmetry in MIRI F1500W and F2100W filters, seen in the two images on the 2nd and 3rd row in the right-most column in Fig. \ref{fig:miri0}. We are only able to identify this point source as the CPD since we know a priori the planet\add{'s} position on the image in our models. Without knowing the planet position in the system, it would not identifiable as a detection, since the CPD and the background CSD are not even resolved. If the planetary system is closer than 100 pc, there is \add{the} possibility that a 1 $M_{\mathrm{Jup}}$ planet with its CPD can be detected by MIRI at > 15 $\mu$m. \add{However, the number of such nearby protoplanetary disks with possible planets in formation is very limited.} Below this wavelength, the 1 $M_{\mathrm{Jup}}$ planet does not show up (i.e. the F1000W image on the top right corner of Fig. \ref{fig:miri0}). \add{Given that the nearest star-forming regions (e.g. the Taurus region) are around 130 pc away, Jupiter-mass planets are unlikely to be detected in disks in these regions. Higher mass planets (5-10 $M_{\mathrm{Jup}}$) might still be detectable by MIRI in these regions. Their fluxes scaled to 130 pc are around 60\% of those at 100 pc, which are still enough for detection.}

The magnitudes of the CPD in the entire system, CPD+planet only case, planet only case and CSD only case for all three inclinations are compared in Table \ref{tab:miri_full} for \textit{JWST}/MIRI. We see a similar trend as the previous instruments that the CPD observability decreases with larger inclinations. For 10 and 5 $M_{\mathrm{Jup}}$ planets, the CSD has little effect on absorbing planet emission. For 1 $M_{\mathrm{Jup}}$ planets at 30 AU, the CPD becomes detectable without the extinction of CSD. The planet-only brightness is \add{a} few orders higher than planets embedded in CPD (except for 1 $M_{\mathrm{Jup}}$ planets, which are not detectable even without absorption of the CPD), indicating that the main absorber of \add{the} planet's emission is the CPD.

From our results, it is recommended that filters above 15 $\mu$m should be used for observing planets of 1 $M_{\mathrm{Jup}}$ and below with MIRI. For bright sources (10 $M_{\mathrm{Jup}}$ and above), we recommend \del{to use}\add{using} detector subarrays (e.g. SUB256) to avoid saturation. Our results do\del{es} not include a realistic simulation of coronagraphy due to the current limitation of \texttt{MIRISim}, but a full coronagraphy simulation should be performed in future observation proposals.

\subsubsection{ELT/MICADO}\label{sec:micado}

\begin{figure*}
  \includegraphics[width=\textwidth]{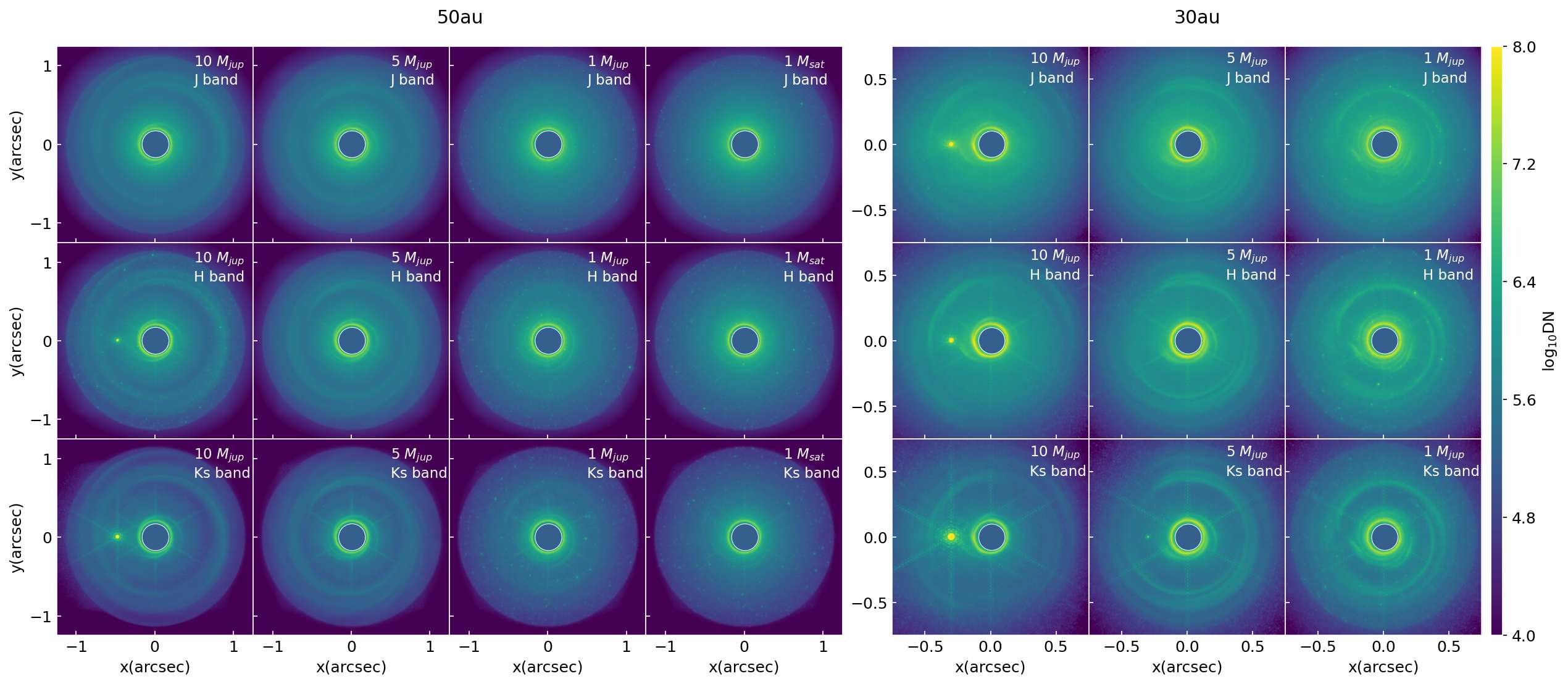}
  \caption{ELT/MICADO synthetic images of face-on systems projected to 100 pc. The columns represent hydrodynamic simulations with different planetary masses (10, 5, 1 $M_{\mathrm{Jup}}$ and 1 $M_{\mathrm{Sat}}$), and the rows represent the 3 filters simulated. The planet is located at 50 AU (\textit{left}) or 30 AU (\textit{right}) from the star at 9 o’clock direction on the images. Although the central star is included in the \textsc{radmc-3d} and telescope simulations, it is masked out with a circle in the final figure to allow focus on the CPD region. At 50 AU separation, only 10 $M_{\mathrm{Jup}}$ planets with their CPDs are detected in H and Ks band. At 30 AU, the CPDs of 10 $M_{\mathrm{Jup}}$ planets are detected in all 3 bands while those for 5 $M_{\mathrm{Jup}}$ planets are detected only in Ks band.}
  \label{fig:micado0}
\end{figure*}

\begin{table*}
\caption{\textit{ELT}/MICADO predicted magnitudes at 100 pc for CPD in the entire system, CPD+planet only, planet only and CSD only in 0$^{\circ}$, 30$^{\circ}$, 60$^{\circ}$ inclinations. The magnitude values for the first 3 groups (CPD in the entire system, CPD+planet only, planet only) are given by flux within an aperture at planet location minus the flux within a same-sized aperture 180$^{\circ}$ from the planet (i.e. anti-planet location). The magnitude values for the last group (CSD only) are given by the flux within a same-sized aperture at the anti-planet location. Non-detections are marked with a '/'.}
\label{tab:micado_full}
\begin{tabular}{ccccccccccc@{\quad \vline \quad}ccc}
\hline
\textbf{model} &
  \textbf{filter} &
  \multicolumn{3}{c}{\textbf{\begin{tabular}[c]{@{}c@{}}CPD brightness in the\\ entire system [mag]\end{tabular}}} &
  \multicolumn{3}{c}{\textbf{\begin{tabular}[c]{@{}c@{}}CPD+planet only\\ brightness [mag]\end{tabular}}} &
  \multicolumn{3}{c}{\textbf{\begin{tabular}[c]{@{}c@{}}planet only brightness\\\relax [mag]\end{tabular}}}\quad  &
  \multicolumn{3}{c}{\textbf{\begin{tabular}[c]{@{}c@{}}CSD only brightness\\\relax [mag]\end{tabular}}} \\ \hline
\multicolumn{2}{c}{\textit{inclination}} &
  0$^{\circ}$ &
  30$^{\circ}$ &
  60$^{\circ}$ &
  0$^{\circ}$ &
  30$^{\circ}$ &
  60$^{\circ}$ &
  0$^{\circ}$ &
  30$^{\circ}$ &
  60$^{\circ}$ &
  0$^{\circ}$ &
  30$^{\circ}$ &
  60$^{\circ}$ \\
\multirow{3}{*}{10jup50au} & J  & /     & /     & /     & 18.57 & 18.07 & 17.46 & 11.10 & 11.10 & 11.12 & 18.95 & 18.98 & 18.88 \\
                           & H  & 13.50 & /     & /     & 15.99 & 16.45 & 16.19 & 10.05 & 10.08 & 10.08 & 18.37 & 18.39 & 18.30 \\
                           & Ks & 11.22 & 14.43 & 18.14 & 12.84 & 14.38 & 15.28 & 9.06  & 9.08  & 9.07  & 17.73 & 17.79 & 17.78 \\[1ex]
\multirow{3}{*}{5jup50au}  & J  & /     & /     & /     & 18.09 & 17.71 & 17.39 & 13.63 & 13.61 & 14.80 & 18.91 & 18.95 & 18.84 \\
                           & H  & /     & /     & /     & 16.65 & 16.41 & 16.22 & 10.68 & 10.65 & 11.12 & 18.33 & 18.36 & 18.27 \\
                           & Ks & /     & /     & /     & 15.87 & 15.71 & 15.71 & 9.88  & 9.81  & 9.94  & 17.71 & 17.77 & 17.76 \\[1ex]
\multirow{3}{*}{1jup50au}  & J  & /     & /     & /     & 17.61 & 17.39 & 17.64 & /     & /     & /     & 18.79 & 18.83 & 18.72 \\
                           & H  & /     & /     & /     & 16.98 & 16.75 & 16.48 & 15.05 & 15.04 & 14.98 & 18.19 & 18.22 & 18.14 \\
                           & Ks & /     & /     & /     & 16.42 & 16.62 & 16.34 & 11.26 & 11.24 & 11.29 & 17.51 & 17.57 & 17.58 \\[1ex]
\multirow{3}{*}{1sat50au}  & J  & /     & /     & /     & /     & 19.52 & 17.99 & /     & /     & /     & 18.72 & 18.73 & 18.62 \\
                           & H  & /     & /     & /     & /     & 18.19 & 19.03 & /     & /     & /     & 18.04 & 18.08 & 17.91 \\
                           & Ks & /     & /     & /     & /     & 18.05 & 17.94 & /     & /     & /     & 17.37 & 17.42 & 17.33 \\[1ex]
\multirow{3}{*}{10jup30au} &
  J &
  12.99 &
  14.84 &
  16.41 &
  16.07 &
  13.77 &
  16.40 &
  11.15 &
  11.16 &
  11.18 &
  17.69 &
  17.72 &
  17.68 \\[1ex]
                           & H  & 13.08 & 11.91 & 16.52 & 12.58 & 11.15 & 14.66 & 10.13 & 10.14 & 10.14 & 17.35 & 17.37 & 17.34 \\
                           & Ks & 10.75 & 10.28 & 13.33 & 10.22 & 10.01 & 12.85 & 9.13  & 9.13  & 9.13  & 17.04 & 17.08 & 17.08 \\[1ex]
\multirow{3}{*}{5jup30au}  & J  & /     & /     & /     & 17.31 & 16.94 & 16.82 & 11.25 & 11.23 & 11.29 & 17.64 & 17.69 & 17.66 \\
                           & H  & /     & /     & /     & 16.48 & 15.48 & 15.90 & 10.27 & 10.27 & 10.28 & 17.29 & 17.34 & 17.28 \\
                           & Ks & 16.51 & 14.20 & /     & 15.49 & 12.08 & 15.59 & 9.60  & 9.60  & 9.60  & 16.97 & 17.03 & 16.98 \\[1ex]
\multirow{3}{*}{1jup30au}  & J  & /     & /     & /     & /     & 18.23 & 17.91 & /     & /     & /     & 17.69 & 17.70 & 17.65 \\
                           & H  & /     & /     & /     & 18.05 & 17.52 & 17.10 & /     & /     & /     & 17.28 & 17.31 & 17.24 \\
                           & Ks & /     & /     & /     & /     & 17.56 & 17.15 & /     & /     & /     & 16.90 & 16.93 & 16.81 \\ \hline
\end{tabular}
\end{table*}

\begin{figure}
  \includegraphics[width=\columnwidth]{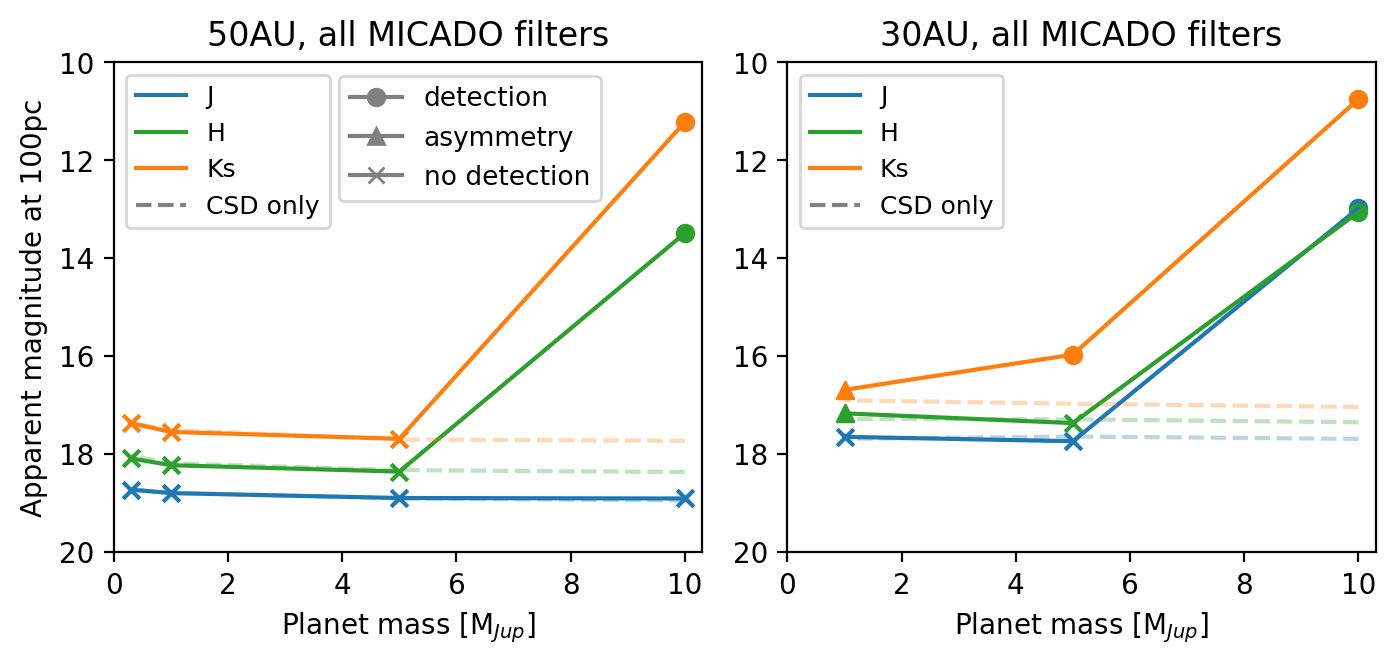}
  \caption{Apparent magnitudes at 100 pc with different planetary masses for the 3 filters in MICADO. Left: 50 AU separation, right: 30 AU separation. The dashed lines represent the magnitude of the circumstellar disk background in the corresponding filter. A circle symbol represents detection, and a cross symbol represents non-detection. A triangle symbol indicates that an asymmetry of flux is observed, but the SNR is not high enough to be count\add{ed} as a detection. Ks band works best for CPD detections, and in the other shorter bands only 10 $M_{\mathrm{Jup}}$ planets are detectable.}
  \label{fig:micado_mass_mag}
\end{figure}

The simulated images of MICADO filters J (1.2 $\mu$m), H (1.6 $\mu$m) and Ks (2.1 $\mu$m) in 0$^{\circ}$ inclination at 100 pc are presented in Fig. \ref{fig:micado0}, while the 30$^{\circ}$ and 60$^{\circ}$ inclination images are in Appendix \ref{sec:app} (Figs. \ref{fig:micado30} \& \ref{fig:micado60}). The CPD brightness with increasing planetary mass for all three MICADO filters is shown in Fig. \ref{fig:micado_mass_mag}. Here, the circumstellar disk is well resolved by MICADO and a triangle symbol indicates that even though the CPD itself \del{it}\add{is} not detected, there is a flux asymmetry possibly coming from the features such as spirals in the circumstellar disk induced by an embedded planet. From Figs. \ref{fig:micado0} and \ref{fig:micado_mass_mag}, it is found that only 10 $M_{\mathrm{Jup}}$ planets in H, Ks bands and 5 $M_{\mathrm{Jup}}$ planets at 30 AU in Ks band are bright enough to be detected (with their CPDs). Even if a planet is hidden by the surrounding disks, the spiral structures induced by gas accretion onto the planet can still be observable in some cases (e.g. the 1 $M_{\mathrm{Jup}}$ in H and Ks band in the last column of Fig. \ref{fig:micado0}). To enable such detections, it is necessary to exploit the full potential of MICADO's high resolution and sensitivity by pushing the limits of its adaptive optics (AO) systems and coronagraphs. 

We also compared the brightness of the CPD in the entire system with the brightness of CPD+planet only case, planet only case and CSD only case for various inclinations in Table \ref{tab:micado_full} for ELT/MICADO. At 30$^{\circ}$ inclination, observability is not significantly reduced. However, at 60$^{\circ}$, only 10 $M_{\mathrm{Jup}}$ planets in Ks band can be detected. Comparing the CPD in \add{the} entire system and \add{the} CPD+planet only case in Table \ref{tab:micado_full}, we see that most CPDs (even for 1 $M_{\mathrm{Jup}}$ planets) become detectable if not absorbed by the CSD. Comparing with planet only case, we see that the CPD is still the main absorber for planets above 1 $M_{\mathrm{Jup}}$. The 1 $M_{\mathrm{Sat}}$ planets are so faint that they cannot be detected by themselves alone, but a surrounding CPD can increase their observability. 

We conclude that the best observability for forming planets is found in Ks band for ELT/MICADO. and AO+coronagraphy is essential to reveal the planet-disk interaction features (e.g. spirals) in the circumstellar disk.

\subsubsection{ELT/METIS}\label{sec:metis}

\begin{figure*}
  \includegraphics[width=\textwidth]{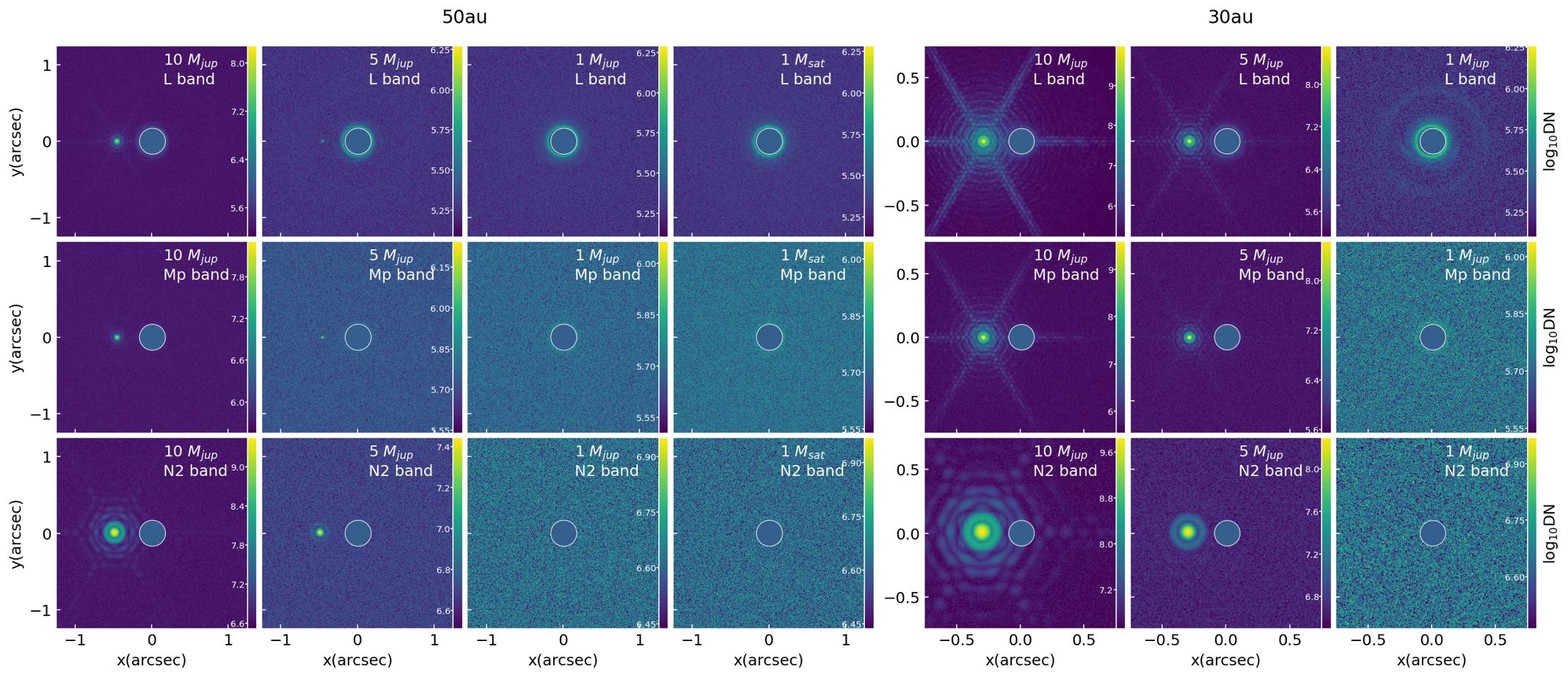}
  \caption{ELT/METIS synthetic images of face-on systems projected to 100 pc. The columns represent hydrodynamic simulations with different planetary masses (10, 5, 1 $M_{\mathrm{Jup}}$ and 1 $M_{\mathrm{Sat}}$), and the rows represent the 3 filters simulated. The planet is located at 50 AU (\textit{left}) or 30 AU (\textit{right}) from the star at 9 o’clock direction on the images. Although the central star is included in the \textsc{radmc-3d} and telescope simulations, it is masked out with a circle in the final figure to allow focus on the CPD region. 10 and 5 $M_{\mathrm{Jup}}$ planets with their CPDs are detected in all 3 bands.}
  \label{fig:metis0}
\end{figure*}

\begin{table*}
\caption{\textit{ELT}/METIS predicted magnitudes at 100 pc for CPD in the entire system, CPD+planet only, planet only and CSD only in 0$^{\circ}$, 30$^{\circ}$, 60$^{\circ}$ inclinations. The magnitude values for the first 3 groups (CPD in the entire system, CPD+planet only, planet only) are given by flux within an aperture at planet location minus the flux within a same-sized aperture 180$^{\circ}$ from the planet (i.e. anti-planet location). The magnitude values for the last group (CSD only) are given by the flux within a same-sized aperture at the anti-planet location. Non-detections are marked with a '/'.}
\label{tab:metis_full}
\begin{tabular}{ccccccccccc@{\quad \vline \quad}ccc}
\hline
\textbf{model} &
  \textbf{filter} &
  \multicolumn{3}{c}{\textbf{\begin{tabular}[c]{@{}c@{}}CPD brightness in the\\ entire system [mag]\end{tabular}}} &
  \multicolumn{3}{c}{\textbf{\begin{tabular}[c]{@{}c@{}}CPD+planet only\\ brightness [mag]\end{tabular}}} &
  \multicolumn{3}{c}{\textbf{\begin{tabular}[c]{@{}c@{}}planet only brightness\\\relax [mag]\end{tabular}}}\quad  &
  \multicolumn{3}{c}{\textbf{\begin{tabular}[c]{@{}c@{}}CSD only brightness\\\relax [mag]\end{tabular}}} \\ \hline
\multicolumn{2}{c}{\textit{inclination}} &
  0$^{\circ}$ &
  30$^{\circ}$ &
  60$^{\circ}$ &
  0$^{\circ}$ &
  30$^{\circ}$ &
  60$^{\circ}$ &
  0$^{\circ}$ &
  30$^{\circ}$ &
  60$^{\circ}$ &
  0$^{\circ}$ &
  30$^{\circ}$ &
  60$^{\circ}$ \\
\multirow{3}{*}{10jup50au} & L  & 14.95 & 13.38 & 17.53 & 14.59 & 13.33 & 17.62 & 5.66  & 5.73  & 5.90  & 21.13 & 21.73 & 20.98 \\
                           & Mp & 13.63 & 12.93 & 16.95 & 13.79 & 12.56 & 16.58 & 4.88  & 4.93  & 5.12  & 18.06 & 18.49 & 19.19 \\
                           & N2 & 8.85  & 9.80  & 15.36 & 9.05  & 9.23  & 13.12 & 4.17  & 4.19  & 4.22  & 14.67 & 14.37 & 14.75 \\[1ex]
\multirow{3}{*}{5jup50au}  & L  & 21.36 & 19.92 & /     & /     & 19.69 & /     & 11.01 & 10.50 & 11.30 & 21.53 & 21.09 & 21.11 \\
                           & Mp & 19.44 & 17.92 & /     & 19.42 & 18.11 & /     & 10.19 & 9.75  & 10.41 & 19.07 & 19.03 & 18.70 \\
                           & N2 & 14.14 & 14.73 & /     & 14.32 & 14.69 & /     & 8.14  & 8.07  & 8.21  & 14.70 & 14.55 & 14.89 \\[1ex]
\multirow{3}{*}{1jup50au}  & L  & /     & /     & /     & /     & /     & /     & /     & /     & /     & 21.17 & 20.88 & 21.48 \\
                           & Mp & /     & /     & /     & /     & /     & /     & /     & /     & /     & 18.78 & 18.72 & 18.71 \\
                           & N2 & /     & /     & /     & /     & /     & /     & 9.75  & 9.75  & 9.74  & 14.62 & 14.61 & 14.55 \\[1ex]
\multirow{3}{*}{1sat50au}  & L  & /     & /     & /     & /     & /     & /     & /     & /     & /     & 21.28 & 21.33 & 21.65 \\
                           & Mp & /     & /     & /     & /     & /     & /     & /     & /     & /     & 18.24 & 19.50 & 18.85 \\
                           & N2 & /     & /     & /     & /     & /     & /     & /     & /     & /     & 14.74 & 15.12 & 14.61 \\[1ex]
\multirow{3}{*}{10jup30au} & L  & 10.76 & 10.77 & 15.65 & 11.47 & 11.65 & 16.30 & 7.33  & 7.14  & 7.11  & 21.44 & 21.22 & 21.60 \\
                           & Mp & 10.45 & 11.22 & 15.21 & 10.45 & 10.45 & 14.77 & 6.57  & 6.42  & 6.36  & 18.32 & 18.99 & 19.06 \\
                           & N2 & 7.83  & 7.64  & 13.77 & 7.74  & 7.89  & 10.62 & 6.50  & 6.47  & 6.69  & 15.28 & 14.79 & 14.86 \\[1ex]
\multirow{3}{*}{5jup30au}  & L  & 13.91 & 14.58 & 19.09 & 14.06 & 14.60 & 19.19 & 7.05  & 7.12  & 7.16  & 21.43 & 21.90 & 21.26 \\
                           & Mp & 12.86 & 13.53 & 17.17 & 13.37 & 13.45 & 18.28 & 6.78  & 6.84  & 6.89  & 18.63 & 18.93 & 18.63 \\
                           & N2 & 11.78 & 11.92 & /     & 11.96 & 12.04 & /     & 6.42  & 6.44  & 6.51  & 14.83 & 14.28 & 14.77 \\[1ex]
\multirow{3}{*}{1jup30au}  & L  & /     & /     & /     & /     & /     & /     & 10.66 & 10.66 & 10.75 & 21.38 & 21.67 & 21.34 \\
                           & Mp & /     & /     & /     & /     & /     & /     & 10.26 & 10.28 & 10.34 & 18.58 & 18.42 & 18.17 \\
                           & N2 & /     & /     & /     & /     & /     & /     & 9.35  & 9.35  & /     & 14.87 & 14.31 & 14.32 \\ \hline
\end{tabular}
\end{table*}

\begin{figure}
  \includegraphics[width=\columnwidth]{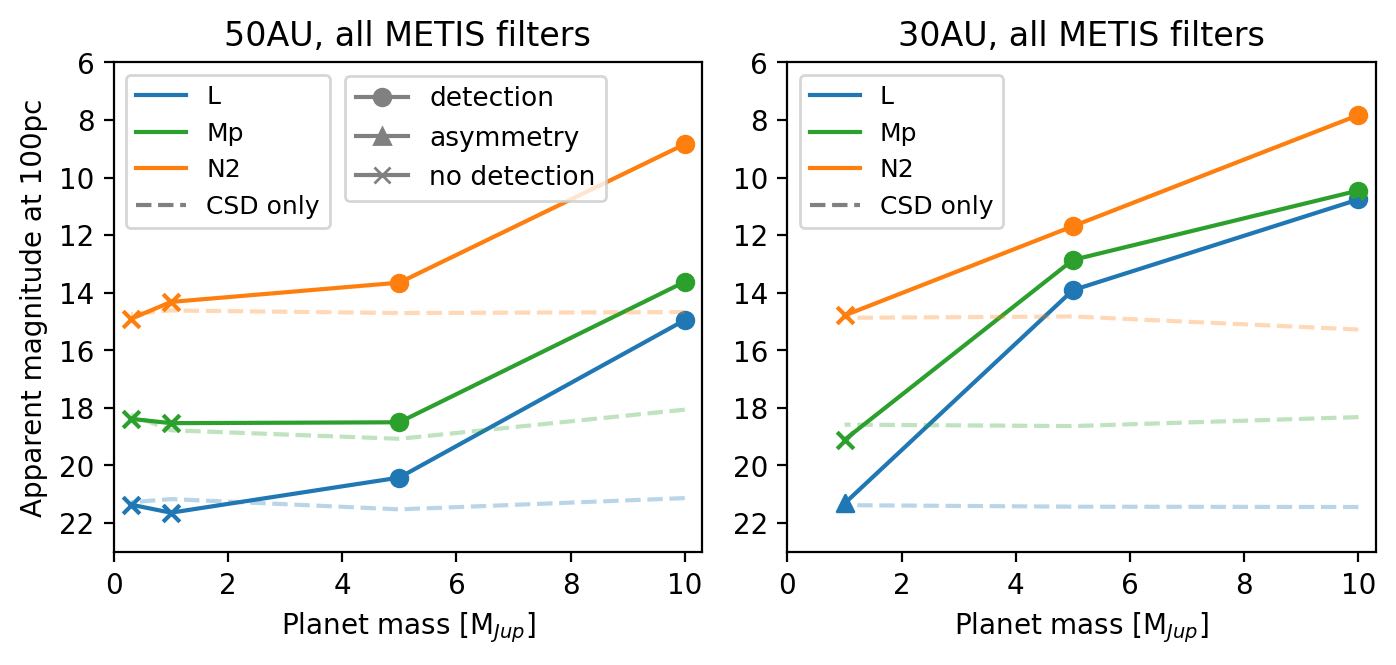}
  \caption{Apparent magnitudes at 100 pc with different planetary masses for the 3 filters in METIS. Left: 50 AU separation between the planet and the star, right: 30 AU separation. The dashed lines represent the magnitude of the circumstellar disk background in the corresponding filter. A circle symbol represents detection, and a cross symbol represents non-detection. A triangle symbol indicates that an asymmetry of flux is observed, but the SNR is not high enough to be count\add{ed} as a detection. Planets with mass above 5 $M_{\mathrm{Jup}}$ (with their CPDs) can be detected in all 3 METIS bands.}
  \label{fig:metis_mass_mag}
\end{figure}

The simulated images of METIS filters L (3.8 $\mu$m), M$_{p}$ (4.8 $\mu$m) and N$_{2}$ (11.8 $\mu$m) in 0$^{\circ}$ inclination at 100 pc are presented in Fig. \ref{fig:metis0}, while the 30$^{\circ}$ and 60$^{\circ}$ inclination images are in Appendix \ref{sec:app} (Figs. \ref{fig:metis30} \& \ref{fig:metis60}). The change in CPD brightness with increasing planetary mass for all three METIS filters \del{are}\add{is} shown in Fig. \ref{fig:metis_mass_mag}. It can be seen that planets with mass above 5 $M_{\mathrm{Jup}}$ (with their CPDs) can be detected in all 3 METIS bands. The circumstellar disks are faint and cannot be detected against the background noise, except for the 1 $M_{\mathrm{Jup}}$ 30 AU case in L band (i.e. the image at the top-right corner of \ref{fig:metis0}). The 1 $M_{\mathrm{Jup}}$ planets and their CPDs can not be detected even in the longest N band at $\sim$ 10 $\mu$m. With the resolution of METIS, it is possible to image the spiral features induced by a forming planet in the circumstellar disk. 

The brightness of the four extinction cases and various inclinations are studied and shown in Table \ref{tab:metis_full}. Like the other instruments, we see that observability is significantly reduced at 60$^{\circ}$ inclination and that the CPD absorbs most of the planet\add{'s} emission.

From our results, we conclude that \add{the} N band of METIS provides the best contrast between the CPD and CSD, and therefore is recommended for observing forming planets and their CPDs.

\subsection{Spectral Energy Distributions}
\label{sec:sed}

\begin{figure*}
  \includegraphics[width=\textwidth]{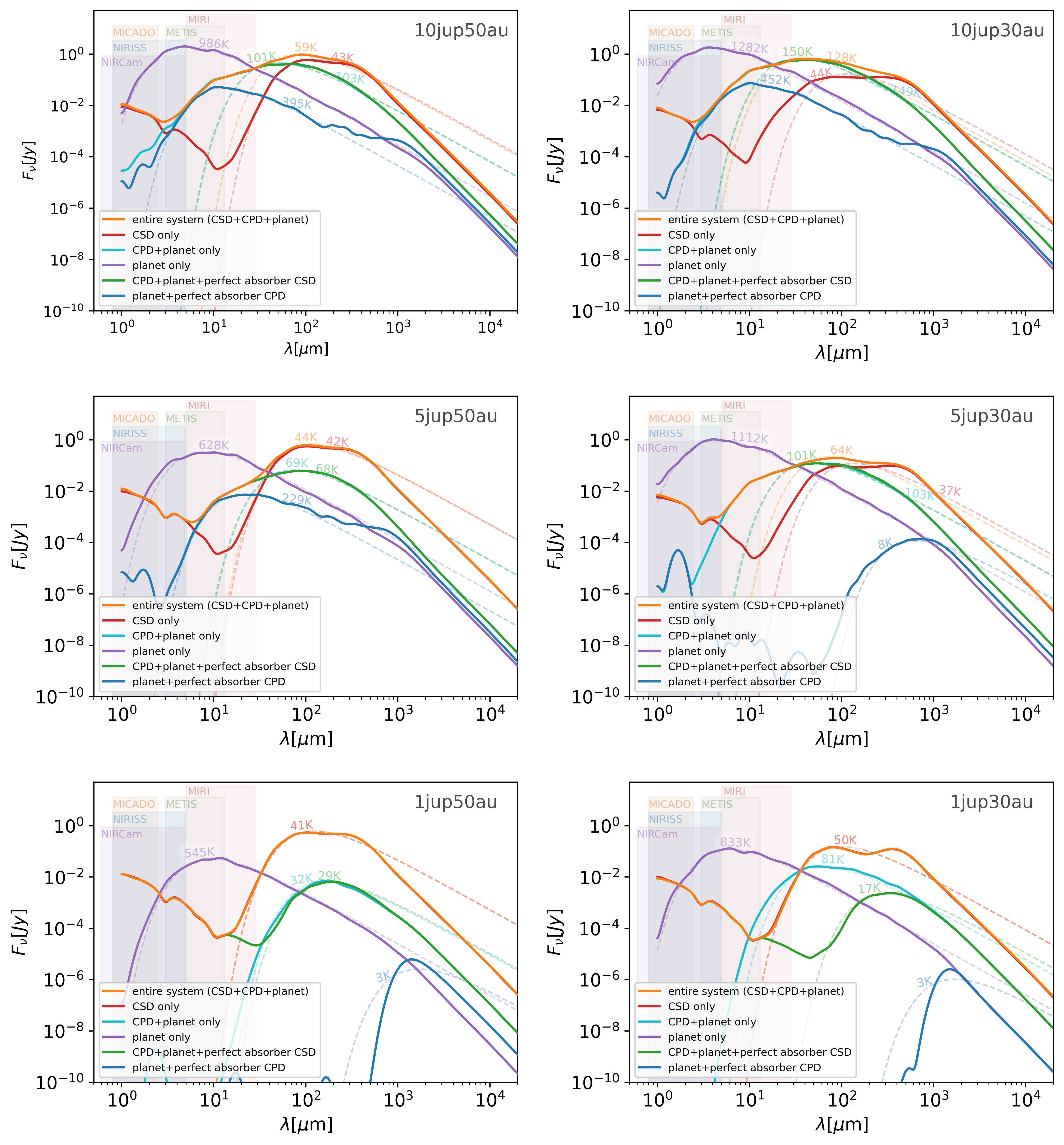}
  \caption{SEDs of the planet forming system with different planetary masses and separation. The 6 curves in each panel correspond to cases: \textit{orange} - entire system (planet + CPD + CSD); \textit{red} - CSD only, i.e CPD region cropped out from the CSD; \textit{cyan} - CPD + planet only; \textit{purple} - planet only; \textit{green} - CPD + planet + perfect absorber CSD, i.e., temperature outside the CPD region is set to 0; \textit{blue} - planet+perfect absorber CPD, i.e., temperature of the CPD outside planet surface is set to 0, and the CSD is cropped out. The CPD is defined as a cylinder with both radius and height equal 0.5$r_{Hill}$. The planet is define as a sphere at the center of CPD with radius = 0.002$r_{Hill}$. Dashed lines are fitted blackbody curves, with their corresponding effective temperature marked alongside. All fluxes are scaled to 100 pc distance. The operating wavelengths of the 5 instruments studied in this paper are marked as shadowed areas in the background.}
  \label{fig:sed}
\end{figure*}

Apart from synthetic images, we also studied the Spectral Energy Distributions (SEDs), separately for the different components of the system\del{ as well}:
\begin{enumerate}
    \item \textit{entire system} - CSD + CPD + planet
    \item \textit{CSD only} -  density and temperature inside the CPD region were set to 0, i.e. the CPD region \del{were}\add{was} cropped out from the CSD
    \item \textit{CPD + planet only} - density and temperature outside the CPD region were set to 0, i.e. the CSD was cropped out
    \item \textit{planet only} - density and temperature outside the planet surface were set to 0
    \item \textit{CPD + planet + perfect absorber CSD} - only temperature outside the CPD region was set to 0
    \item \textit{planet only + perfect absorber CPD} - \add{the} temperature outside planet surface was set to 0 and density outside CPD was set to 0. 
\end{enumerate}
The SEDs of the parameter space studied (different planetary masses and orbital separations) are shown in Fig. \ref{fig:sed}, without the inclusion of the stellar black body.
The wavelength coverage of the 5 instruments (NIRCam, NIRISS, MIRI, MICADO and METIS) simulated in this paper are plotted as shaded areas in the background as a reference. 

Both the CSD and the CPD material affect the overall flux observed from the forming planet in two ways: they either contribute to the flux by thermal emission and scattering of photons on the dust grains, or they reduce the emission of the forming planet due to absorption. It can be seen by comparing the orange, cyan and green curves in Fig. \ref{fig:sed} that the main contribution of CSD to the observed brightness is below 3 $\mu$m and above 30-50 $\mu$m. Between these wavelengths, orange and green curves overlap, which means that the CSD is acting as a perfect absorber, due to its low temperature. The upward slope in the short wavelengths is due to \add{the} scattering of photons on the dust grains (the stellar black body is not included in these SEDs). Most cyan and green lines overlap except for 1 $M_{\mathrm{Jup}}$ cases, which means that the CSD only absorbs significantly when the planet mass is small and \add{the} planetary gap is not too deep and wide. 
The contribution of the CPD to the overall flux can be seen from comparing the cyan, purple and blue curves in Fig. \ref{fig:sed}: Taking the 10 $M_{\mathrm{Jup}}$ 50 AU case as an example, the CPD contributes negatively to the overall flux below $\sim$30 $\mu$m and positively above this wavelength. It absorbs significantly below 10 $\mu$m (comparing blue and purple curves), meanwhile it re-emits above this wavelength (comparing cyan and blue curves). Below $\sim$35 $\mu$m the planet itself is the brightest, which means in this wavelength regime the disks around the planet absorb significantly, making a forming planet detection difficult in the near/mid IR. The CPD is the brightest in the wavelength range between 50 $\mu$m till approx. 200 $\mu$m. However, in these wavelengths the difference between the CSD and CPD is small, so it would be difficult to distinguish these two disks. While the inner CPDs close to the planet in all cases have higher temperature\add{s} than 1000 K, the effective temperatures of the CPDs are never above 150 K. This is due to the optical thickness of the CPD.
The blue curves for 5 $M_{\mathrm{Jup}}$ planets at 30 AU and 1 $M_{\mathrm{Jup}}$ planets with perfect absorber CPDs show very low effective temperatures (below 10 K). This means that almost all emission from the planet \del{are}\add{is} absorbed by the perfect-absorber CPD. In the two 1 $M_{\mathrm{Jup}}$ panels on the bottom row, the orange and red curves almost completely overlapped, meaning that the planet and its CPD cannot be distinguished from the CSD at all (these are non-detection cases).

Fig. \ref{fig:sed} also shows what wavelength ranges works best to distinguish the setup of a system, and which range is the best to detect the CPD on the "background" of the CSD. To distinguish whether there is only a planet or a planet surrounded by a CPD, one should use wavelengths below 30 $\mu$m. If the CPD is cold (perfect absorber), then longer wavelengths up until 1000 $\mu$m could work. An emitting CPD diverges from a perfect absorber CPD from 10 $\mu$m onward (cyan and blue lines in Fig. \ref{fig:sed}), so wavelengths above 10 $\mu$m can be used to determine whether the CPD is an absorber or emitter. The best contrast between CPD and CSD is found in the sub-mm and radio wavelengths for 1 $M_{\mathrm{Jup}}$ and 1 $M_{\mathrm{Sat}}$ planets. For massive, gap opening planets (above 5 $M_{\mathrm{Jup}}$), the mid IR range between 10-30 $\mu$m also provides good contrast. Below $\sim$3 $\mu$m, it is hard to distinguish the components of the system since this regime is dominated by scattering which highly depends on the dust grain properties. The transition regions between the curves lie at slightly different wavelengths for different simulations, since the planet and CPD have different temperature and densities structures due to their different characteristics (mass, temperature, etc.).


\section{Discussion}

Our models and methods are subject to limitations. In these hydrodynamic simulations, dust was not separately treated but was assumed to have a fixed gas-to-dust ratio of 0.01. \add{A} different assumption about the circumstellar disk mass could change the observability of planets since a more massive CSD could form a more massive CPD and lead to even more absorption. On the other hand, a more massive CSD could also enhance the accretion rate onto the planet and increase the CPD brightness through accretional luminosity \citep{Zhu2015, Szulagyi2017}. Moreover, the hydrodynamic simulations did not include magnetic fields and disk self-gravity. The assumption about the chemical composition of the dust particles and the choice of opacity tables also can affect the results. 

In the \textsc{radmc-3d} post-processing step, the choice of dust opacity table could influence the results, since this choice affects e.g. the amount of extinction. We have used a photon number of 10$^{7}$ and taken the median from 5 repeated runs for each image to reduce the photon noise. In our manipulation of the density- and temperature fields to make the cuts for the different extinction cases, the result CPD and planet brightness depend on the radius where the edge of the CPD or planet is defined. If the planet radius \add{was taken} too large, its surface would have a much lower temperature and \add{this would lead} to an underestimation of planet brightness. If the planet radius \add{was taken} too small, the cutting would be too close to the hot planet interior and \add{leading to an overestimation of} the planet-only brightness. Both in the hydrodynamic simulations and the \textsc{radmc-3d} post-processing we assumed a Sun equivalent star. Different stellar types would change a bit the observability of the CPD and the reported magnitudes as well.

The telescope simulators we used \del{has}\add{have} their limitations too. First of all, if the simulators will include coronagraphs, that would change the size and shape of the PSF and affect the detectability of the CPD. The simulated instrument PSFs used in these packages also deviates from the real PSFs. A common deficit is that due to the limited size of simulated PSF, the truncated PSF edges are visible in the image if the sources are bright. The flux normalization would also make the flux more concentrated in the PSF center than it should be, which then causes the planet to appear brighter at the center. Our synthetic images do\del{es} not include advanced PSF subtraction strategies, which are expect\add{ed} to be applied when processing real direct imaging data. Thus, our results only represent a bottom line of the capability of these instrument\add{s} for detecting forming planets. For ELT/MICADO and METIS, since there \del{is}\add{are} no existing data reduction pipelines, the manual calculation of flux conversion factors could also introduce differences to the final results, before such a full pipeline is made.

Compar\add{ed} to the previous papers in the series, this work confirmed partly their findings and complemented the previous studies with predictions for near-future telescopes.
We arrived at a similar conclusion with \citet{Szulagyi2019} and \citet{Szulagyi2021} that only very massive planets (above 10 $M_{\mathrm{Jup}}$ and 5 $M_{\mathrm{Jup}}$ in some filters and closer orbital separations) can be detected in near-IR. For planets with small masses (below 1 $M_{\mathrm{Jup}}$), the sub-mm wavelength with ALMA is more suitable, as it is predicted in \citet{Szulagyi2018} that 1 $M_{\mathrm{Sat}}$ planets with their CPDs are already observable by ALMA. Compar\add{ed} to \citet{Szulagyi2019}, the results of this paper \del{is}\add{are} also similar, even though done for different instruments. The SEDs showed that \add{the} best contrast between CPD and CSD is found in sub-mm/radio wavelength if the planet mass is a prior unknown, and in the mid-IR region between 10-30 $\mu$m if the planet is known to have mass larger than 5 $M_{\mathrm{Jup}}$, in agreement with \citet{Szulagyi2019}.

\del{Comparing}\add{Compared} to \citet{Sanchis2020}, our results also showed that the brightness for a given planet's CPD increases with wavelength in the near IR. We did not \del{found}\add{find} the decrease of brightness at the 10-$\mu$m silicate feature, since that is only observed for planets below 2 $M_{\mathrm{Jup}}$ in \citet{Sanchis2020} and those cases are not detected by the instruments studied in this paper. Unlike what we found that CPDs significantly absorb the planet's photons below 10 $\mu$m for all planetary masses, \citet{Sanchis2020} arrived at a different conclusion, namely that the circumplanetary materials do\del{es} not absorb significantly for planets above 5 $M_{\mathrm{Jup}}$. That is true only for the CSD absorption but not for the CPD absorption in our case. The difference can be partly explained by the different modeling methods adopted. In \citet{Sanchis2020}, the planet luminosity was calculated from evolutionary models and the extinction was derived from column density above the planet assuming interstellar medium (ISM) extinction curve. In this paper, however, the planet temperature and extinction by disks were calculated by the radiative hydrodynamic simulation and radiative transfer post-processing.

\section{Conclusions}

In this paper of the paper series, we investigated the observability of forming planets and their CPDs with \textit{JWST} and the ELT using 3D hydrodynamic simulations, radiative transfer post-processing and telescope simulator pipelines. We created synthetic images in 0$^{\circ}$, 30$^{\circ}$ and 60$^{\circ}$ inclinations for NIRCam, NIRISS AMI, MIRI, MICADO and METIS for planets with various masses (10 $M_{\mathrm{Jup}}$, 5 $M_{\mathrm{Jup}}$, 1 $M_{\mathrm{Jup}}$ and 1 $M_{\mathrm{Sat}}$) and orbital separations (30 or 50 AU) in a range of near-IR and mid-IR filters, and provided their predicted brightness. 

From the synthetic images, the longer wavelengths (mid-IR and beyond) seem to be the best to detect CPDs and their forming planets. Even in these cases, only very massive planet's CPD can be found with \textit{JWST} \& ELT. This science goal can be best achieved by \textit{JWST}/MIRI among all instruments studied in this paper. On ELT, METIS seems to be the best suited for forming planets \& CPD hunting. The specific findings from the images are summarized below:
\begin{itemize}
    \item CPDs of planets with 10 $M_{\mathrm{Jup}}$ and above are observable by all instruments (at least in some of their filters). Detectability of CPDs is significantly reduced at 60$^{\circ}$ inclination, but there are still cases \del{which}\add{that} can be detected.
    \item Below 3 $\mu$m, in the J, H, K band, both NIRCam and MICADO can detect 10 $M_{\mathrm{Jup}}$ planets with their CPDs. For 5 $M_{\mathrm{Jup}}$ planets, only those at 30 AU orbital separation are detectable by NIRCam in all 3 bands, or by MICADO in \add{the} K band. 
    \item In the L and M band (3-5 $\mu$m), both NIRCam and METIS can detect all 10 and 5 $M_{\mathrm{Jup}}$ planets with their CPDs. NIRCam benefits from its high sensitivity and requires less observation time for\del{ a} detection.
    \item For NIRISS AMI, 10 $M_{\mathrm{Jup}}$ planets with their CPDs can be detected in all filters except the case at 50 AU separation in F277W. 5 $M_{\mathrm{Jup}}$ planets with their CPDs are only detected at 30 AU in F480M. Direct imaging with NIRISS is not recommended.    
    \item In the mid-IR, both MIRI and METIS can detect 10 and 5 $M_{\mathrm{Jup}}$ planets with their CPDs in their whole wavelength coverage. For both MIRI and METIS, \add{the} N band ($\sim $10 $\mu$m) works best for CPD detection.
    \item For planetary mass smaller than 5 $M_{\mathrm{Jup}}$, wavelengths above 15 $\mu$m are needed. MIRI uniquely provides imaging capability in this wavelength range.
\end{itemize}

We also studied the extinction effect of the CPD and CSD separately with CPD+planet only and planet only cut-out cases. With simulated SEDs, we showed the absorption, emission and scattering contribution by the planet, CPD and CSD components to the total brightness separately. We found that the CSD only contributes to the observed brightness below 3 $\mu$m or above 30-50 $\mu$m, and between these wavelengths, it acts as a perfect absorber. The absorption by CSD is only significant for planets with mass below 1 $M_{\mathrm{Jup}}$. In most cases, the CPD is the main absorber of the planet's emission, not the CSD. The CPD absorbs significantly below 10 $\mu$m and re-emits above this wavelength. The absorption of these two disks makes detecting a forming planet difficult in near- \& mid-IR. From the SEDs we concluded that for unknown planetary mass the best contrast between CPD and CSD is found in the sub-mm and radio wavelengths, while for planets with known mass above 5 $M_{\mathrm{Jup}}$ the contrast is best between 10-30 $\mu$m, in accordance with \citet{Szulagyi2019}.

In conclusion, longer wavelengths (mid-IR and above) are the best range to look for CPDs. This can be achieved by MIRI in mid-IR, or even more in sub-mm/radio by ALMA \citep{Szulagyi2018,Szulagyi2019}. While we tried to cover a large parameter space, target-specific simulations and advanced image processing algorithms will be needed to uncover even more realistic detection limits for the CPDs in a specific system.

\section*{Acknowledgements}

\add{We thank the anonymous referee for their comments.} We thank \del{to }Gabriele Cugno for the discussion about which units and extinction information are best usable for observers. These results are part of a project that has received funding from the European Research Council (ERC) under the European Union’s Horizon 2020 research and innovation program (grant agreement number 948467). Furthermore, we appreciate the financial support through the Swiss National Science Foundation (SNSF) Ambizione grant PZ00P2\_174115. Computations partially have been done on the ‘Piz Daint’ machine hosted at the Swiss National Computational Centre and partially carried out on ETH Z\"urich’s Euler cluster.

\section*{Data Availability}

The mock images data underlying this article will be shared \del{on}\add{at} reasonable request to the corresponding author.



\bibliographystyle{mnras}
\bibliography{bibliography} 

\begin{thebibliography}{}
\makeatletter
\relax
\def\mn@urlcharsother{\let\do\@makeother \do\$\do\&\do\#\do\^\do\_\do\%\do\~}
\def\mn@doi{\begingroup\mn@urlcharsother \@ifnextchar [ {\mn@doi@}
  {\mn@doi@[]}}
\def\mn@doi@[#1]#2{\def\@tempa{#1}\ifx\@tempa\@empty \href
  {http://dx.doi.org/#2} {doi:#2}\else \href {http://dx.doi.org/#2} {#1}\fi
  \endgroup}
\def\mn@eprint#1#2{\mn@eprint@#1:#2::\@nil}
\def\mn@eprint@arXiv#1{\href {http://arxiv.org/abs/#1} {{\tt arXiv:#1}}}
\def\mn@eprint@dblp#1{\href {http://dblp.uni-trier.de/rec/bibtex/#1.xml}
  {dblp:#1}}
\def\mn@eprint@#1:#2:#3:#4\@nil{\def\@tempa {#1}\def\@tempb {#2}\def\@tempc
  {#3}\ifx \@tempc \@empty \let \@tempc \@tempb \let \@tempb \@tempa \fi \ifx
  \@tempb \@empty \def\@tempb {arXiv}\fi \@ifundefined
  {mn@eprint@\@tempb}{\@tempb:\@tempc}{\expandafter \expandafter \csname
  mn@eprint@\@tempb\endcsname \expandafter{\@tempc}}}

\bibitem[\protect\citeauthoryear{{Aoyama}, {Ikoma}  \& {Tanigawa}}{{Aoyama}
  et~al.}{2018}]{Aoyama2018}
{Aoyama} Y.,  {Ikoma} M.,   {Tanigawa} T.,  2018, \mn@doi [\apj]
  {10.3847/1538-4357/aadc11}, \href
  {https://ui.adsabs.harvard.edu/abs/2018ApJ...866...84A} {866, 84}

\bibitem[\protect\citeauthoryear{{Aoyama}, {Marleau}, {Ikoma}  \&
  {Mordasini}}{{Aoyama} et~al.}{2021}]{Aoyama2021}
{Aoyama} Y.,  {Marleau} G.-D.,  {Ikoma} M.,   {Mordasini} C.,  2021, \mn@doi
  [\apjl] {10.3847/2041-8213/ac19bd}, \href
  {https://ui.adsabs.harvard.edu/abs/2021ApJ...917L..30A} {917, L30}

\bibitem[\protect\citeauthoryear{{Bell} \& {Lin}}{{Bell} \&
  {Lin}}{1994}]{BellLin1994}
{Bell} K.~R.,  {Lin} D.~N.~C.,  1994, \mn@doi [\apj] {10.1086/174206}, \href
  {https://ui.adsabs.harvard.edu/abs/1994ApJ...427..987B} {427, 987}

\bibitem[\protect\citeauthoryear{Benisty et~al.,}{Benisty
  et~al.}{2021}]{Benisty2021}
Benisty M.,  et~al., 2021, \mn@doi [The Astrophysical Journal Letters]
  {10.3847/2041-8213/ac0f83}, 916, L2

\bibitem[\protect\citeauthoryear{{Biller} et~al.,}{{Biller}
  et~al.}{2014}]{Biller2014}
{Biller} B.~A.,  et~al., 2014, \mn@doi [\apjl] {10.1088/2041-8205/792/1/L22},
  \href {https://ui.adsabs.harvard.edu/abs/2014ApJ...792L..22B} {792, L22}

\bibitem[\protect\citeauthoryear{Boccaletti et~al.,}{Boccaletti
  et~al.}{2015}]{Boccaletti2015}
Boccaletti A.,  et~al., 2015, \mn@doi [Publications of the Astronomical Society
  of the Pacific] {10.1086/682256}, 127, 633–645

\bibitem[\protect\citeauthoryear{{Bohren} \& {Huffman}}{{Bohren} \&
  {Huffman}}{1984}]{Bohren1984}
{Bohren} C.~F.,  {Huffman} D.~R.,  1984, \nat, \href
  {https://ui.adsabs.harvard.edu/abs/1984Natur.307R.575B} {307, 575}

\bibitem[\protect\citeauthoryear{{Brandl} et~al.,}{{Brandl}
  et~al.}{2008}]{Brandl2008metis}
{Brandl} B.~R.,  et~al., 2008, in {McLean} I.~S.,  {Casali} M.~M.,  eds,
  Society of Photo-Optical Instrumentation Engineers (SPIE) Conference Series
  Vol. 7014, Ground-based and Airborne Instrumentation for Astronomy II. p.
  70141N (\mn@eprint {arXiv} {0807.3271}), \mn@doi{10.1117/12.789241}

\bibitem[\protect\citeauthoryear{{Christiaens}, {Cantalloube}, {Casassus},
  {Price}, {Absil}, {Pinte}, {Girard}  \& {Montesinos}}{{Christiaens}
  et~al.}{2019}]{Christiaens2019}
{Christiaens} V.,  {Cantalloube} F.,  {Casassus} S.,  {Price} D.~J.,  {Absil}
  O.,  {Pinte} C.,  {Girard} J.,   {Montesinos} M.,  2019, \mn@doi [\apjl]
  {10.3847/2041-8213/ab212b}, \href
  {https://ui.adsabs.harvard.edu/abs/2019ApJ...877L..33C} {877, L33}

\bibitem[\protect\citeauthoryear{{Commer{\c{c}}on}, {Teyssier}, {Audit},
  {Hennebelle}  \& {Chabrier}}{{Commer{\c{c}}on} et~al.}{2011}]{Commerson2011}
{Commer{\c{c}}on} B.,  {Teyssier} R.,  {Audit} E.,  {Hennebelle} P.,
  {Chabrier} G.,  2011, \mn@doi [\aap] {10.1051/0004-6361/201015880}, \href
  {https://ui.adsabs.harvard.edu/abs/2011A&A...529A..35C} {529, A35}

\bibitem[\protect\citeauthoryear{{Currie} et~al.,}{{Currie}
  et~al.}{2019}]{Currie2019}
{Currie} T.,  et~al., 2019, \mn@doi [\apjl] {10.3847/2041-8213/ab1b42}, \href
  {https://ui.adsabs.harvard.edu/abs/2019ApJ...877L...3C} {877, L3}

\bibitem[\protect\citeauthoryear{{Davies} et~al.,}{{Davies}
  et~al.}{2010}]{Davies2010micado}
{Davies} R.,  et~al., 2010, in {McLean} I.~S.,  {Ramsay} S.~K.,   {Takami} H.,
  eds,  Society of Photo-Optical Instrumentation Engineers (SPIE) Conference
  Series Vol. 7735, Ground-based and Airborne Instrumentation for Astronomy
  III. p. 77352A (\mn@eprint {arXiv} {1005.5009}), \mn@doi{10.1117/12.856379}

\bibitem[\protect\citeauthoryear{{Dipierro}, {Pinilla}, {Lodato}  \&
  {Testi}}{{Dipierro} et~al.}{2015}]{Dipierro2015}
{Dipierro} G.,  {Pinilla} P.,  {Lodato} G.,   {Testi} L.,  2015, \mn@doi
  [\mnras] {10.1093/mnras/stv970}, \href
  {https://ui.adsabs.harvard.edu/abs/2015MNRAS.451..974D} {451, 974}

\bibitem[\protect\citeauthoryear{Dong, Zhu  \& Whitney}{Dong
  et~al.}{2015a}]{Dong2015a}
Dong R.,  Zhu Z.,   Whitney B.,  2015a, \mn@doi [\apj]
  {10.1088/0004-637x/809/1/93}, 809, 93

\bibitem[\protect\citeauthoryear{Dong, Hall, Rice  \& Chiang}{Dong
  et~al.}{2015b}]{Dong2015b}
Dong R.,  Hall C.,  Rice K.,   Chiang E.,  2015b, \mn@doi [\apj]
  {10.1088/2041-8205/812/2/l32}, 812, L32

\bibitem[\protect\citeauthoryear{Dong, Fung  \& Chiang}{Dong
  et~al.}{2016}]{Dong2016}
Dong R.,  Fung J.,   Chiang E.,  2016, \mn@doi [\apj]
  {10.3847/0004-637x/826/1/75}, 826, 75

\bibitem[\protect\citeauthoryear{{Doyon} et~al.,}{{Doyon}
  et~al.}{2012}]{Doyon2012nrs}
{Doyon} R.,  et~al., 2012, in {Clampin} M.~C.,  {Fazio} G.~G.,  {MacEwen}
  H.~A.,   {Oschmann} Jacobus~M. J.,  eds,  Society of Photo-Optical
  Instrumentation Engineers (SPIE) Conference Series Vol. 8442, Space
  Telescopes and Instrumentation 2012: Optical, Infrared, and Millimeter Wave.
  p. 84422R, \mn@doi{10.1117/12.926578}

\bibitem[\protect\citeauthoryear{{Dullemond}, {Juhasz}, {Pohl}, {Sereshti},
  {Shetty}, {Peters}, {Commercon}  \& {Flock}}{{Dullemond}
  et~al.}{2012}]{Dullemond2012}
{Dullemond} C.~P.,  {Juhasz} A.,  {Pohl} A.,  {Sereshti} F.,  {Shetty} R.,
  {Peters} T.,  {Commercon} B.,   {Flock} M.,  2012, {RADMC-3D: A multi-purpose
  radiative transfer tool} (\mn@eprint {ascl} {1202.015})

\bibitem[\protect\citeauthoryear{{Eisner}}{{Eisner}}{2015}]{Eisner2015}
{Eisner} J.~A.,  2015, \mn@doi [\apjl] {10.1088/2041-8205/803/1/L4}, \href
  {https://ui.adsabs.harvard.edu/abs/2015ApJ...803L...4E} {803, L4}

\bibitem[\protect\citeauthoryear{{Ertel}, {Wolf}  \& {Rodmann}}{{Ertel}
  et~al.}{2012}]{Ertel2012}
{Ertel} S.,  {Wolf} S.,   {Rodmann} J.,  2012, \mn@doi [\aap]
  {10.1051/0004-6361/201219236}, \href
  {https://ui.adsabs.harvard.edu/abs/2012A&A...544A..61E} {544, A61}

\bibitem[\protect\citeauthoryear{{Gardner} et~al.,}{{Gardner}
  et~al.}{2006}]{Gardner2006}
{Gardner} J.~P.,  et~al., 2006, \mn@doi [\ssr] {10.1007/s11214-006-8315-7},
  \href {https://ui.adsabs.harvard.edu/abs/2006SSRv..123..485G} {123, 485}

\bibitem[\protect\citeauthoryear{{Gonzalez}, {Pinte}, {Maddison}, {M{\'e}nard}
  \& {Fouchet}}{{Gonzalez} et~al.}{2012}]{Gonzalez2012}
{Gonzalez} J.~F.,  {Pinte} C.,  {Maddison} S.~T.,  {M{\'e}nard} F.,   {Fouchet}
  L.,  2012, \mn@doi [\aap] {10.1051/0004-6361/201218806}, \href
  {https://ui.adsabs.harvard.edu/abs/2012A&A...547A..58G} {547, A58}

\bibitem[\protect\citeauthoryear{Haffert, Bohn, de Boer, Snellen, Brinchmann,
  Girard, Keller  \& Bacon}{Haffert et~al.}{2019}]{Haffert2019}
Haffert S.~Y.,  Bohn A.~J.,  de Boer J.,  Snellen I. A.~G.,  Brinchmann J.,
  Girard J.~H.,  Keller C.~U.,   Bacon R.,  2019, \mn@doi [Nature Astronomy]
  {10.1038/s41550-019-0780-5}, 3, 749–754

\bibitem[\protect\citeauthoryear{Hilbert et~al.,}{Hilbert
  et~al.}{2019}]{Hilbert2019}
Hilbert B.,  et~al., 2019, spacetelescope/mirage: First github release,
  \mn@doi{10.5281/zenodo.3519262}, \url
  {https://doi.org/10.5281/zenodo.3519262}

\bibitem[\protect\citeauthoryear{{Isella}, {Benisty}, {Teague}, {Bae},
  {Keppler}, {Facchini}  \& {P{\'e}rez}}{{Isella} et~al.}{2019}]{Isella2019}
{Isella} A.,  {Benisty} M.,  {Teague} R.,  {Bae} J.,  {Keppler} M.,  {Facchini}
  S.,   {P{\'e}rez} L.,  2019, \mn@doi [\apjl] {10.3847/2041-8213/ab2a12},
  \href {https://ui.adsabs.harvard.edu/abs/2019ApJ...879L..25I} {879, L25}

\bibitem[\protect\citeauthoryear{{Keppler} et~al.,}{{Keppler}
  et~al.}{2018}]{Keppler2018}
{Keppler} et~al., 2018, \mn@doi [A\&A] {10.1051/0004-6361/201832957}, 617, A44

\bibitem[\protect\citeauthoryear{{Klaassen} et~al.,}{{Klaassen}
  et~al.}{2021}]{Klaassen2021}
{Klaassen} P.~D.,  et~al., 2021, \mn@doi [\mnras] {10.1093/mnras/staa3416},
  \href {https://ui.adsabs.harvard.edu/abs/2021MNRAS.500.2813K} {500, 2813}

\bibitem[\protect\citeauthoryear{{Kley}}{{Kley}}{1989}]{Kley1989}
{Kley} W.,  1989, \aap, \href
  {https://ui.adsabs.harvard.edu/abs/1989A&A...208...98K} {208, 98}

\bibitem[\protect\citeauthoryear{Kraus \& Ireland}{Kraus \&
  Ireland}{2011}]{Kraus2012}
Kraus A.~L.,  Ireland M.~J.,  2011, \mn@doi [\apj] {10.1088/0004-637x/745/1/5},
  745, 5

\bibitem[\protect\citeauthoryear{{Lebreton}, {Beichman}, {Bryden},
  {Defr{\`e}re}, {Mennesson}, {Millan-Gabet}  \& {Boccaletti}}{{Lebreton}
  et~al.}{2016}]{Lebreton2016}
{Lebreton} J.,  {Beichman} C.,  {Bryden} G.,  {Defr{\`e}re} D.,  {Mennesson}
  B.,  {Millan-Gabet} R.,   {Boccaletti} A.,  2016, \mn@doi [\apj]
  {10.3847/0004-637X/817/2/165}, \href
  {https://ui.adsabs.harvard.edu/abs/2016ApJ...817..165L} {817, 165}

\bibitem[\protect\citeauthoryear{{Leschinski} et~al.,}{{Leschinski}
  et~al.}{2019}]{Leschinski2019}
{Leschinski} K.,  et~al., 2019, in {Molinaro} M.,  {Shortridge} K.,   {Pasian}
  F.,  eds,  Astronomical Society of the Pacific Conference Series Vol. 521,
  Astronomical Data Analysis Software and Systems XXVI. p.~527

\bibitem[\protect\citeauthoryear{{Marleau} et~al.,}{{Marleau}
  et~al.}{2021}]{Marleau2021}
{Marleau} G.~D.,  et~al., 2021, arXiv e-prints, \href
  {https://ui.adsabs.harvard.edu/abs/2021arXiv211112099M} {p. arXiv:2111.12099}

\bibitem[\protect\citeauthoryear{{Mayer}, {Peters}, {Pineda}, {Wadsley}  \&
  {Rogers}}{{Mayer} et~al.}{2016}]{Mayer2016}
{Mayer} L.,  {Peters} T.,  {Pineda} J.~E.,  {Wadsley} J.,   {Rogers} P.,  2016,
  \mn@doi [\apjl] {10.3847/2041-8205/823/2/L36}, \href
  {https://ui.adsabs.harvard.edu/abs/2016ApJ...823L..36M} {823, L36}

\bibitem[\protect\citeauthoryear{{Mendigut{\'\i}a}, {Oudmaijer}, {Schneider},
  {Hu{\'e}lamo}, {Baines}, {Brittain}  \& {Aberasturi}}{{Mendigut{\'\i}a}
  et~al.}{2018}]{Mendigutia2018}
{Mendigut{\'\i}a} I.,  {Oudmaijer} R.~D.,  {Schneider} P.~C.,  {Hu{\'e}lamo}
  N.,  {Baines} D.,  {Brittain} S.~D.,   {Aberasturi} M.,  2018, \mn@doi [\aap]
  {10.1051/0004-6361/201834233}, \href
  {https://ui.adsabs.harvard.edu/abs/2018A&A...618L...9M} {618, L9}

\bibitem[\protect\citeauthoryear{{M{\"u}ller} et~al.,}{{M{\"u}ller}
  et~al.}{2018}]{Muller2018}
{M{\"u}ller} A.,  et~al., 2018, \mn@doi [\aap] {10.1051/0004-6361/201833584},
  \href {https://ui.adsabs.harvard.edu/abs/2018A&A...617L...2M} {617, L2}

\bibitem[\protect\citeauthoryear{{Osorio} et~al.,}{{Osorio}
  et~al.}{2014}]{Osorio2014}
{Osorio} M.,  et~al., 2014, \mn@doi [\apjl] {10.1088/2041-8205/791/2/L36},
  \href {https://ui.adsabs.harvard.edu/abs/2014ApJ...791L..36O} {791, L36}

\bibitem[\protect\citeauthoryear{Perez, Dunhill, Casassus, Roman,
  Szul{\'{a}}gyi, Flores, Marino  \& Montesinos}{Perez
  et~al.}{2015}]{Perez2015}
Perez S.,  Dunhill A.,  Casassus S.,  Roman P.,  Szul{\'{a}}gyi J.,  Flores C.,
   Marino S.,   Montesinos M.,  2015, \mn@doi [\apj]
  {10.1088/2041-8205/811/1/l5}, 811, L5

\bibitem[\protect\citeauthoryear{Reggiani et~al.,}{Reggiani
  et~al.}{2014}]{Reggiani2014}
Reggiani M.,  et~al., 2014, \mn@doi [\apj] {10.1088/2041-8205/792/1/l23}, 792,
  L23

\bibitem[\protect\citeauthoryear{{Rieke}, {Kelly}  \& {Horner}}{{Rieke}
  et~al.}{2005}]{Rieke2005nrc}
{Rieke} M.~J.,  {Kelly} D.,   {Horner} S.,  2005, in {Heaney} J.~B.,
  {Burriesci} L.~G.,  eds,  Society of Photo-Optical Instrumentation Engineers
  (SPIE) Conference Series Vol. 5904, Cryogenic Optical Systems and Instruments
  XI. pp~1--8, \mn@doi{10.1117/12.615554}

\bibitem[\protect\citeauthoryear{{Rieke} et~al.,}{{Rieke}
  et~al.}{2015}]{Rieke2015miri}
{Rieke} G.~H.,  et~al., 2015, \mn@doi [\pasp] {10.1086/682252}, \href
  {https://ui.adsabs.harvard.edu/abs/2015PASP..127..584R} {127, 584}

\bibitem[\protect\citeauthoryear{{Sallum} et~al.,}{{Sallum}
  et~al.}{2015}]{Sallum2015}
{Sallum} S.,  et~al., 2015, \mn@doi [\nat] {10.1038/nature15761}, \href
  {https://ui.adsabs.harvard.edu/abs/2015Natur.527..342S} {527, 342}

\bibitem[\protect\citeauthoryear{{Sanchis}, {Picogna}, {Ercolano}, {Testi}  \&
  {Rosotti}}{{Sanchis} et~al.}{2020}]{Sanchis2020}
{Sanchis} E.,  {Picogna} G.,  {Ercolano} B.,  {Testi} L.,   {Rosotti} G.,
  2020, \mn@doi [\mnras] {10.1093/mnras/staa074}, \href
  {https://ui.adsabs.harvard.edu/abs/2020MNRAS.492.3440S} {492, 3440}

\bibitem[\protect\citeauthoryear{{Sivaramakrishnan} et~al.,}{{Sivaramakrishnan}
  et~al.}{2010}]{Sivaramakrishnan2010nrsami}
{Sivaramakrishnan} A.,  et~al., 2010, in {Oschmann} Jacobus~M. J.,  {Clampin}
  M.~C.,   {MacEwen} H.~A.,  eds,  Society of Photo-Optical Instrumentation
  Engineers (SPIE) Conference Series Vol. 7731, Space Telescopes and
  Instrumentation 2010: Optical, Infrared, and Millimeter Wave. p. 77313W,
  \mn@doi{10.1117/12.858161}

\bibitem[\protect\citeauthoryear{{Stolker}, {Dominik}, {Min}, {Garufi},
  {Mulders}  \& {Avenhaus}}{{Stolker} et~al.}{2016}]{Stolker2016}
{Stolker} T.,  {Dominik} C.,  {Min} M.,  {Garufi} A.,  {Mulders} G.~D.,
  {Avenhaus} H.,  2016, \mn@doi [\aap] {10.1051/0004-6361/201629098}, \href
  {https://ui.adsabs.harvard.edu/abs/2016A&A...596A..70S} {596, A70}

\bibitem[\protect\citeauthoryear{{Szul{\'a}gyi}}{{Szul{\'a}gyi}}{2017}]{Szulagyi2017}
{Szul{\'a}gyi} J.,  2017, \mn@doi [\apj] {10.3847/1538-4357/aa7515}, \href
  {https://ui.adsabs.harvard.edu/abs/2017ApJ...842..103S} {842, 103}

\bibitem[\protect\citeauthoryear{{Szul{\'a}gyi} \& {Ercolano}}{{Szul{\'a}gyi}
  \& {Ercolano}}{2020}]{Szulagyi2020}
{Szul{\'a}gyi} J.,  {Ercolano} B.,  2020, \mn@doi [\apj]
  {10.3847/1538-4357/abb5a2}, \href
  {https://ui.adsabs.harvard.edu/abs/2020ApJ...902..126S} {902, 126}

\bibitem[\protect\citeauthoryear{{Szul{\'{a}}gyi} \& {Garufi}}{{Szul{\'{a}}gyi}
  \& {Garufi}}{2021}]{Szulagyi2021}
{Szul{\'{a}}gyi} J.,  {Garufi} A.,  2021, \mn@doi [\mnras]
  {10.1093/mnras/stab1723}, \href
  {https://ui.adsabs.harvard.edu/abs/2021MNRAS.506...73S} {506, 73}

\bibitem[\protect\citeauthoryear{{Szul{\'a}gyi} \& {Mordasini}}{{Szul{\'a}gyi}
  \& {Mordasini}}{2017}]{SzulagyiMordasini2017}
{Szul{\'a}gyi} J.,  {Mordasini} C.,  2017, \mn@doi [\mnras]
  {10.1093/mnrasl/slw212}, \href
  {https://ui.adsabs.harvard.edu/abs/2017MNRAS.465L..64S} {465, L64}

\bibitem[\protect\citeauthoryear{Szul{\'{a}}gyi, Morbidelli, Crida  \&
  Masset}{Szul{\'{a}}gyi et~al.}{2014}]{Szulagyi2014}
Szul{\'{a}}gyi J.,  Morbidelli A.,  Crida A.,   Masset F.,  2014, \mn@doi
  [\apj] {10.1088/0004-637x/782/2/65}, 782, 65

\bibitem[\protect\citeauthoryear{{Szul{\'a}gyi}, {Masset}, {Lega}, {Crida},
  {Morbidelli}  \& {Guillot}}{{Szul{\'a}gyi} et~al.}{2016}]{Szulagyi2016}
{Szul{\'a}gyi} J.,  {Masset} F.,  {Lega} E.,  {Crida} A.,  {Morbidelli} A.,
  {Guillot} T.,  2016, \mn@doi [\mnras] {10.1093/mnras/stw1160}, \href
  {https://ui.adsabs.harvard.edu/abs/2016MNRAS.460.2853S} {460, 2853}

\bibitem[\protect\citeauthoryear{Szul{\'{a}}gyi, Plas, Meyer, Pohl, Quanz,
  Mayer, Daemgen  \& Tamburello}{Szul{\'{a}}gyi et~al.}{2018}]{Szulagyi2018}
Szul{\'{a}}gyi J.,  Plas G. v.~d.,  Meyer M.~R.,  Pohl A.,  Quanz S.~P.,  Mayer
  L.,  Daemgen S.,   Tamburello V.,  2018, \mnras, 473, 3573

\bibitem[\protect\citeauthoryear{Szul{\'{a}}gyi, Dullemond, Pohl  \&
  Quanz}{Szul{\'{a}}gyi et~al.}{2019}]{Szulagyi2019}
Szul{\'{a}}gyi J.,  Dullemond C.~P.,  Pohl A.,   Quanz S.~P.,  2019, \mnras,
  487, 1248

\bibitem[\protect\citeauthoryear{{Tanigawa}, {Ohtsuki}  \&
  {Machida}}{{Tanigawa} et~al.}{2012}]{Tanigawa2012}
{Tanigawa} T.,  {Ohtsuki} K.,   {Machida} M.~N.,  2012, \mn@doi [\apj]
  {10.1088/0004-637X/747/1/47}, \href
  {https://ui.adsabs.harvard.edu/abs/2012ApJ...747...47T} {747, 47}

\bibitem[\protect\citeauthoryear{Wagner et~al.,}{Wagner
  et~al.}{2018}]{Wagner2018}
Wagner K.,  et~al., 2018, \mn@doi [The Astrophysical Journal Letters]
  {10.3847/2041-8213/aad695}, 863, L8

\bibitem[\protect\citeauthoryear{Zhang et~al.,}{Zhang et~al.}{2018}]{Zhang2018}
Zhang S.,  et~al., 2018, \mn@doi [\apj] {10.3847/2041-8213/aaf744}, 869, L47

\bibitem[\protect\citeauthoryear{{Zhu}}{{Zhu}}{2015}]{Zhu2015}
{Zhu} Z.,  2015, \mn@doi [\apj] {10.1088/0004-637X/799/1/16}, \href
  {https://ui.adsabs.harvard.edu/abs/2015ApJ...799...16Z} {799, 16}

\bibitem[\protect\citeauthoryear{{Zhu}, {Ju}  \& {Stone}}{{Zhu}
  et~al.}{2016}]{Zhu2016}
{Zhu} Z.,  {Ju} W.,   {Stone} J.~M.,  2016, \mn@doi [\apj]
  {10.3847/0004-637X/832/2/193}, \href
  {https://ui.adsabs.harvard.edu/abs/2016ApJ...832..193Z} {832, 193}

\bibitem[\protect\citeauthoryear{{de Juan Ovelar}, {Min}, {Dominik},
  {Thalmann}, {Pinilla}, {Benisty}  \& {Birnstiel}}{{de Juan Ovelar}
  et~al.}{2013}]{Ovelar2013}
{de Juan Ovelar} M.,  {Min} M.,  {Dominik} C.,  {Thalmann} C.,  {Pinilla} P.,
  {Benisty} M.,   {Birnstiel} T.,  2013, \mn@doi [\aap]
  {10.1051/0004-6361/201322218}, \href
  {https://ui.adsabs.harvard.edu/abs/2013A&A...560A.111D} {560, A111}

\makeatother
\end{thebibliography}


\appendix

\section{Simulated images at different inclinations}\label{sec:app}

\begin{figure*}
  \includegraphics[width=\textwidth]{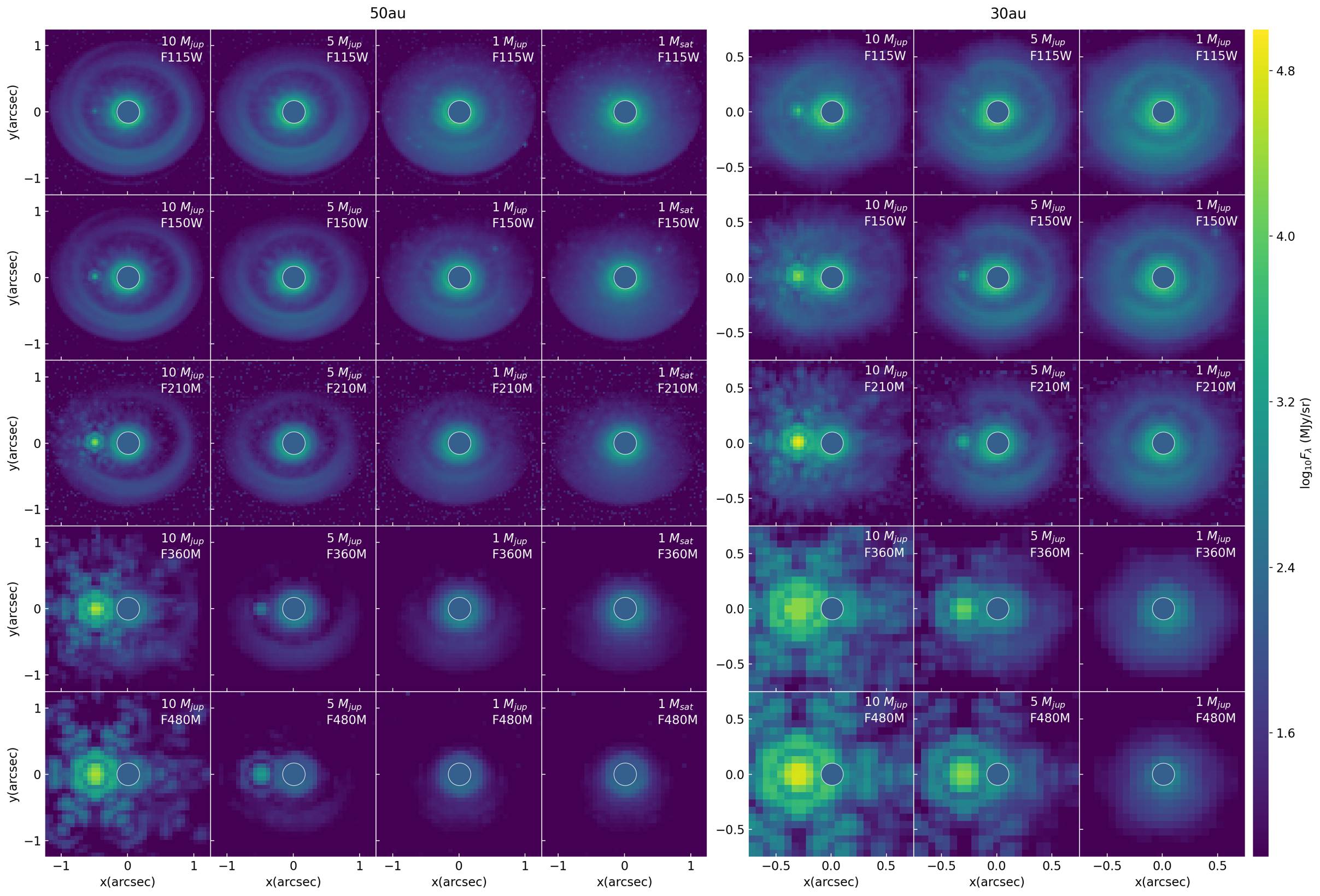}
  \caption{\textit{JWST}/NIRCam synthetic images projected to 100 pc, same as Fig. \ref{fig:nrc0} but in 30$^{\circ}$ inclination. The columns represent hydrodynamic simulations with different planetary masses (10, 5, 1 $M_{\mathrm{Jup}}$ and 1 $M_{\mathrm{Sat}}$), and the rows represent the 5 filters simulated. The planet is located at 50 AU (\textit{left}) or 30 AU (\textit{right}) from the star at 9 o’clock direction on the images. The central star is masked out with a circle.}
  \label{fig:nrc30}
\end{figure*}

\begin{figure*}
  \includegraphics[width=\textwidth]{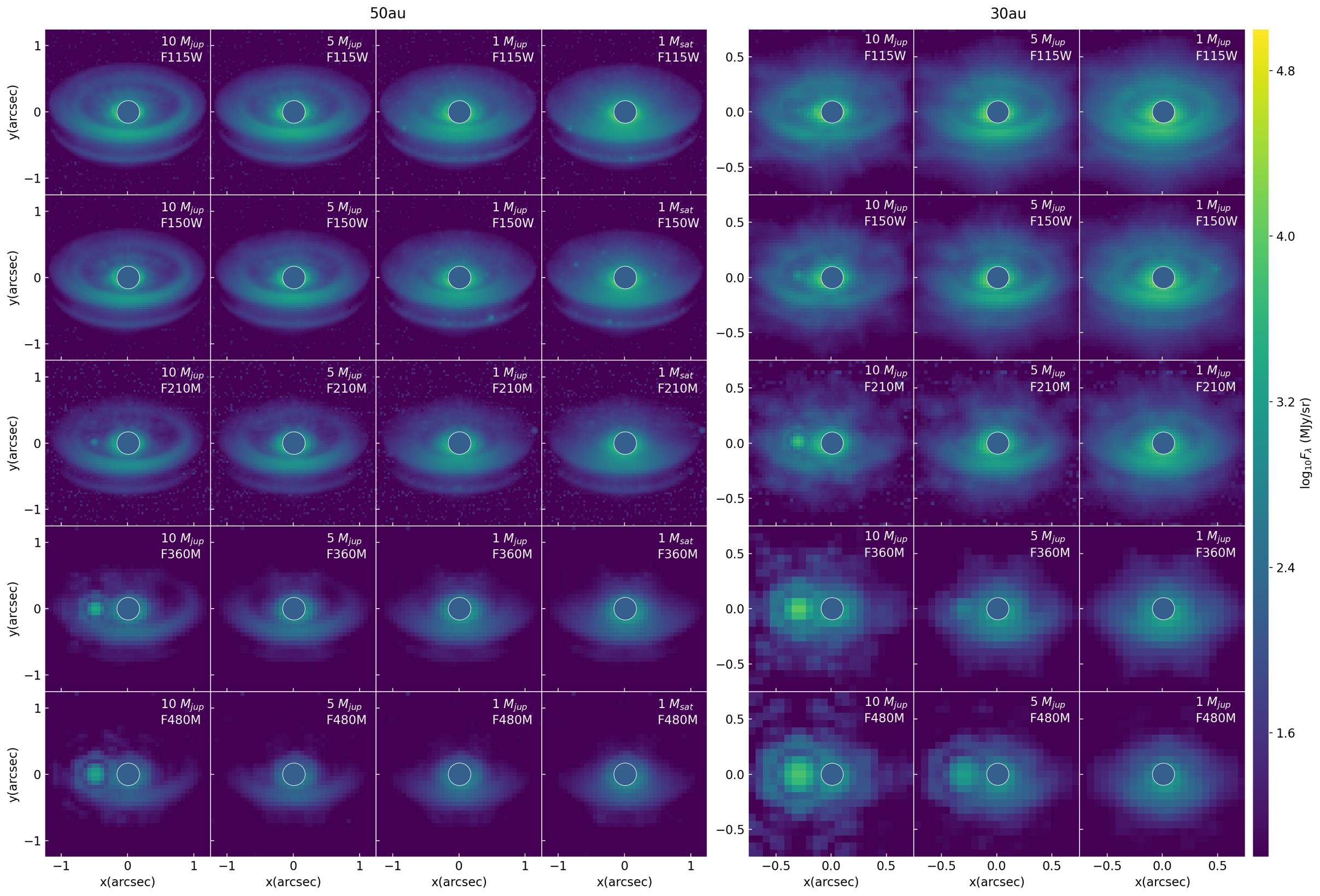}
  \caption{\textit{JWST}/NIRCam synthetic images projected to 100 pc, same as Fig. \ref{fig:nrc0} but in 60$^{\circ}$ inclination.}
  \label{fig:nrc60}
\end{figure*}

\begin{figure*}
  \includegraphics[width=\textwidth]{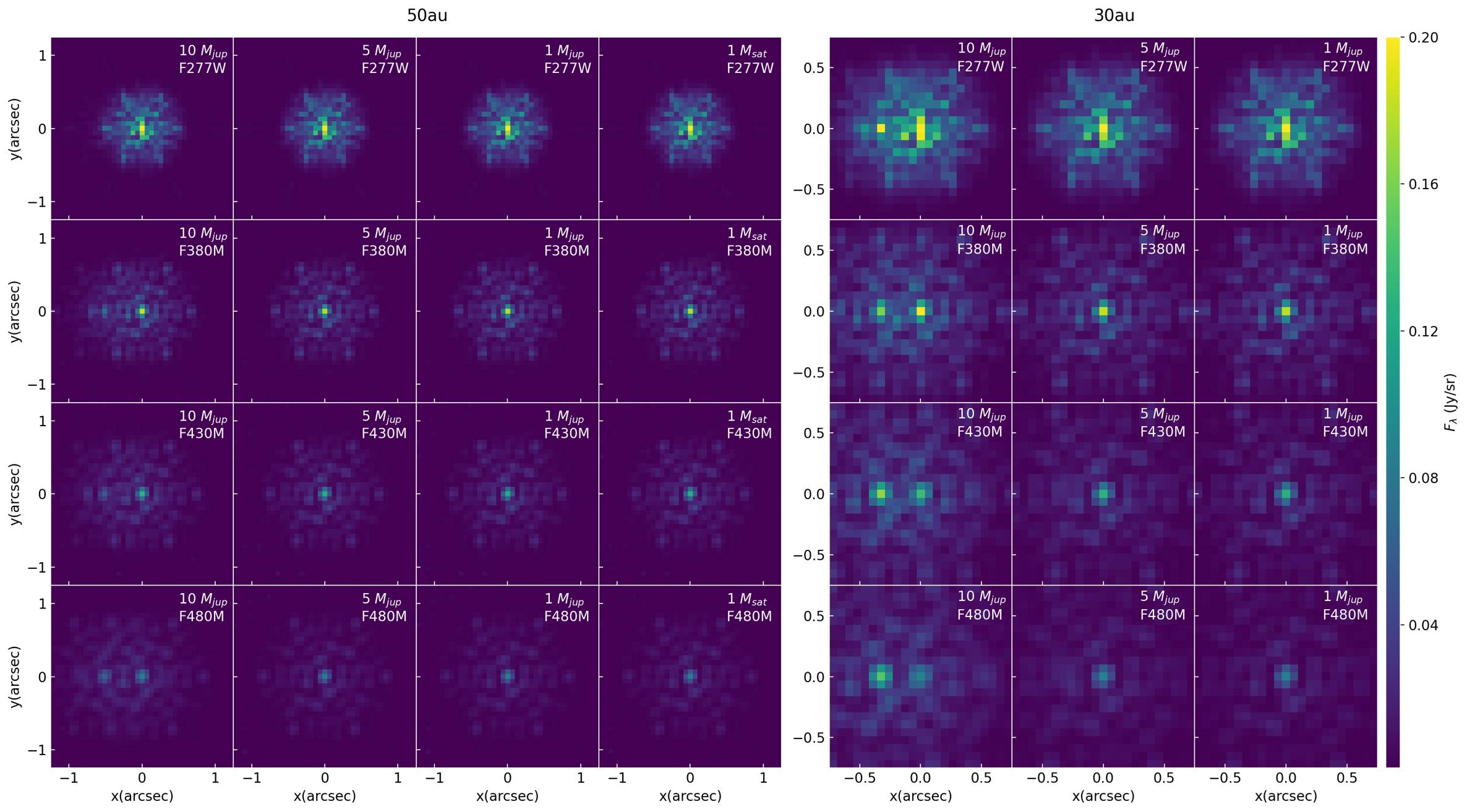}
  \caption{\textit{JWST}/NIRISS AMI synthetic images projected to 100 pc, same as Fig. \ref{fig:nrsami0} but in 30$^{\circ}$ inclination. The columns represent hydrodynamic simulations with different planetary masses (10, 5, 1 $M_{\mathrm{Jup}}$ and 1 $M_{\mathrm{Sat}}$), and the rows represent the 5 filters simulated. The planet is located at 50 AU (\textit{left}) or 30 AU (\textit{right}) from the star at 9 o’clock direction on the images. The central star is masked out with a circle.}
  \label{fig:nrsami30}
\end{figure*}

\begin{figure*}
  \includegraphics[width=\textwidth]{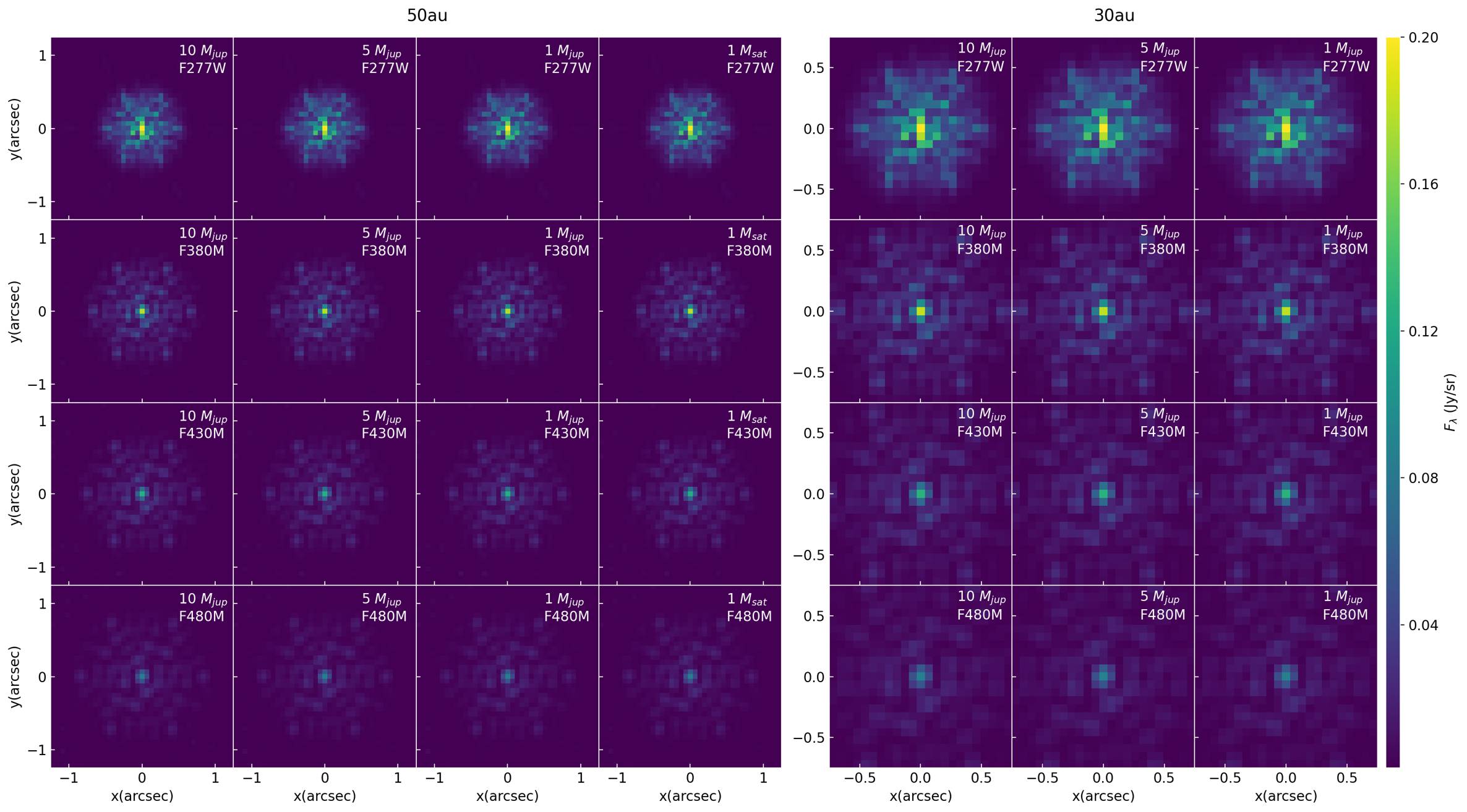}
  \caption{\textit{JWST}/NIRISS AMI synthetic images projected to 100 pc, same as Fig. \ref{fig:nrsami0} but in 60$^{\circ}$ inclination.}
  \label{fig:nrsami60}
\end{figure*}

\begin{figure*}
  \includegraphics[width=\textwidth]{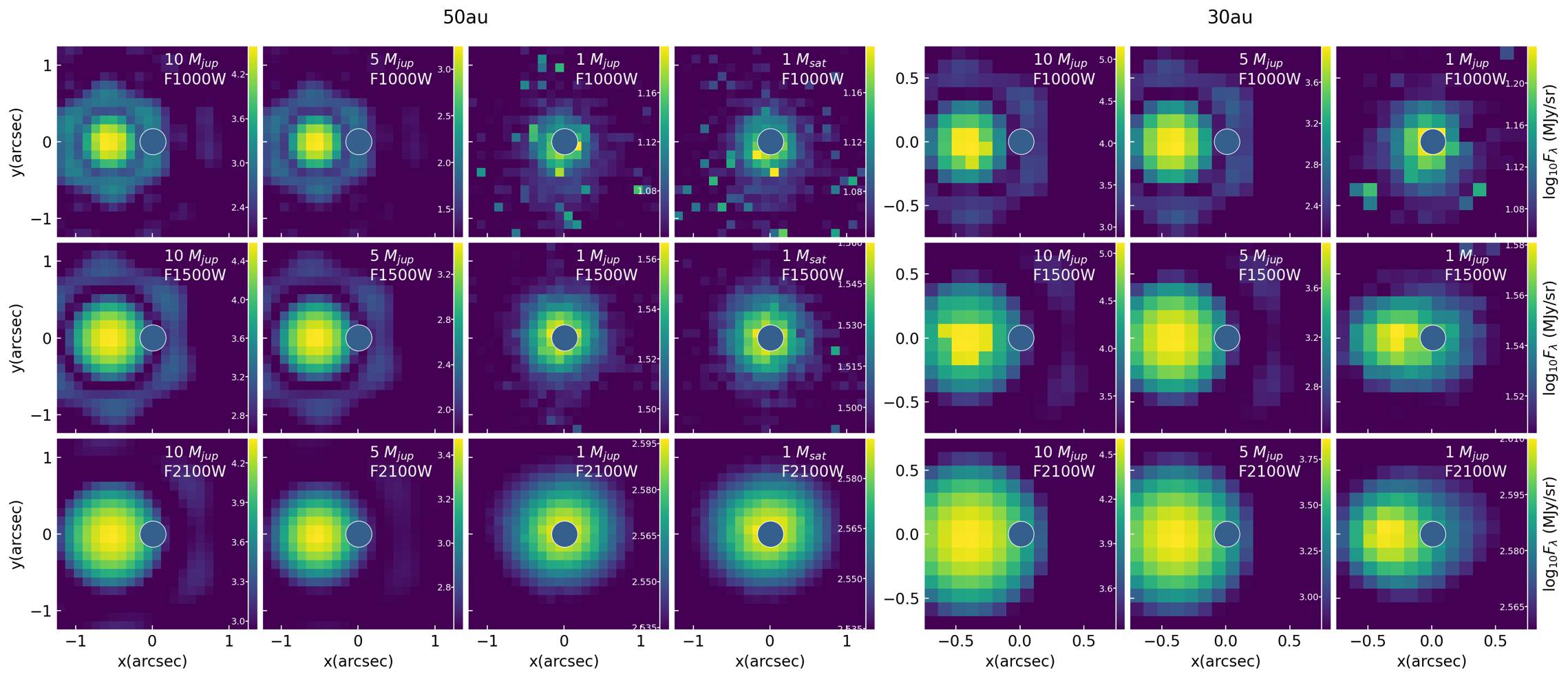}
  \caption{\textit{JWST}/MIRI synthetic images projected to 100 pc, same as Fig. \ref{fig:miri0} but in 30$^{\circ}$ inclination. The columns represent hydrodynamic simulations with different planetary masses (10, 5, 1 $M_{\mathrm{Jup}}$ and 1 $M_{\mathrm{Sat}}$), and the rows represent the 5 filters simulated. The planet is located at 50 AU (\textit{left}) or 30 AU (\textit{right}) from the star at 9 o’clock direction on the images. The central star is masked out with a circle.}
  \label{fig:miri30}
\end{figure*}

\begin{figure*}
  \includegraphics[width=\textwidth]{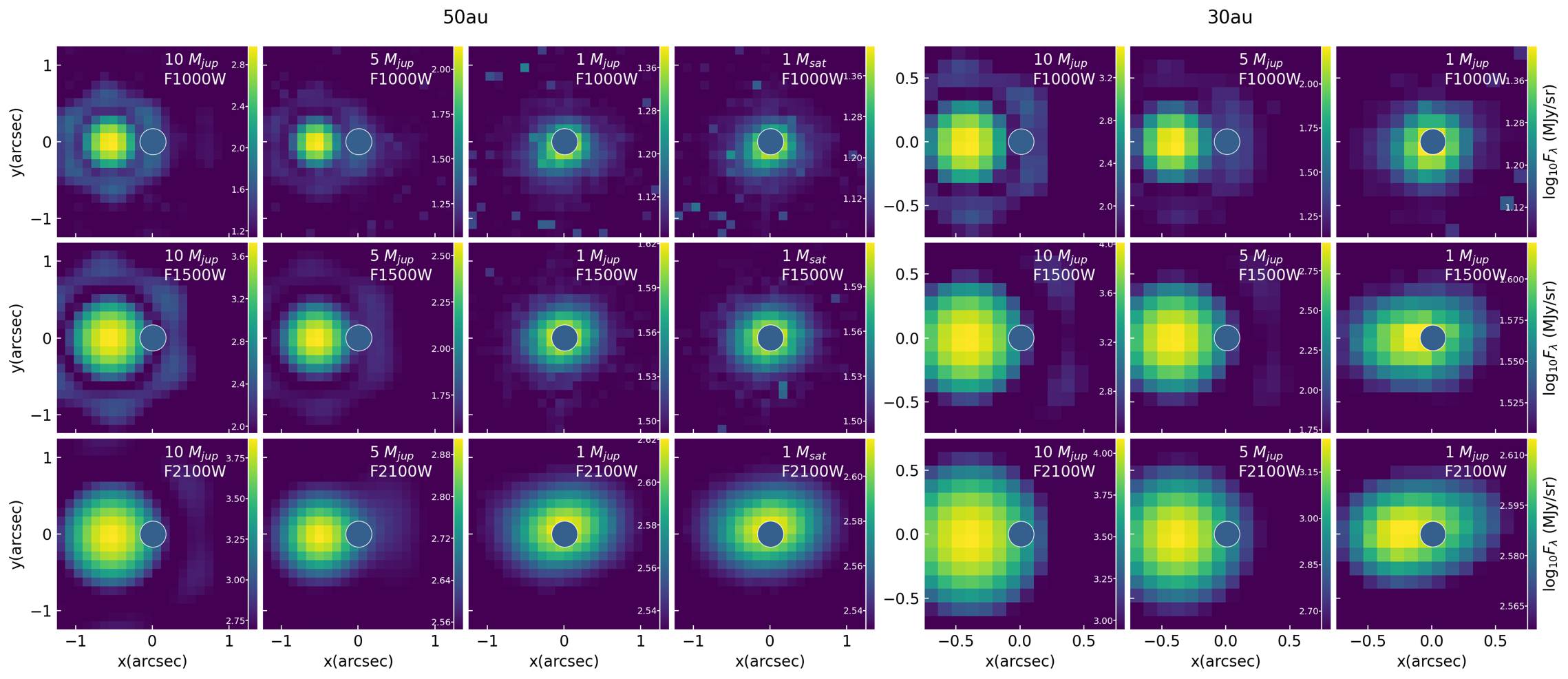}
  \caption{\textit{JWST}/MIRI synthetic images projected to 100 pc, same as Fig. \ref{fig:miri0} but in 60$^{\circ}$ inclination.}
  \label{fig:miri60}
\end{figure*}

\begin{figure*}
  \includegraphics[width=\textwidth]{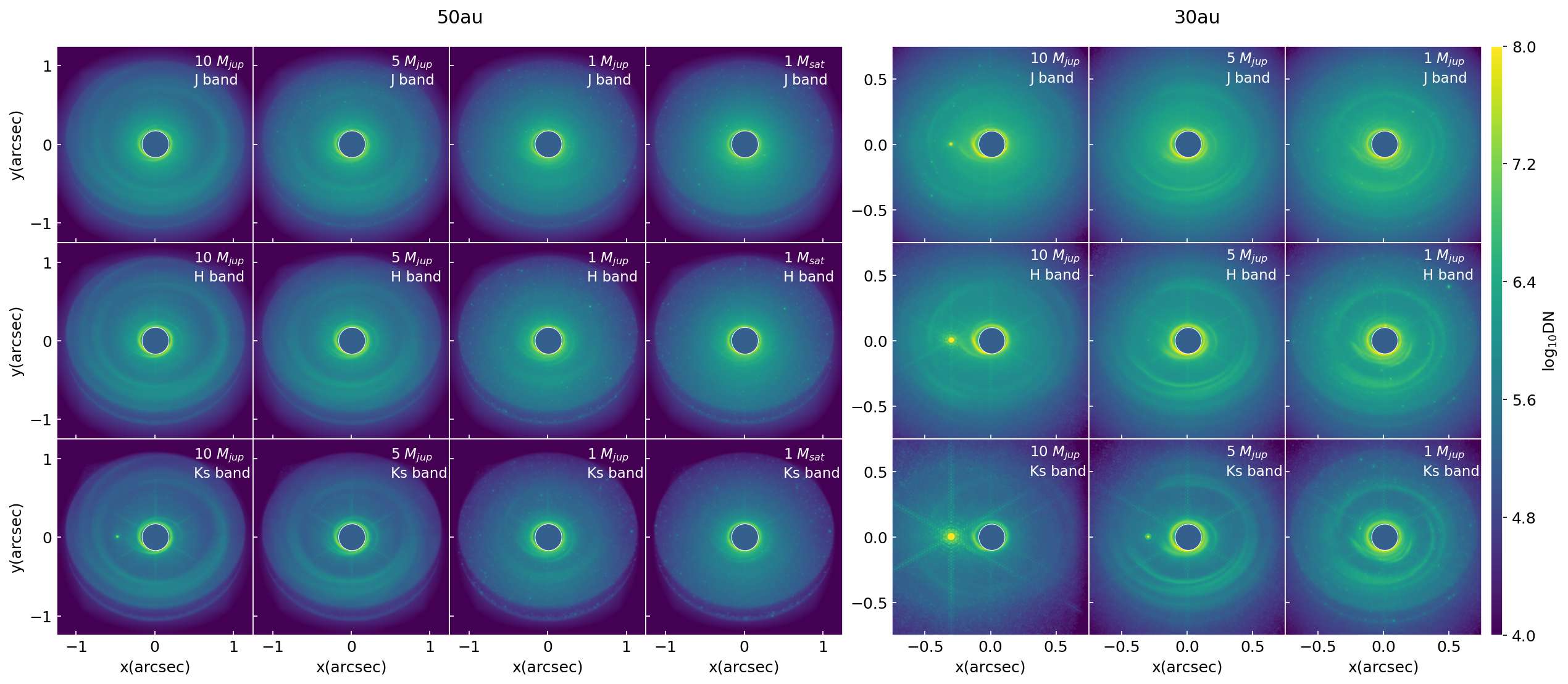}
  \caption{ELT/MICADO synthetic images projected to 100 pc, same as Fig. \ref{fig:micado0} but in 30$^{\circ}$ inclination. The columns represent hydrodynamic simulations with different planetary masses (10, 5, 1 $M_{\mathrm{Jup}}$ and 1 $M_{\mathrm{Sat}}$), and the rows represent the 5 filters simulated. The planet is located at 50 AU (\textit{left}) or 30 AU (\textit{right}) from the star at 9 o’clock direction on the images. The central star is masked out with a circle.}
  \label{fig:micado30}
\end{figure*}

\begin{figure*}
  \includegraphics[width=\textwidth]{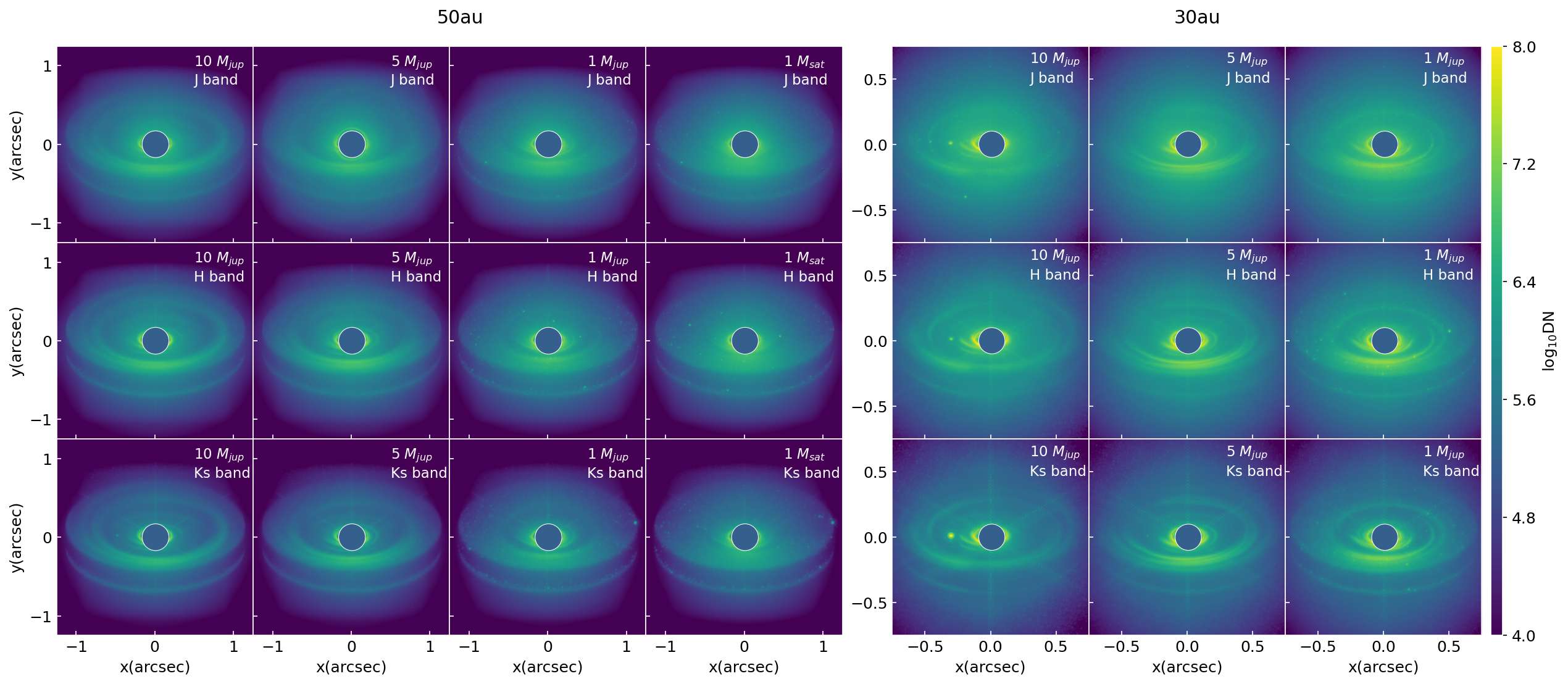}
  \caption{ELT/MICADO synthetic images projected to 100 pc, same as Fig. \ref{fig:micado0} but in 60$^{\circ}$ inclination.}
  \label{fig:micado60}
\end{figure*}

\begin{figure*}
  \includegraphics[width=\textwidth]{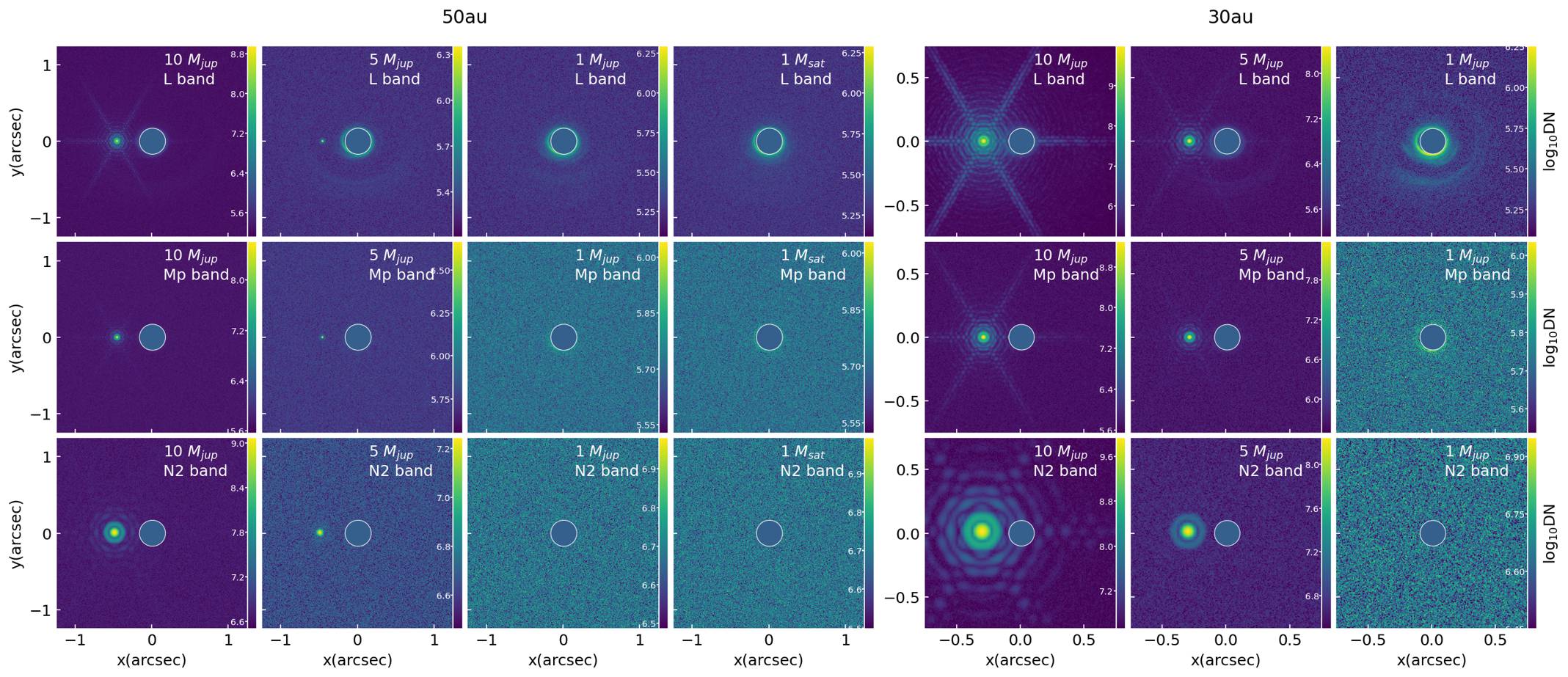}
  \caption{ELT/METIS synthetic images projected to 100 pc, same as Fig. \ref{fig:metis0} but in 30$^{\circ}$ inclination. The columns represent hydrodynamic simulations with different planetary masses (10, 5, 1 $M_{\mathrm{Jup}}$ and 1 $M_{\mathrm{Sat}}$), and the rows represent the 5 filters simulated. The planet is located at 50 AU (\textit{left}) or 30 AU (\textit{right}) from the star at 9 o’clock direction on the images. The central star is masked out with a circle.}
  \label{fig:metis30}
\end{figure*}

\begin{figure*}
  \includegraphics[width=\textwidth]{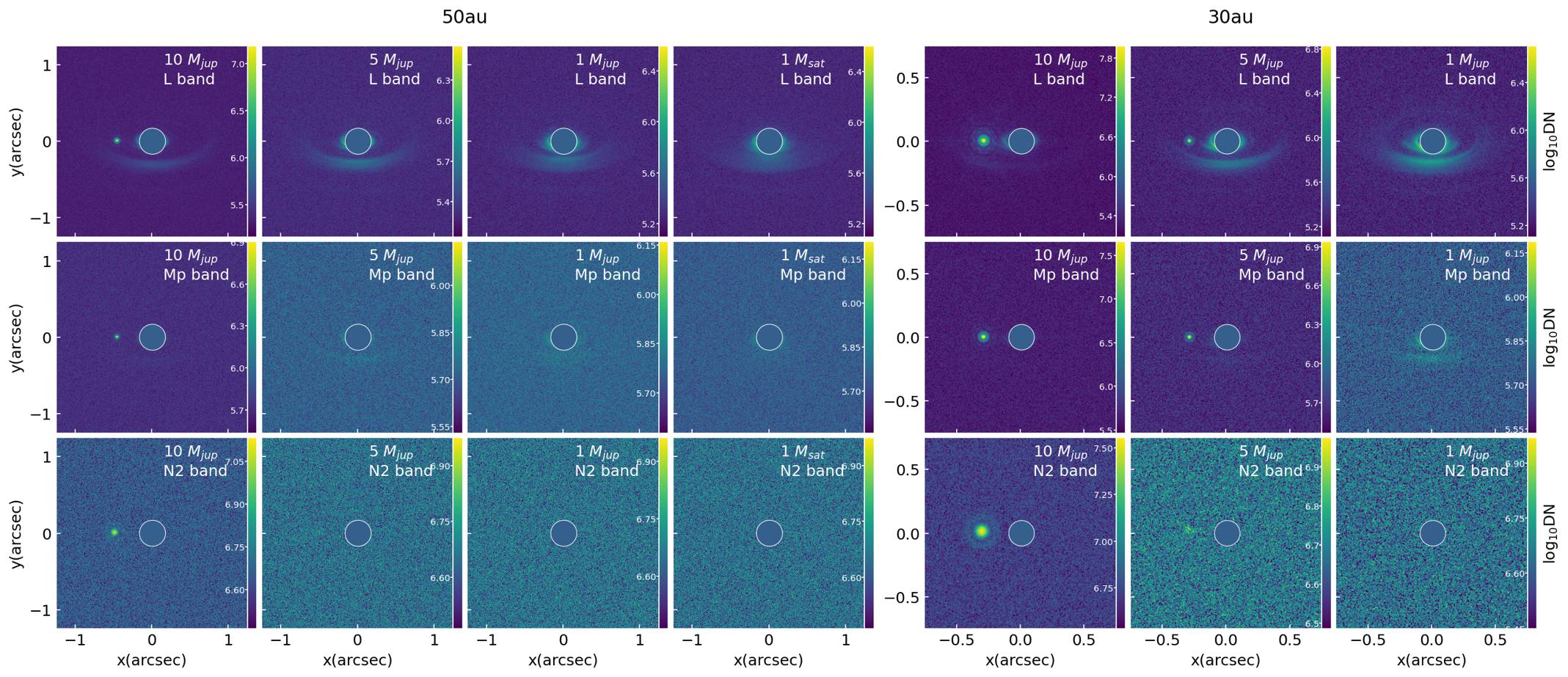}
  \caption{ELT/METIS synthetic images projected to 100 pc, same as Fig. \ref{fig:metis0} but in 60$^{\circ}$ inclination.}
  \label{fig:metis60}
\end{figure*}

Similar to Fig. \ref{fig:nrc0}, Figs. \ref{fig:nrc30}, \ref{fig:nrc60} show the simulated NIRCam images in 30$^{\circ}$ and 60$^{\circ}$ inclinations in the 5 filters for the parameter space studied. The simulated NIRISS AMI images in 30$^{\circ}$ and 60$^{\circ}$ inclinations are showed in Figs. \ref{fig:nrsami30} and \ref{fig:nrsami60}. The simulated MIRI images in 30$^{\circ}$ and 60$^{\circ}$ are shown in Figs. \ref{fig:miri30} and \ref{fig:miri60}. The simulated MICADO images in 30$^{\circ}$ and 60$^{\circ}$ inclinations are shown in Figs. \ref{fig:micado30} and \ref{fig:micado60}. The METIS simulated images in 30$^{\circ}$ and 60$^{\circ}$ inclinations are shown in Figs. \ref{fig:metis30} and \ref{fig:metis60}.

\bsp	
\label{lastpage}
\end{document}